\documentclass[journal=jacsat,manuscript=article]{achemso}

\usepackage[version=3]{mhchem} 
\usepackage{xcolor}
\usepackage{amsmath}
\usepackage{siunitx}
\usepackage{color,soul}
\usepackage{overpic}

\definecolor{DarkGreen}{rgb}{0.56, 0.74, 0.56}

\author{Debarshi Banerjee}
\affiliation[ICTP]{International Centre for Theoretical Physics (ICTP), Strada Costiera 11, 34151 Trieste, Italy}
\alsoaffiliation[SISSA]{Scuola Internazionale Superiore di Studi Avanzati (SISSA), via Bonomea 265, 34136 Trieste, Italy}

\author{Khatereh Azizi}
\affiliation[IPM]{School of Nano Science, Institute for Research in Fundamental Sciences (IPM), Tehran 19395-5531, Iran}
\alsoaffiliation[ICTP]{International Centre for Theoretical Physics (ICTP), Strada Costiera 11, 34151 Trieste, Italy}

\author{Colin K. Egan}
\affiliation[ICTP]{International Centre for Theoretical Physics (ICTP), Strada Costiera 11, 34151 Trieste, Italy}

\author{Edward Danquah Donkor}
\affiliation[ICTP]{International Centre for Theoretical Physics (ICTP), Strada Costiera 11, 34151 Trieste, Italy}
\alsoaffiliation[SISSA]{Scuola Internazionale Superiore di Studi Avanzati (SISSA), via Bonomea 265, 34136 Trieste, Italy}

\author{Cesare Malosso}

\author{Solana Di Pino}
\affiliation[UBAQI]{Departamento de Quimica Inorganica, Analitica y Quimica Fisica/INQUIMAE. Facultad de Ciencias Exactas y Naturales, Universidad de Buenos Aires.  Pabellón II Ciudad Universitaria 1428 Buenos Aires - Argentina}

\author{Gonzalo Díaz Mirón}
\affiliation[ICTP]{International Centre for Theoretical Physics (ICTP), Strada Costiera 11, 34151 Trieste, Italy}

\author{Martina Stella}
\affiliation[ICTP]{International Centre for Theoretical Physics (ICTP), Strada Costiera 11, 34151 Trieste, Italy}

\author{Giulia Sormani}
\affiliation[ICTP]{International Centre for Theoretical Physics (ICTP), Strada Costiera 11, 34151 Trieste, Italy}

\author{Germaine Neza Hozana}
\affiliation[ICTP]{International Centre for Theoretical Physics (ICTP), Strada Costiera 11, 34151 Trieste, Italy}
\alsoaffiliation{Dipartimento di Fisica, Universit\'a degli Studi di Trieste, Via Alfonso Valerio 2, 34127, Trieste, Italy}

\author{Marta Monti}
\affiliation[ICTP]{International Centre for Theoretical Physics (ICTP), Strada Costiera 11, 34151 Trieste, Italy}

\author{Uriel N. Morzan}
\affiliation[UBADF]{Instituto de Fisica de Buenos Aires. Universidad de Buenos Aires. Facultad de Ciencias Exactas y Naturales Pabellón I Ciudad Universitaria 1428 Buenos Aires - Argentina}

\author{Alex Rodriguez}
\affiliation{Dipartimento di Matematica e Geoscienze, Universit\'a degli Studi di Trieste, via Alfonso Valerio 12/1, 34127, Trieste, Italy}
\alsoaffiliation[ICTP]{International Centre for Theoretical Physics (ICTP), Strada Costiera 11, 34151 Trieste, Italy}

\author{Giuseppe Cassone}
\affiliation{Institute for Chemical-Physical Processes, National Research Council (IPCF-CNR), Via S. d'Alcontres 37, 98158 Messina, Italy}

\author{Asja Jelic}
\affiliation[ICTP]{International Centre for Theoretical Physics (ICTP), Strada Costiera 11, 34151 Trieste, Italy}

\author{Damian Scherlis}
\affiliation[UBAQI]{Departamento de Quimica Inorganica, Analitica y Quimica Fisica/INQUIMAE. Facultad de Ciencias Exactas y Naturales, Universidad de Buenos Aires.  Pabellón II Ciudad Universitaria 1428 Buenos Aires - Argentina}

\author{Ali Hassanali}
\affiliation[ICTP]{International Centre for Theoretical Physics (ICTP), Strada Costiera 11, 34151 Trieste, Italy}

\title[]
  {
  Aqueous Solution Chemistry In Silico and the Role of Data Driven Approaches
  }

\begin{document}

\begin{tocentry}

Some journals require a graphical entry for the Table of Contents.
This should be laid out ``print ready'' so that the sizing of the
text is correct.




\end{tocentry}

\begin{abstract}
The use of computer simulations to study the properties of aqueous systems is, today more than ever, an active area of research. In this context, during the last decade there has been a tremendous growth in the use of data-driven approaches to develop more accurate potentials for water as well as to characterize its complexity in chemical and biological contexts. We highlight the progress, giving a historical context, on the path to the development of many-body and reactive potentials to model aqueous chemistry, including the role of machine learning strategies. We focus specifically on conceptual and methodological challenges along the way in performing simulations that seek to tackle problems in modeling the chemistry of aqueous solutions. In conclusion, we summarize our perspectives on the use and integration of advanced data-science techniques to provide chemical insights in physical chemistry and how this will influence computer simulations of aqueous systems in the future.
\end{abstract}

\section{Introduction}

Liquid water is one of the key ingredients for life\cite{ball2008water}. It forms a central lubricant for biological materials and is perhaps the most ubiquitous solvent in physical, chemical, engineering and technological applications\cite{chemrevphotosynthesis2014,chemrevinterfaces2016,HBN_watersplitting}. Many of the unique properties of water arise from its hydrogen-bond network\cite{stillinger1974improved,tet4,stillinger1972molecular,rahman1973hydrogen} and how it changes in different thermodynamic conditions. Dissecting the microscopic structure of water and aqueous solutions in terms of both the static and dynamical properties of hydrogen bonding has been the subject of numerous experimental and theoretical studies and rather lively controversies\cite{gallo2016water,pettersson2018two,ChemRev3}. 


Computer simulations of varying levels of complexity have played an important role in providing molecular-level insights into the structural, dynamical and electronic properties of both bulk water and solutes in aqueous solutions\cite{marx2010aqueous,rscparrinello2014,ChemRev3}. Over the last five decades, the typical models that are used to simulate aqueous systems can be broadly separated into two categories namely molecular mechanics (MM)\cite{Dykstra1993} empirical potentials and first-principles \emph{ab-initio} (AIMD) approaches\cite{tuckerman2005}. While MM based water models provide a powerful way to explore, for example, the phase diagram of water\cite{gallo2016water,tet4} and perform simulations of complex biological systems\cite{scheraga2007,jungwirthtobias2006,laagehynes2017}, they typically do not allow for chemistry to occur. Although being more computationally prohibitive, electronic structure-based AIMD simulations overcome this limitation allowing for modeling chemical reactions where bond breaking or formation occurs. One of the most popular electronic structure approaches for modeling the properties of water has been Density Functional Theory (DFT)\cite{marx2010aqueous,rscparrinello2014,gillanalfemichaelides2016}, although over the last decade more advanced approaches have also been employed, using, for example, quantum Monte-Carlo\cite{zen_ab_2015} and quantum chemistry techniques such as Møller–Plesset perturbation theory (MP2)\cite{joostmp2}.


In both the MM and AIMD approaches to study aqueous systems, there are several challenges that have emerged if one is interested in producing meaningful simulations that can be interpreted and compared with experiments. The first is the quality of the electronic structure theory which ultimately controls the underlying potential energy surface associated with the hydrogen bonds needed to reproduce structural, dynamical and spectroscopic properties of water\cite{gillanalfemichaelides2016,reddy2016accuracy,lambros_how_2020}. Secondly, there is the problem of sampling arising from the fact that it takes a long time for the system to hop over free-energy barriers\cite{bonomi2009plumed,barduccibonomoparrinello2011}. This is particularly true for modeling chemical reactions such as water ionization in different environments\cite{Geissler2001,sirkin2018,DiPino_angewandte2023} with AIMD, as well as exploring the complex phase diagram of water, for example, under super cooled conditions\cite{debenedettiscience2020}. Finally, in many particle systems involving the coupling of both the solute and its aqueous environment, identifying the relevant and important degrees of freedom (commonly referred to as order parameters, reaction coordinates or collective variables) along which physical, chemical or biological processes occur, is far from trivial\cite{Warshel2002,vallsonparrinello2016}.


Herein, we will provide an overview of the most recent advances in the field aimed at overcoming these challenges. Figure 1 provides a schematic outline of our review. In the first part, we begin with a brief historical perspective on the challenges in modeling liquid water from first principles electronic structure simulations. We thereafter discuss the advances made in the development of reactive, many-body and machine-learning (ML) based potentials which are opening new turf in studying the physical chemistry of liquid water (Figure 1a,b). We focus specifically on highlighting important advances that have been made in the applications of these techniques using them to also underscore the challenges in modeling aqueous chemistry in the bulk, at interfaces and under confinement. Furthermore, a critical aspect associated with understanding aqueous chemistry is the determination and exploration of relevant order parameters of reaction coordinates. Due to the collective nature associated with processes involving chemistry in water, identifying these coordinates requires going beyond chemical imagination; we highlight in this context the importance of data-driven approaches (Figure 1c). Finally, we conclude with some perspectives on the future of data-driven approaches in empowering conceptual advances in the chemical physics and physical chemistry of aqueous systems.

\section{Towards Advanced Data Science Techniques}

The present section is essentially divided into three parts, as illustrated in Figure 1. We begin by providing an overview of the main approaches that have been used to simulate aqueous chemistry focusing on DFT, Many-Body and reactive potentials. As indicated in the Introduction, our focus is to highlight key applications related to aqueous chemistry that underscore the challenges. This subsequently serves as motivation for the second part where we focus on recent developments in the use of machine-learning potentials for modeling water and its constituent ions in different contexts. In the last part of this section, we discuss advances and challenges in data-mining for water-chemistry.

\begin{figure}[H]
        \includegraphics[width=16cm]{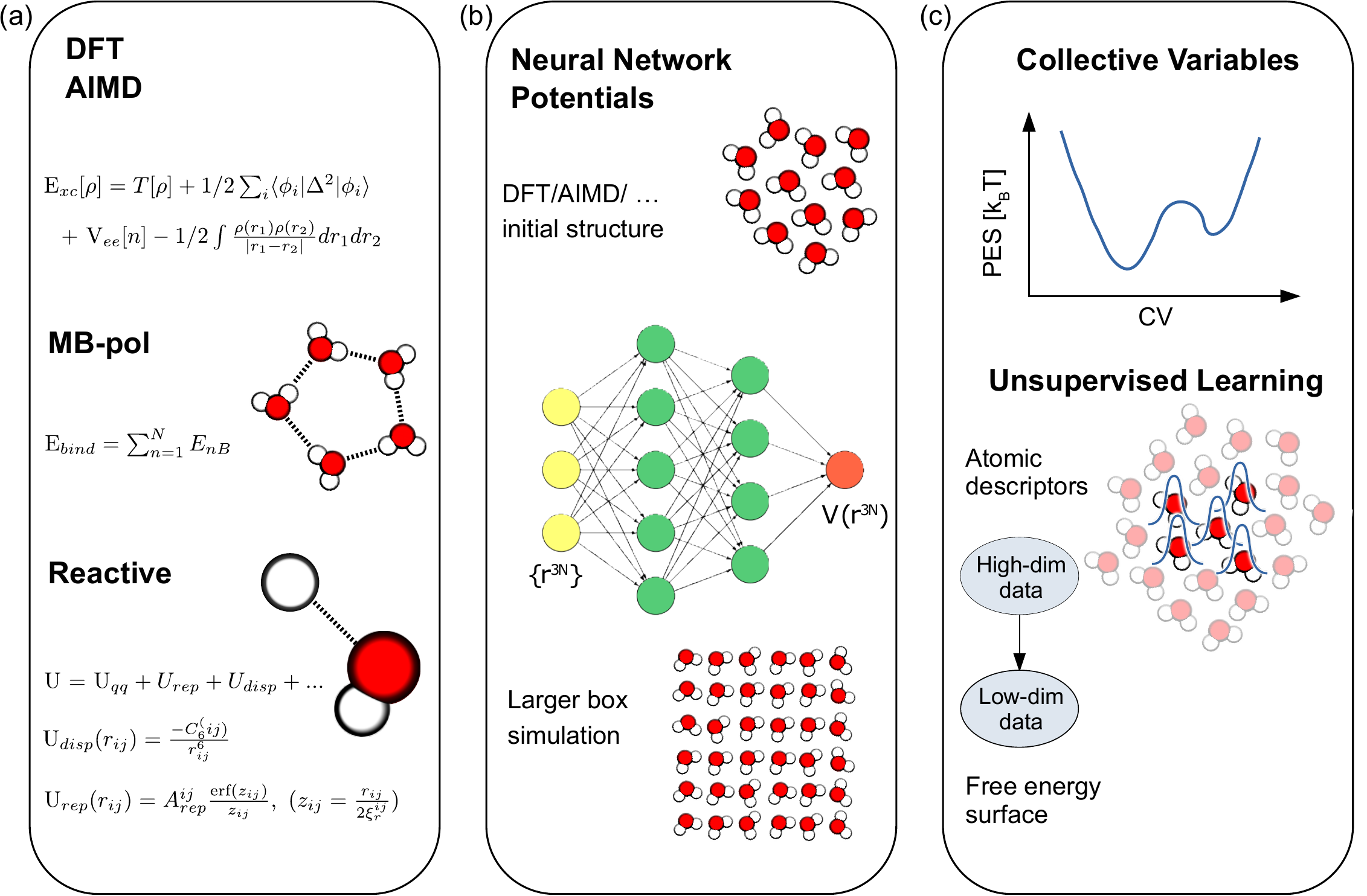}
		\caption{Schematic summary outlining the content of our review. Panel (a) illustrates the main potential energy functions used to simulate aqueous systems discussed in this review. Panel (b) highlights recent developments in the training of neural-network based potentials facilitating larger system sizes and longer simulation times. Panel (c) shows the use of data-driven approaches to enable the interpretation and analysis of simulations where aqueous chemistry occurs.}
		\label{FIG1}
\end{figure}

\subsection{DFT Water, Many-Body and Reactive Potentials}

\subsubsection{DFT Water}

DFT has become one of the most popular electronic structure methods for simulating complex chemical and biological systems from first principles\cite{parryang1995,kohnparr1996} due to its favourable compromise between computational efficiency and accuracy.  The standard way in which one can couple DFT with finite temperature molecular dynamics is using Born-Oppenheimer (BO) or Car-Parrinello (CP) MD. For more details on the theory of DFT, BOMD and CPMD, the interested reader is referred to specific books and other reviews on this subject and references therein\cite{rscparrinello2014,tuckermaniftimieminary2005,marx2000modern}.

DFT-based \emph{ab initio} molecular dynamics simulations (AIMD) of bulk liquid water and its constituent ions, the proton and hydroxide,  offered a first glimpse into the coupling of nuclear and electronic degrees of freedom in water\cite{TuckermanLaasonenSprikParrinello1995,TuckermanLaasonenSprikParrinello1995JCP,TuckermanMarxKleinParrinello1997,MarxTuckermanHutterParrinello1999}. In particular, such techniques have opened up a microscopic window into the celebrated Grotthuss mechanism\cite{Agmon1995} involving the inter-conversion of covalent and hydrogen bonds in the water network. These simulations have since guided the interpretation of many IR and Raman-based spectroscopy experiments of acidic and basic water~\cite{williamtokmakoff2018,fournierbroadband_2018,daly2017,kozari2021}. 

\begin{figure}[H]
\includegraphics[width=0.85\textwidth]{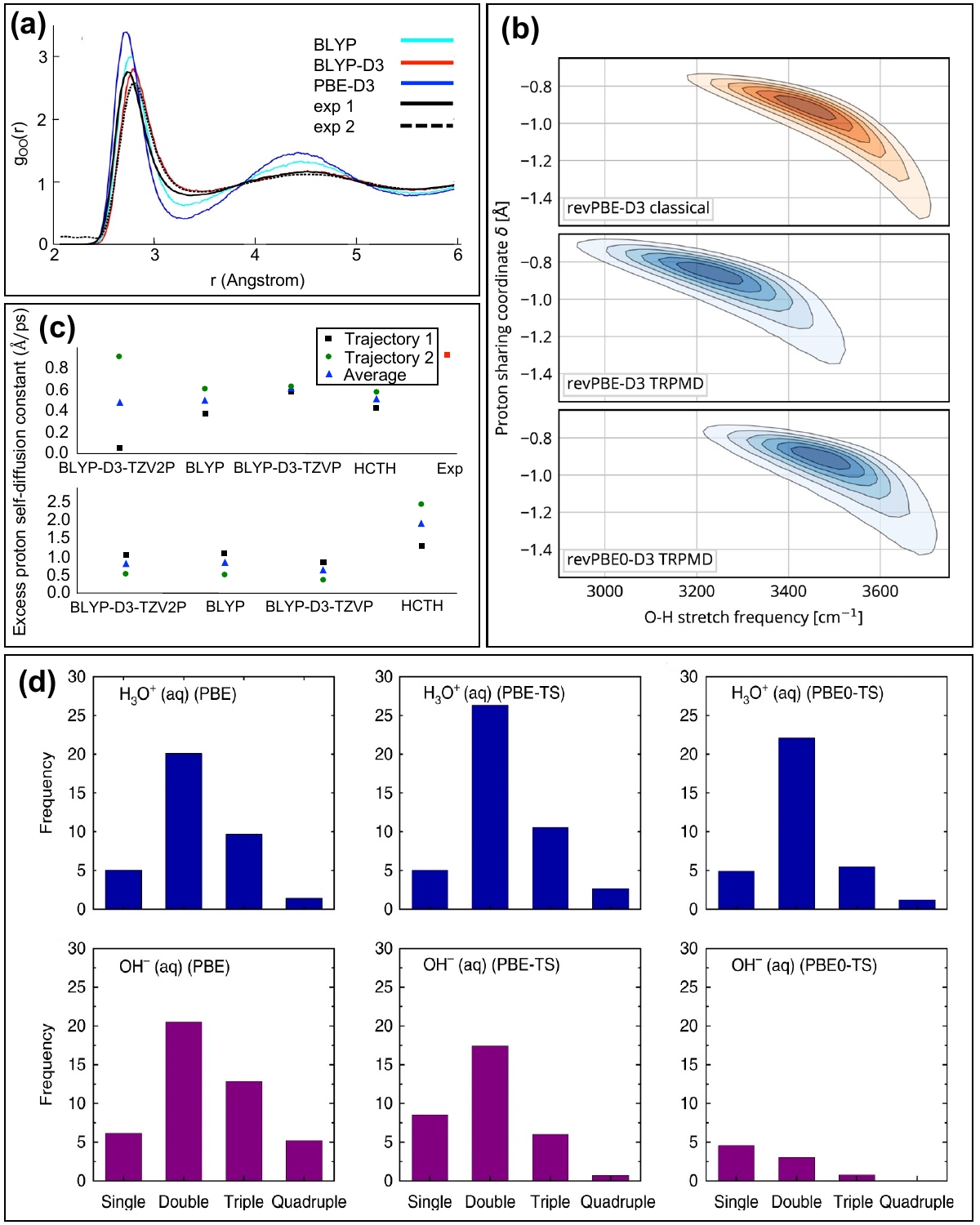}
		\caption{Panel (a) shows the pair-correlation function of bulk water obtained from AIMD simulations with different functionals with and without empirical dispersion corrections compared to experimental predictions\cite{skinner_jcp13,Soper_2007}. This figure is adapted from Reference~\citenum{gillanalfemichaelides2016}. Panel (b) shows the comparison of the high frequency part of the IR spectra of bulk water using different methods with and without the inclusion of nuclear quantum effects. The correlation between these high frequency vibrational modes and the proton delocalization along the hydrogen bonds is shown. This figure is adapted from Reference~\citenum{marselekmarkland2017}. Panel (c) illustrates the sensitivity of the extracted diffusion constant of the excess proton to different methods and simulation protocols. This figure is adapted from Reference~\citenum{tse_analysis_2015}. Panel (d) summarizes the effect of the quality in the electronic structure description on the mechanism of proton or proton-hole jumps in water. This figure is adapted from Reference~\citenum{Chen2018}.}
		\label{FIG2}
\end{figure}

Although DFT is an exact theory, within the Kohn-Sham formalism, the functional form for the exchange-correlation (XC) energy is not known and various approximations need to be made to treat this term. As a result, this poses limits on the accuracy that DFT can achieve for the prediction of electronic structure properties. Numerous studies over the last decades have highlighted how the choice of the XC functional can, in fact, hugely impact the description of the structural and dynamical properties of water.
For a detailed review on the effect of DFT on water in different conditions, the reader is referred to a perspective by Michaelides and co-workers\cite{gillanalfemichaelides2016}. Panel (a) in Figure \ref{FIG2} shows the radial distribution functions (RDF) obtained for two different popular functionals with and without the inclusion of dispersion interaction corrections (e.g., Grimme's semiempirical D3\cite{Grimme2010}), compared to experimental measurements\cite{skinner_jcp13,Soper_2007}. In summary, it can be seen that accounting for dispersion interactions plays an important role in correcting for approximations made in standard generalized-gradient approximation (GGA) functionals, an aspect that has been reinforced in numerous studies\cite{TuckermanLaasonenSprikParrinello1995,vondeleparrinello2004,KuoMundy2004,kuhneparrinello2009}. Specifically, standard GGA functionals yield a much more glassy liquid.  Besides the inclusion of dispersion corrections\cite{gillanalfemichaelides2016,mohan2017,serra2011,chuntavernelliroth2012}, the inclusion of exact Hartree-Fock exchange through the use of hybrid functionals has also been shown to give more accurate estimates of water polarizability\cite{distasio2014}. However, despite the generally favourable scaling of DFT and especially in cases where sophisticated functionals and extra dispersion corrections are employed, the computational cost of the simulations poses severe limits on both the size of the system investigated (relatively small box sizes ($<$2 nm)) and the length of the dynamics (short simulation times (~$\sim$ 100 ps)).



Besides the challenges of dealing with the quality of the electronic structure, another issue in simulating the physical and chemical behavior of water is the high zero-point energy (ZPE) of the O-H covalent bonds which is almost about an order of magnitude larger than thermal energy at room temperature. This is reflected in the O-H pair correlation function of water which tends to be overlocalized in simulations where the nuclei are treated classically. Car and co-workers were the first to elucidate the importance of nuclear quantum effects (NQEs) on the structural properties of the hydrogen bonds in water by using path-integral molecular dynamics (PIMD) together with CPMD simulations\cite{morronecar2008}. Examining the O-H RDF, one observes that in the PIMD simulations, the proton is much more delocalized, leading to a broader first peak consistent with the experiments. 
In this context, path-integral approaches coupled with BOMD or CPMD simulations have become somewhat routine, especially in combination with generalized-Langevin based thermostats\cite{ceriotti2011accelerating,ceriottimarkland2016,Markland2018}. These and other types of simulations that include NQEs have shown that these effects are important for understanding the structural, dynamical and even electronic properties of bulk water as well as chemical and biologically relevant solutes\cite{law2015role,lawhassanali2018,sappati2016nuclear,rossi2016anharmonic,alfredo2016,litman2019elucidating,Ceriotti_Nature2018,Cassone_JPCL2020,Shiga_JCP2021}.

The broader O-H bond length distribution found in PIMD was associated with enhanced transient autoionization events where charged-pairs of waters' constituent ions apparently form due to extreme events involving protons delocalizing along hydrogen bonds\cite{CeriottiCunyParrinelloManolopoulos2013}. The extent of these proton fluctuations are however, very sensitive to the quality in the underlying electronic structure. Specifically, the details of the DFT functional along with dispersion corrections compete with NQEs in highly non-trivial ways, for example, in affecting spectroscopic properties of hydrogen bonds. Figure 2b nicely demonstrates this effect by illustrating how the high frequency modes of the O-H vibrations change as a function of combining hybrid DFT functionals with nuclear quantum dynamics. Marselek and Markland show that standard GGA functionals, even when including dispersion corrections, exaggerate the fraction of these transient autoionization events leading to larger red shifts in the O-H vibrational stretch frequencies\cite{marselekmarkland2017}. Upon using hybrid functionals together with dispersion corrections, the inclusion of quantum effects almost perfectly reproduces the experimental spectra.

The transient autoionization events previously discussed are, of course, the initial seeds for water dissociation that leads to the creation of hydronium (H$_3$O)$^{+}$ and hydroxide (OH)$^{-}$ ions. The preceding issues regarding the choice of the quality in the electronic structure description within the framework of DFT have been shown to play a critical role in affecting structural and dynamical properties of these ionic topological defects. In Figure \ref{FIG2}c, Voth and co-workers perform a systematic analysis on the factors affecting the diffusion constant of the excess proton in water\cite{tse_analysis_2015}. The reported values of the diffusion constant are extremely sensitive to the inclusion of dispersion corrections, the choice of the basis set, density functional employed and, finally, the initial conditions. Furthermore, Car and co-workers recently studied the role of including exact exchange on the diffusion mechanism of the excess proton and hydroxide in liquid water\cite{Chen2018}. Previous studies from some of us, had suggested that the Grotthuss mechanism involves concerted proton hopping events for both the proton and hydroxide ion\cite{HassanaliGibertiCunyKuhneParrinello2013}. Specifically, it was found that protons and proton-holes (hydroxide ions) diffuse along water wires involving concerted double jumps. In Figure \ref{FIG2}d, Car and co-workers demonstrate that while the use of GGA functionals display consistent results for the excess proton compared to the hybrids, the story is much more complicated for the hydroxide ion. In particular, they show that the migration of hydroxide is dominated by a stepwise mechanism rather than concerted hopping. The qualitative difference between the effect of the DFT electronic structure on the proton and hydroxide is rooted in the fact that the use of hybrid functionals along with \emph{ab initio} dispersion corrections sensitively affects the solvent environment of the hydroxide, favoring a hyper-coordinated versus a three-coordinated solvation structure.

\subsubsection{Many-Body Polarizable Potentials}

Although DFT-based MD simulations (discussed above) provide a very practical way to study aqueous solutions and have proven useful in obtaining new insights into mechanisms that cannot be observed experimentally, the quality of the underlying potential energy surface corresponding to widely-used XC functionals does not reach the chemical accuracy required in several applications \cite{gillanalfemichaelides2016,wang2010assessment,howard2015assessing}. For example, even though a specific XC functional may work for bulk water at ambient conditions, studies have shown that there is no guarantee of transferability to other regions of the phase diagram\cite{gillanalfemichaelides2016}. At the same time, due to the computational bottleneck of converging the electronic structure, one is often limited to rather small system sizes and simulation times (on the order of 100 atoms and picoseconds). For example, while the simulations shown in Figure~\ref{FIG2} are certainly commendable, they are limited to timescales on the order of 100s of picoseconds.  Alternatively, one can avoid the explicit calculation of the electronic structure of a system through the use of model potentials, often called molecular mechanics (MM) force fields.  This approach is widely-used due to its low cost compared to DFT, but is often hindered by limits in accuracy and transferability.  

Over the last decade, the Paesani group has made significant strides to fill some of these gaps in DFT and MM approaches through the development of the MB-pol many-body potential for water\cite{babin_v_2013_b,babin_v_2014,medders_g_2014,reddy2016accuracy,paesani_f_2016}.  The idea behind these many-body potentials follows from two major principles: 1) a large difference in the complexity of close-range interactions compared to long-range interactions, 2) the dominance of the first three terms in the many-body expansion (MBE) for aqueous systems.  In brief, long-range interactions such as permanent electrostatics, classical polarization and London dispersion interactions can be accurately accounted for through standard MM-based force field approaches\cite{Dykstra1993,stones_book,jensen2017introduction,cisneros2016modeling,lambros_how_2020}. At close range, interactions become significantly more complicated, including repulsive interactions due to Pauli exclusion, charge penetration effects between overlapping orbitals of different molecules, and charge transfer effects.\cite{stones_book,umeyama1977origin,chen1996energy,mo2000energy,glendening2005natural,van2003testing,khaliullin2007unravelling,khaliullin2009electron,khaliullin2013microscopic,kuhne2014nature,mao2017energy,das2019development,egan2020nature,palos2022assessing}  DFT and other electronic structure methods capture the full extent of all long-range and close-range intermolecular interactions (within the accuracy of their underlying approximations). However, obtaining simple analytic functions which provide an accurate representation is extremely challenging in general.



The Paesani group managed to overcome this difficulty by exploiting the rapid convergence of the MBE in aqueous systems.  The total binding energy of a system of $N$ particles (e.g. water molecules) can be written as a MBE involving various $n$-body terms:\cite{hankins1970water}
\begin{equation} \label{eq:mbe_energy}
\begin{split}
E_{\text{bind}}
&= E_{\text{1B}} + E_{\text{2B}} + E_{\text{3B}} + \cdots + E_{\text{NB}} \\
&= \sum_{n=1}^N E_{\text{nB}}
\end{split}
\end{equation}

In the context of many-body interactions in water, $E_{\text{1B}}$ corresponds to the sum of distortion (stretching and bending) energies of individual water molecules, $E_{\text{2B}}$ corresponds to interactions between pairs of water molecules (e.g.\ hydrogen bonding), and each $E_{\text{nB}}$ term for $n > 2$ corresponds to (excess) polarization effects (including charge transfer) between $n$ water molecules.

Typical rigid, nonpolarizable MM force fields\cite{mark2001structure,zielkiewicz2005structural,cisneros2016modeling} such as the SPC\cite{berendsen1987missing,berweger1995force} and TIP\cite{jorgensen1983comparison,mahoney2000five,horn2004development} families of models only include explicit 2B potentials, often in the form of the Lennard-Jones potential describing long-range dispersion and close-range repulsive interactions, and pairwise Coulomb potentials capturing permanent electrostatic interactions. Although classical induction can be added on top of pairwise potentials to include higher-body interactions, they are inadequate for capturing the reorganization of electronic structure that occurs at close-range. A critical improvement to classical potentials can be made by recognizing that the MBE for water-water interactions converges rapidly, with the first three terms in Eq.~\ref{eq:mbe_energy} containing the vast majority of the total binding energy.\cite{xantheas1994ab,cui2006theoretical,gora2011interaction,heindel2020many}  Building on earlier ideas used in the development of, for example, the CC-pol\cite{bukowski2007predictions,bukowski2008polarizable,cencek2008accurate,gora2014predictions,jankowski2015ab} and WHBB\cite{huang2006ab,wang2009full,wang2011flexible,babin2012toward} families of models for interactions within gas phase water clusters, the Paesani group combined a long-range potential (consisting of dispersion, permanent electrostatics and classical induction) with highly-accurate models for close-range 2B and 3B interactions fitted to coupled-cluster (CCSD(T)\cite{purvis_gd_1982,raghavachari_k_1989} and CCSD(T)-F12b\cite{adler_tb_2007,knizia_g_2009}) energies.  The resulting model, MB-pol, is able to reproduce thermodynamic and dynamical properties of condensed phase water across much of the phase diagram as discussed below. There is, of course, a computational cost associated with the use of many-body potentials, lying around one order of magnitude above classical polarizable potentials, but several orders of magnitude below AIMD\cite{reddy2016accuracy}.

\begin{figure}[H]
        \includegraphics[width=0.85\textwidth]{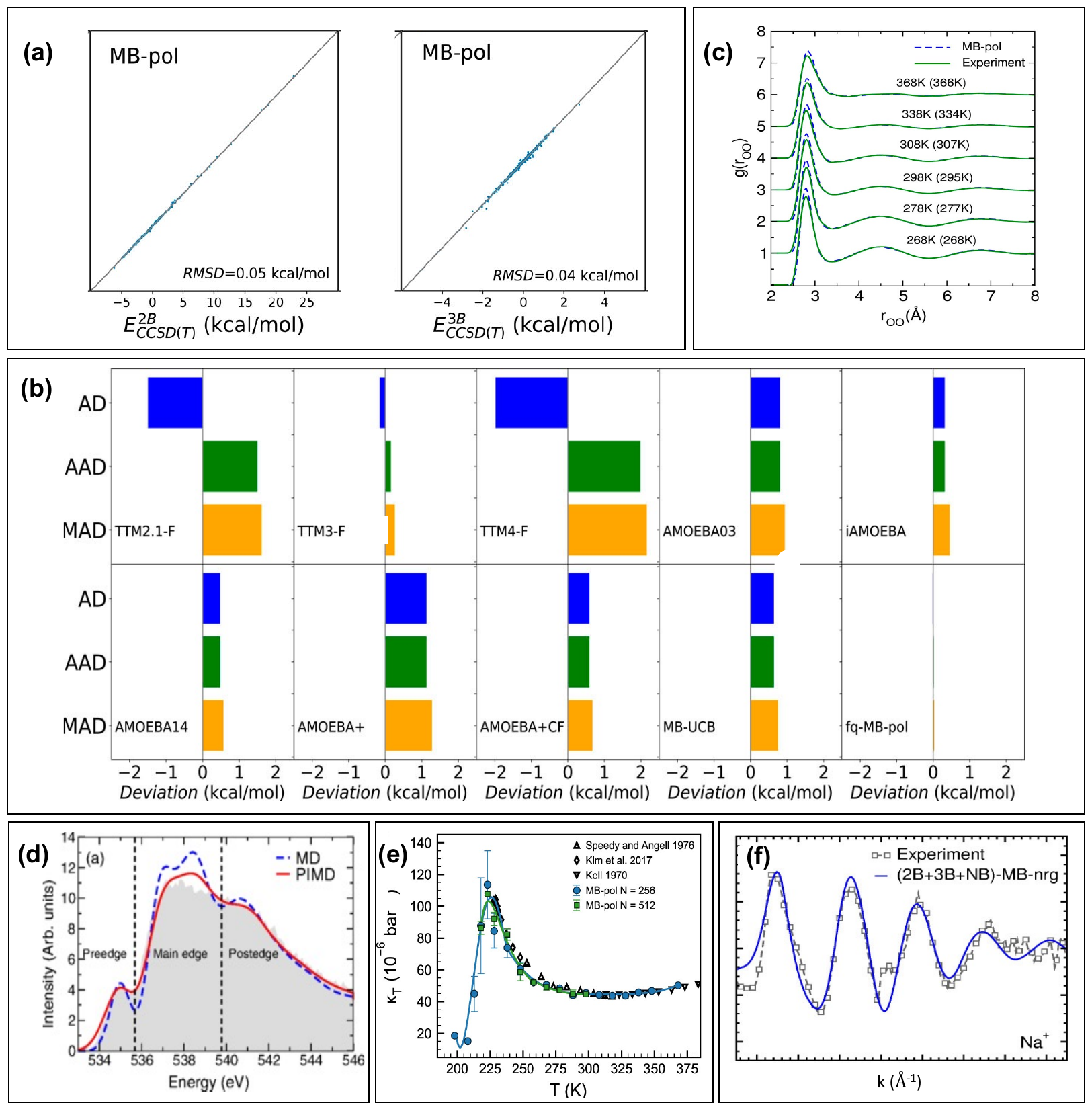}
		\caption{Panel (a) shows the comparison of the 2-body and 3-body energies obtained with the MB-pol potential and coupled-cluster theory (CCSD(T)) computed on water dimers and trimers. Panel (b) shows the average (AD), average absolute (AAD) and maximum absolute (MAD) deviations of the \emph{total} binding energy per water molecule calculated with a number of classical potentials relative to those calculated with MB-pol for configurations obtained from a molecular dynamics simulation of MB-pol at room temperature. Both panels (a) and (b) are reproduced from Reference~\citenum{lambros_how_2020}. Panel (c) shows the pair-correlation function (O-O radial distribution function) across a range of temperatures obtained with the MB-pol model from molecular dynamics simulations. This panel was reproduced from reference~\citenum{reddy2016accuracy}. Panel (d) shows the X-ray absorption spectrum computed with many-body perturbation theory on samples of trajectories with the MB-pol potential with (PIMD) and without (MD) the inclusion of nuclear quantum effects. This figure is reproduced from Reference~\citenum{sun2018electron}. Panel (e) shows the isothermal compressibility of water across a range of temperatures into the supercooled regime obtained with MB-pol and compared with experiments. This figure is reproduced from Reference~\citenum{gartner2022anomalies}. Panel (f) shows the K-edge EXAFS spectrum of hydrated sodium ion compared to that obtained with a new generation of MB-pol extended to include ions. This figure is reproduced from Reference~\citenum{zhuang2022hydration}.}
		\label{FIG3}
\end{figure}


For neutral bulk water, MB-pol has proven to be among  the most accurate potentials currently available,\cite{reddy2016accuracy,nguyen2018comparison,lambros_how_2020,zhu2023mb} and extensions such as MB-nrg potentials\cite{bajaj2016toward,riera_m_2017,paesani2019chemical,bizzarro2019nature,zhuang2019many,egan2020nature,caruso2021data,caruso2022accurate,zhuang2022hydration} for modelling ions in solution also display comparable accuracy. Comparisons between MB-pol and CCSD(T) for 2- and 3-body energies of clusters taken from MD simulations demonstrates the quality of the model potential (Figure~\ref{FIG3}a-b). The RMSD obtained for these two energies are 10 times smaller than thermal energy at room temperature. It is worth noting that errors obtained with classical polarizable models such as TTM and AMOEBA have been reported to range between up to 2-to-3 times larger than thermal energy\cite{lambros_how_2020}. This feature is manifested even at higher temperature and in the condensed phase of liquid water.  Note that while low-order many-body energies can be seen as gas phase properties, evaluating the model accuracy on dimers and trimers \emph{extracted from MD simulations} allows for validation of condensed phase energetics, given the rapid convergence of the MBE in water.  As an illustration of this point, Figure~\ref{FIG3}b shows the average (AD), average absolute (AAD) and maximum absolute (MAD) deviations of the \emph{total} binding energy per water molecule calculated with a number of classical potentials relative to those calculated with MB-pol for configurations obtained from a molecular dynamics simulation of MB-pol at 300K\cite{lambros_how_2020}. 

Figure 3c shows the pair-correlation function of MB-pol liquid water across a range of temperatures compared to experiments. Although there is a slight tendency for MB-pol to overestimate the height of the first peak, very likely due to the fact that these simulations were conducted without nuclear quantum effects, there is a consistency in the distributions from the supercooled regime to below the boiling temperature\cite{reddy2016accuracy}.  Another key experimental probe for the structure of liquid water is X-ray absorption spectroscopy (XAS).\cite{wernet2004structure,fransson2016x}  In Reference~\citenum{sun2018electron} it was shown that Bethe-Salpeter electronic structure calculations on top of approximate quantum dynamics with the MB-pol potential correctly reproduces the experimentally measured XAS, further validating the liquid structure resulting from the model potential.  In particular, the calculations reproduce the maximum amplitude of the pre-edge feature at 533 eV (although slightly overestimating its width) indicating that MB-pol accurately captures the experimentally measured disorder in the hydrogen bonding network (see Figure~\ref{FIG3}d).  It is important to note that condensed phase structure and thermodynamics were not directly included in the development of MB-pol, so accuracy in reproducing these properties reflects the importance of close-range low-order many-body effects in aqueous systems.

As alluded to earlier in the review, water is characterized by different types of anomalies which are not found in simple liquids. These include for example, the density maximum at 4$^\circ$ C\cite{kell1967precise} and the minimum in isothermal compressibility at 46.5$^\circ$ C\cite{skinner2014structure}. The MB-pol model has been successful at reproducing many of these anomalies where standard DFT-based methods at the GGA level of accuracy typically fail~\cite{Gillan_JCP2016} and effective MM potentials have experienced significant challenges. In part, this originates from the lack of transferability of effective MM potentials as they are often parameterized to work at a specific region of the phase diagram\cite{mahoney2001quantum}. 
In Figure 3e we show the isothermal compressibility of liquid water from ambient temperatures to the supercooled regime, comparing MB-pol to experimental measurements.  The agreement is extremely impressive.  In addition to the well-established minimum at 46.5$^\circ$ C, MB-pol also reproduces the value of the measured compressibility at the tentative maximum at around -44$^o$ C (albeit with large error bars due to long relaxation times near the Widom line).\cite{kim2017maxima,gartner2022anomalies} The quality of the potential energy surface sampled by MB-pol is clearly superior to the vast majority of MM models and commonly used XC functionals on the market.

Among the many interesting properties of water, perhaps the most challenging is determining spectroscopic properties both in the bulk and at interfaces.  Specifically, vibrational spectra in liquid water are sensitive to both the details of microscopic structure, including properties of the hydrogen bonding network, as well as the potential energy surface along specific vibrational modes.  Early studies validating the MB-pol potential demonstrated that the model successfully reproduces experimentally measured features of the bulk vibrational (infrared and Raman) spectra, within the limits of the approximate quantum dynamics method used, demonstrating that liquid water simulations with MB-pol produce accurate hydrogen bond dynamics.\cite{medders2015infrared,medders2015interplay} 
Additionally, the ability of MB-pol to reproduce the features of the experimental SFG spectrum of the air-water interface (within the limits of the approximate quantum dynamics used) is a noteworthy achievement.  In particular, the MB-pol calculations predicted the lack of a positive feature around 3000 cm$^{-1}$ prior to the correction of experimental artifacts between 2011 and 2015.\cite{medders2016dissecting}  These examples further show that the bottom up approach involving converging the MBE expansion ensures the transferability of the potential. 



Finally, we would like to showcase recent extensions of MB-pol (MB-nrg) to go beyond bulk water and allow for modeling solutions such as ions in water.\cite{bajaj2016toward,riera_m_2017,paesani2019chemical,bizzarro2019nature,zhuang2019many,egan2020nature,caruso2021data,caruso2022accurate,zhuang2022hydration} An instructive example is the dilute aqueous Na$^+$ solution, for which pairwise classical potentials tend to over-structure the solvation shells around Na$^+$, while classical polarizable models under-structure the Na$^+$ solvation, predicting an almost continuous transition between the first and second shells\cite{zhuang2022hydration}.  The accurate description of close-range quantum mechanical effects, such as charge transfer and charge penetration, at the 2B and 3B levels provided by a full many-body potential\cite{egan2020nature} allows MB-nrg to predict the correct solvation structure, and thus reproduce the experimentally measured extended X-ray absorption fine structure (EXAFS) spectrum with unparalleled accuracy\cite{zhuang2022hydration}. Figure 3f compares the K-edge EXAFS spectra for water solvating the Na$^+$ ion obtained from MB-nrg and experiments. Interestingly, in the case of Na$^+$ solvation, it was found that the MB-nrg potential without 3B corrections produced a solvation structure similar to the classical polarizable potential. EXAFS spectra computed for solutions of halides and other alkali ions using the MB-nrg potentials are presented in several references, demonstrating the applicability of the approach to each system.\cite{zhuang2019many,caruso2021data,caruso2022accurate,zhuang2022hydration} An MB-nrg model for treating solvated hydrocarbons such as methane, has also been recently reported.\cite{riera2020data,robinson2022behavior}  Inspired by both experiments\cite{pruteanu2017immiscible} and previous DFT-based AIMD simulations\cite{pruteanu2020squeezing} it was shown that under pressure, methane picks up a dipole moment which leads to its enhanced solubility. In this case, the use of MB-nrg potentials allows for running larger and longer simulations than the far more expensive AIMD approaches permit, offering the possibility of obtaining accurate predictions on the corresponding thermodynamics\cite{robinson2022behavior}.

\subsubsection{Reactive Empirical Potentials}

Although many-body polarizable potentials, such as MB-pol, allow for an extremely accurate characterization of the structural, dynamical and spectroscopic properties of bulk water, they do not allow for bond-breaking and bond-formation. One of the most fundamental processes in aqueous chemistry that requires chemistry to occur in water is the dissociation of water into its constituent ions, proton and hydroxide\cite{eigen1958self,eigen1964proton}. This equilibrium determines the pH of water and has numerous implications for the biochemistry of biomolecules in solution. A lot of our understanding of water dissociation and the diffusion of protons and hydroxide ions in solution has come from molecular simulations that allow for reactive chemistry to occur. In this regard, DFT based AIMD simulations of liquid water and its ionic products have been instrumental over the last three decades in providing a molecular lens into the structure of the excess proton in water. 
These simulations have guided and motivated state-of-the-art spectroscopy experiments\cite{tokmakoff2015,williamtokmakoff2018}. As alluded to earlier, for chemical reactions such as those involving the breaking of covalent bonds, the standard DFT simulations are hindered by short timescales and small system sizes. 

\begin{figure}[H]
        \includegraphics[width=\textwidth]{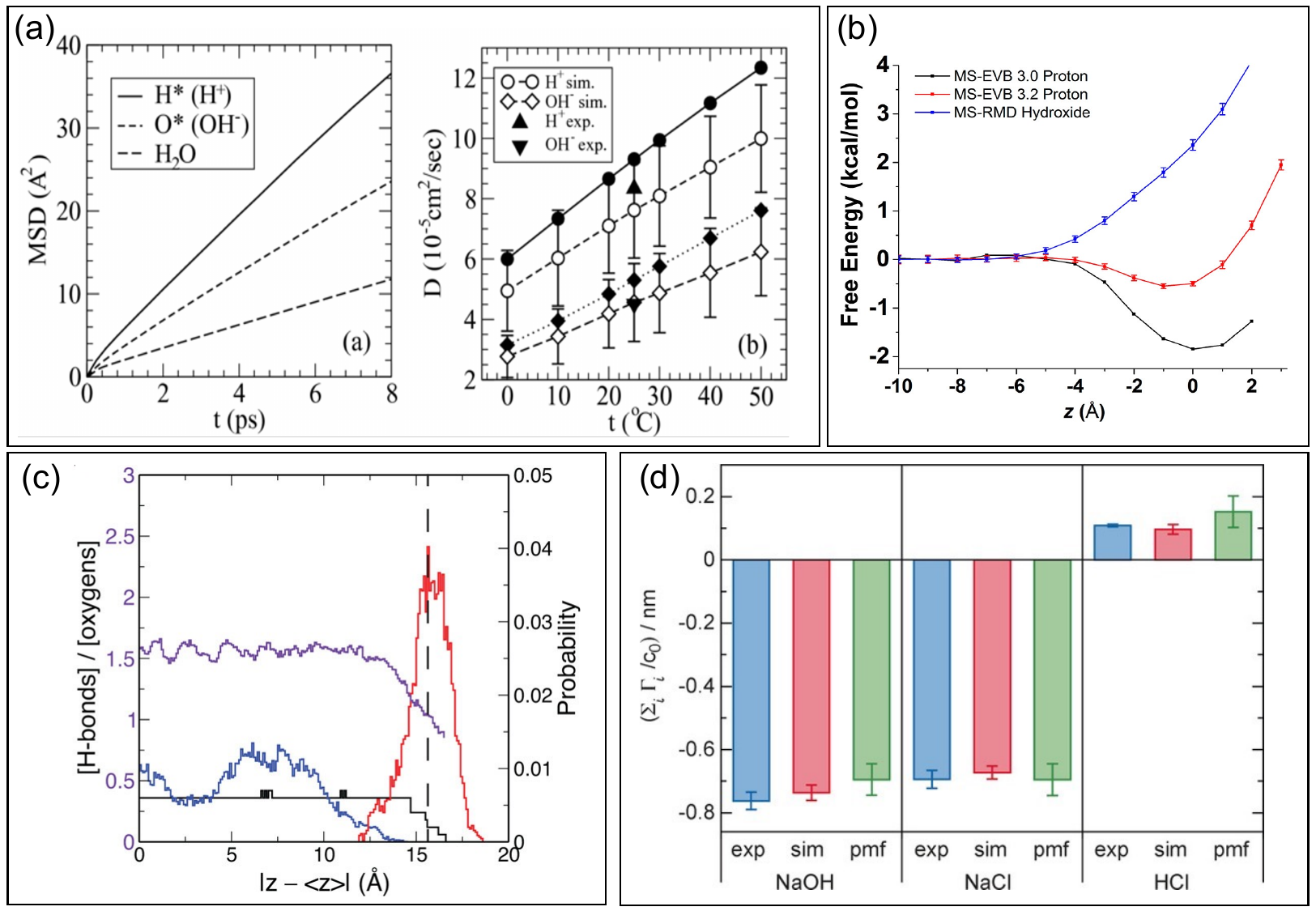}
		\caption{Panel (a) shows the mean square displacement (MSD) of water and its constituent ions obtained with the OSS2 potential in the left panel and the estimates of the diffusion constant of the proton and hydroxide at a range of temperatures, in the right panel. This figure is reproduced from Reference~\citenum{lee_proton_2011}. Panel (b) shows the potentials of mean force for the excess proton and hydroxide ion to bind to the surface of water obtained using MS-EVB simulations. This figure is reproduced from Reference~\citenum{TseLindbergKumarVoth2015}. Panel (c) shows probability distributions of the water (violet), the proton (blue) and hydroxide ion (red) obtained from MD simulations using LEWIS based potentials. This figure is reproduced from Reference~\citenum{baihertzfeld2016}. Panel (d) compares the surface activities of different salt solutions obtained with three different methods: experiments, equilibrium molecular simulations and free energy calculations (pmf). This figure is reproduced from Reference~\citenum{jansjavkatnetz2017}.}
		\label{FIG4}
\end{figure}

Besides the ML-based force fields designed to overcome these challenges that will be discussed in the next section, there are also empirical potentials that have been developed to study water dissociation. Attempts to construct dissociative water potentials date back to work almost four decades ago by Stillinger and Rahman\cite{stillinger_computer_1981,stillinger_polarization_1982,stillinger_polarization_2008,halley_polarizable_1993}. These initial studies laid the groundwork for one of the most popular dissociative water potentials, that describes both intramolecular and intermolecular interactions associated with water and its constituent ions, namely the OSS family\cite{larssinger1998}. The functional forms of these potentials include 2-body radial, 3-body angular contributions, and polarizable oxygen ions. The first form of the OSS potential was shown to be rather promising at characterizing potential energy surfaces in protonated clusters. Over the last two decades it has been systematically improved to allow for modeling proton and hydroxide transfer in the bulk\cite{LeeRasiah2010,lee_proton_2011,lee_note_2013}. The left panel of Figure \ref{FIG4}a illustrates the mean square displacement (MSD) and, thus, the inferred diffusion constants for the excess proton, hydroxide, and neutral water, obtained from the OSS2 potential by Rasaiah and co-workers\cite{lee_proton_2011}. It is first important to note that the potential successfully predicts the relative differences in the diffusion constants for the three species, namely that D$_{H^+}$ $>$  D$_{OH^-}$ $>$ D$_{H_2O}$, consistent with experiments. Instead, the right panel of Figure 4a shows the magnitude of the diffusion constants for the proton and hydroxide, at different temperatures again compared with experimental results. While the trends in the change of the diffusion constants as a function of temperature are consistently reproduced by the model, the differences appear to get more pronounced at larger temperatures. Being able to quantitatively predict these types of dynamical properties of water's constituent ions at different thermodynamic conditions remains an open challenge.

Another family of dissociative potentials that have caught some momentum in the literature are those developed by Garofalini and co-workers to deal with water dissociation in the bulk and near inorganic metal oxide interfaces such as silica\cite{mahadevangarofalini2007}. Similar to the OSS potential, water dissociation is facilitated by incorporating intramolecular interactions, which encompass 2 and 3-body terms. This approach has been employed in recent studies to investigate the mechanisms involved in water dissociation. Garofalini and co-workers show that liquid water is characterized by a larger fraction of transient autoionization events, similar to previous studies where this was observed upon inclusion of nuclear quantum effects\cite{Garofalini_JPCB2023}. They also point out the challenge in identifying the relevant collective variables that ultimately lead to separation of the hydronium and hydroxide, a topic that will be discussed at greater length later in the review. Over the last decade, Wiedemair and co-workers have extended the original potential proposed by Garofalini, in an effort to improve various dynamical properties by replacing the non-Coulombic Morse-like potential with either a Lennard-Jones or Buckhingham potential, allowing for it to be of greater practical use in other common MD softwares\cite{wiedamair2017}. 



In the last two decades, the Voth group has pioneered the development of reactive potentials based on the multi-state empirical valence bond formalism (MS-EVB). The EVB-type potentials have been typically trained with different levels of DFT simulations and have allowed for examining problems such as proton transfer in bulk liquid water with impressive detail and accuracy\cite{PaveseChawlaLuLobaughVoth1997,PaveseVoth1998,LapidAgmonPetersenVoth2005,knight2012curious,WuChenWangPaesaniVoth2008,markovitch2008special}. Since this has already been tackled in several previous reviews\cite{knight2012curious,MarxChandraTuckerman2010}, here we focus on highlighting a specific problem that has been the subject of raging debate and regards the propensity of waters constituent ions for the air-water interface and the apparent negative charge near hydrophobic interfaces\cite{BeattieDjerdjevWarr2009,poli2020charge}. Figure \ref{FIG4}b illustrates potential of mean force (PMF) calculations performed by the Voth group, which show that the excess proton has a slight propensity for the surface of water while the hydroxide ion is repelled\cite{TseLindbergKumarVoth2015,voth2020}. On the other hand, previous studies using DFT-based AIMD simulations reached different conclusions regarding the propensity of the hydronium and hydroxide ions for the surface of water -- the latter, hydroxide ion, is instead found to be weakly attracted while the proton has no affinity for the air-water interface\cite{Mundy20092,baermundy2014}. Such a dual -- acidic or basic -- behavior of the surface of water has also recently been the subject of a large-scale QM/MM study, attempting at resolving the controversial finding that the excess proton presents a higher affinity for the surface compared to the hydroxide~\cite{Rashid_PCCP23}. The origin of this behavior is attributed to differences in the local solvation structures which are challenging to sample with standard DFT based simulations. 

The Hertzfeld group have also developed LEWIS based potentials, which include valence electrons as semi-classical particles interacting with each other through pairwise potentials\cite{baihertzfeld2016,baihertzfeld2017}. The LEWIS model tends to predict a much higher affinity of hydroxide ions for the surface of water compared to both MS-EVB and DFT based simulations (see Figure \ref{FIG4}c)\cite{TseLindbergKumarVoth2015,MundyKuoTuckermanLeeTobias2009}.  Besides, potentials from the Netz group, by including several extra bond-stretching and angle potentials, as well as altering the point charges, have been interfaced with the SPC/E water model. While this does not allow for studying the Grotthuss mechanism, thermodynamic properties such as solvation energies and surface activities are accurately reproduced\cite{jansjavkatnetz2017}. This is illustrated in Figure \ref{FIG4}d which compares the surface activities extracted from experiments, simulations and through PMF calculations, for HCl, NaOH and NaCl. One observes that the proton is surface active, while the hydroxide is not, consistent with the predictions of the PMFs by Voth shown in Figure 4c. Needless to say, it is clear that this is an extremely challenging problem and, depending on the flavor of the underlying potential that is used, can lead to very different qualitative and quantitative conclusions.

Finally, another dissociative potential of relevance is the ReaxFF force-field developed by van Duin, Goddard III and co-workers~\cite{reaxff_2001, reaxff_2008, reaxff_nature_review}. This is a bond-order based force field, typically employed for molecular dynamics simulations involving chemical reactions. Conventional force fields face limitations in representing chemical reactions due to their explicit bond definition requirements. In contrast, ReaxFF employs bond orders rather than explicit bonds, enabling continuous bond formation and breaking, making it more suitable for modeling chemical reactions. 
The first-generation ReaxFF water force field (water-2010) underestimated the bulk density of liquid water by $\sim8\%$ at ambient conditions\cite{vanDuin2013} (an issue also occurring with many DFT XC functionals), but also incorrectly predicted the order of the diffusion coefficients as: D(H$_2$O) $<$ D(H$_3$O$^+$) $<$ D(OH$^-$)\cite{reaxff_diffusion_book}.

The second-generation of ReaxFF force field \cite{reaxff2}(water-2017) was developed by fitting energies and structures to QM data (as in water-2010) and, in addition, by explicitly including both the experimental density of water and the correct diffusion coefficients of H$_2$O, H$_3$O$^+$, OH$^-$ in the training procedure. This resolved the aforementioned issues and resulted in a more accurate density of bulk water, as well as the correct order of the diffusion coefficients: D(H$_2$O) $<$ D(OH$^-$) $<$ D(H$_3$O$^+$). This force field  has also been shown to describe water dissociation and the Grotthuss mechanism underlying the propagation of protons. However, in comparison to water-2010, the water-2017 force field slightly misjudges the location and intensity of the first peak in the O-O and H-O RDFs, even though they are comparable to what is predicted by common classical water models, such as SPC/E, TIP3P, and TIP4P-2005\cite{reaxff2}. It was hypothesized that this was caused by including the density of bulk water in the training data, which might have resulted in an overestimation of the intermolecular interactions. The reactive nature of this force field has allowed it to be used to simulate water dissociation in liquid-solid interfaces\cite{reaxff_solid_dissoc}. 



\subsection{Aqueous Chemistry with Machine-Learning Force Fields}

One of the challenges with developing accurate potential energy surfaces (PES) for use in empirical potentials like the ones described earlier is the construction of an appropriate functional form for the interacting particles. The reactive potentials discussed in the previous section can often have very complicated functional forms. 
{This has triggered the use of models for which the knowledge of the functional form is not needed, with Neural-Networks (NN) being a natural choice, as they are universal approximators of any mathematical function\cite{HORNIK1989}. Therefore, in principle, by properly optimizing the internal weights of the NN,} one can fit any PES as a function of the atomic positions without having to explicitly separate out the reactive or bonded/non-bonded nature of the interactions. 

The vast majority of NN potentials for water used in molecular dynamics applications builds on the seminal work by Behler and Parrinello\cite{behler2007generalized} who developed a generalized NN representation of high-dimensional PES that was trained on DFT data. The reader is referred to several other detailed reviews on the topic\cite{gastegger2020molecular, Noe_2020, Behler_2022}. Here, we outline the essential principles of the method. The idea is to express the total energy of a system as a sum of individual atomic contributions $E_{i}$ which depend on the local chemical environment. Each atom is dressed with its own NN so that the total energy of the system is given by: $E = \sum_{i=1}^N E_{i}$. In order to be physically meaningful, each $E_i$ must be invariant under the permutation of identical atoms in the local environment and under the rotation/translation of their coordinates. However, learning these symmetries (although theoretically possible) would be extremely costly since the algorithm would need a lot of data points during the training. Therefore, it is a common practice to employ some kind of symmetry functions ($G_{i}$) that encode the local atomic structure and satisfy these symmetry conditions, providing a direct link between the atomic energies $E_{i}$ and the local atomic environments. Specifically, in the Behler-Parrinello NN (BPNN) framework, the atomic coordinates are mapped onto a set of two- and three-body symmetry functions.

The structure of the atomic NN shown in Figure~\ref{FIG1}b is inspired by the architecture of artificial neurons which consists of nodes organized in various layers. The symmetry function coordinates described earlier are used as input for the first layer, while the last layer produces as output the atomic energy. The weights of the connecting nodes are optimized to reduce the error of these energies with respect to electronic structure calculations\cite{Behler_2022}.

Several versions of neural network potentials of water have been developed over the last decade which began by originally fitting the potential energy surface of the water dimer based on environment-dependent atomic energies and charges\cite{morawietzbehler2012}. Over time, this potential was refined to handle water clusters highlighting the importance of dispersion interactions\cite{morawietzbehler2013}.  These NN potentials clearly allow for the simulation of systems over much longer timescales. Naturally, however, the quality of the potential depends on where exactly the training data set comes from. For example, training an NN potential using DFT-based molecular dynamics without dispersion corrections leads to overstructured and glassy water dynamics\cite{morawietzbehler2016}. 

\begin{figure}[H]
        \includegraphics[width=\textwidth]{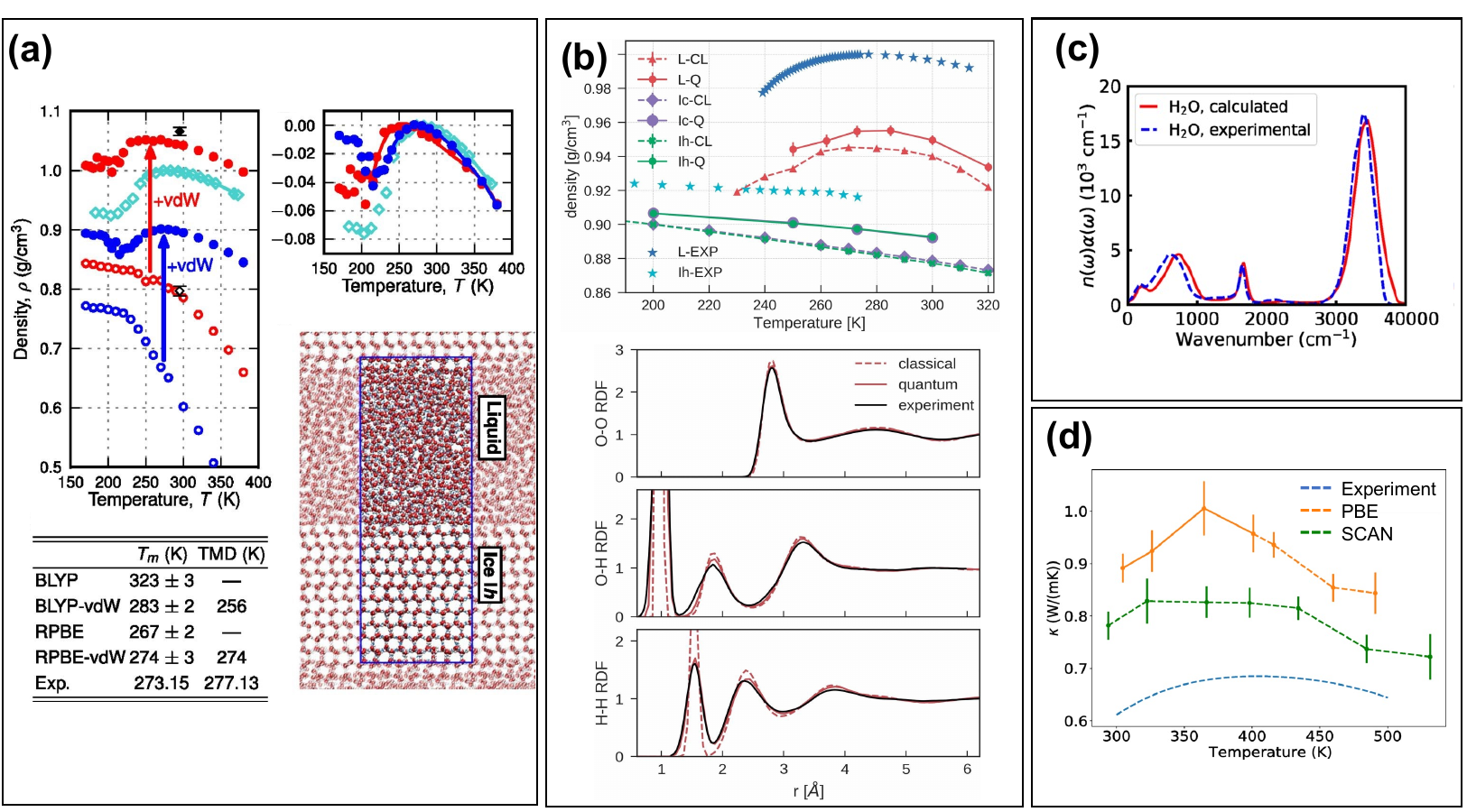}
		\caption{Panel (a) shows the outcome of examining the co-existence of the liquid and solid phase of water using a NN potential trained with different levels of theory, details which can be found in the text of the review. This figure is reproduced from Reference~\citenum{morawietzbehler2016}. Panel (b) shows the density-temperature map of water as well as two phases of ice comparing classical and quantum simulations again with a NN potential. Also shown are the pair-correlation functions of liquid water. This figure is reproduced from Reference~\citenum{chengceriottibehler2019}. Panel (c) compares the high-frequency Raman spectra of liquid water computed from a DNN to that obtained from experiments. This figure is reproduced from Reference~\citenum{zhangcar2020}. Panel (d) shows the thermal conductivity obtained from different DNN simulations compared to that seen in experiments. This figure is reproduced from Reference~\citenum{Baroni2021}.}
		\label{FIG5}
\end{figure}

Figure~\ref{FIG5}a illustrates a powerful example of using different types of NN potentials to study the behavior of the density maximum of liquid water, as well as the thermodynamics associated with the ice-water equilibrium. Behler and Dellago developed a series of NN potentials with which they demonstrated that the inclusion of dispersion corrections with different DFT functionals, reflects the correct existence of a density maximum\cite{morawietzbehler2016}. In this work, they also computed melting temperatures for systems like the one shown in Figure~\ref{FIG5}a, consisting of over 2000 water molecules and with numerous simulations on the nanosecond timescale. Table C in Figure~\ref{FIG5}a shows that the melting temperatures are significantly improved upon the inclusion of dispersion corrections, in comparison with the experiments. Again, these conclusions are based on the possibility of conducting large-scale and long-time (several nanoseconds) simulations with the NN potentials.

The use of NN-potentials for water has opened up exciting applications in which they are coupled with path integral molecular dynamics (PIMD). 
A nice example of this synergy is a recent application involving Ceriotti and Behler that uses a NN-potential of water combined with path integral techniques to predict the thermodynamic properties of liquid water, as well as hexagonal and cubic ice\cite{chengceriottibehler2019}
By training a NN with a hybrid functional, these simulations rigorously take into account nuclear quantum effects, the disorder of the protons, and anharmonic fluctuations. 
The upper panel of Figure~\ref{FIG5}b shows the density isobars computed with the NN for liquid water, hexagonal, and cubic ice. 
The temperature of maximum density is in perfect agreement with the experiments, while the density isobars for all studied systems are within 3\% of the experiments. 
The lower panel of Figure \ref{FIG5}b, instead, shows the coupling of the NN with path integral simulations being used to determine the role of NQEs in controlling the structure of liquid water. 
The PIMD simulations coupled with the NN show excellent agreement with experiments for all three pair-correlation functions. 
Furthermore, as pointed out earlier, quantum effects are essential for capturing the delocalization of the proton along the hydrogen bond as seen in the bottom two panels of the RDFs.

In recent years, Car and others have extended the scope of standard NN potentials with the development of deep neural network potentials (DNN)\cite{DeepNN2018Car,wang2018deepmd,zhang2021deepmd,han2018deep}. 
DNN potentials have emerged in order to tackle some critical limitations with the classical NN methodology, including the somewhat ad-hoc choice of symmetry functions. 
Furthermore, DNNs have demonstrated a substantial enhancement in the efficiency of the learning phase for potential functions. 
This improvement is achieved through the incorporation of optimal loss functions and local decomposition methods, which enable training DNNs on relatively small systems while maintaining their applicability to much larger systems. 
It is noteworthy that the computational cost of employing DNNs scales linearly with system size, primarily because they are highly parallelizable due to the earlier mentioned local decomposition approach. 
Their adoption has expanded the scope of applications in simulating various aqueous systems. 
Notably, DNNs exhibit remarkable accuracy in reproducing average energy, density, and RDFs across a range of water and ice systems when compared to AIMD simulations \cite{DeepNN2018Car}.

Car and co-workers have recently also used the DNN framework to not only learn structural properties, but also the environment-dependent polarizability tensor of water molecules\cite{zhangcar2020}. 
This, in turn, has allowed determining the Raman spectrum of liquid water\cite{pccprobertocar2020}. Figure~\ref{FIG5}c shows the Raman spectrum for bulk water obtained from a DNN which is used to sample the configurations during a molecular dynamics simulation, but also predicts the Wannier functions necessary for computing Raman intensities. 
The DNN approach was also used to develop a model that faithfully reproduces the potential energy surface of water compared to the SCAN (Strongly-Constrained and Appropriately-Normed) approximation of DFT over a wide range of temperature and pressure conditions (up to 50 GPa and 2400K). 
The computational efficiency of this model enables the prediction of the phase diagram of water through MD simulations, in good agreement with experimental data \cite{deepmd_water_phase_diagram}. 
Another recent application of this methodology was investigation of the possibility of a second critical point in water under supercooling. 
Debendetti and Car trained a DNN to represent the ab-initio potential energy surface coming from DFT calculations using the SCAN functional\cite{gartnerLLCP2020}. 
Although the calculations are rather limited in terms of both system size and timescales, they strongly indicate the validity and the existence of the liquid-liquid critical point (LLCP), consistent with previous classical point-charge models\cite{debenedettiscience2020}. 
Furthermore, Paesani and co-workers have significantly expanded the scope of MB-pol by coupling it with DNN, which has recently allowed the exploration of the phase diagram of water\cite{Bore2023}.



Finally, highlighting further opportunities and possibilities in the future, Car and Baroni present a very recent study on using a  DNN of liquid water to predict the viscosities and thermal conductivities from first principles using the Green-Kubo theory\cite{Malosso2022,Baroni2021}. 
Figure~\ref{FIG5}d shows the comparison of the behavior of thermal conductivity for bulk water across a range of temperatures with different functionals. 
While the temperature dependence is in moderately good agreement with the experiments and there is a shift in the right direction when going from standard GGAs to meta-GGAs, it is clear that the quality of the generated DNN ultimately depends on the electronic structure theory used in the training. 

Fully dissociative NN and DNN allowing for water ionization and simulating the excess proton and hydroxide simultaneously and, in general, chemical reactions, have only recently been gathering momentum. 
Behler and co-workers have pioneered some of the early NNs stemming from neutral water that have displayed encouraging results on capturing the potential energy surface of gas-phase protonated clusters\cite{natarajan2015,schranmarx2020}.
Interestingly, NN potentials have been developed to allow simulations of sodium hydroxide solutions (NaOH) at different concentrations, which have also been coupled with PIMD simulations to investigate the role of nuclear quantum effects\cite{matti2016,ceriottibehler2018}. 
Recently, Markland and co-workers have demonstrated very promising steps in the development of an NN potential that allows for simultaneous modeling of the proton and hydroxide ion in water\cite{atsango_developing_2023}. 
In the last year, Car and co-workers have reported a water dissociable DNN potential which is used to compute the pKw of water\cite{marcos2023}. 
The findings are extremely encouraging and open up new possibilities for studying water chemistry at interfaces, under confinement, and in organic and salt solutions.  
For example, a recent AIMD study by Di Pino and co-workers showed that the pKw of water is rather insensitive to confinement until one gets to very small cluster sizes where the local solvation of the hydronium and hydroxide changes\cite{DiPino_angewandte2023}. 
We expect that over the next decade, there will be new advancements on the methodological side that will allow for routine application of DNNs to study water ionization in the bulk as well as at interfaces and under confinement such as in this recent application.

Another interesting family of models not relying on NN are the General-purpose gradient-domain machine learning potentials (GDML), which have originally shown tremendous promise for medium to large-size single molecules.\cite{gdml} In this approach, the coordinates are mapped into the eigenvalues of the Coulomb Matrix, which contains the inverse distances between all distinct pairs of atoms in the system. The functional relationship between atomic coordinates and interatomic forces is then obtained by employing a generalization of the Kernel Ridge Regression technique. GDML models using small training sets have been shown to perform similarly to potentials requiring large training sets. However, GDML applicability is limited to the same system it was trained on.\cite{mbGDML} A recent  GDML-driven many-body expansion framework (mbGDML) has enabled size transferability, opening the doors for the simulation of bulk water. Predictions on static clusters of up to 16 monomers had energy errors of less than 0.251 kcal mol${^{-1}}$ per monomer, and the simulated RDFs with mbGDML for bulk water agree remarkably well with the reference experimental data.\cite{mbGDML}

\subsection{Unravelling Complexity in Aqueous Systems}

As alluded to earlier, for many problems that involve probing fluctuations of the hydrogen bond network -- such as exploring the phase diagram in the supercooled regime or studying the breaking of bonds in the ionization of water -- there is a computational challenge of dealing with slow timescales of the underlying activated processes.  In the literature, this challenge is commonly termed as the sampling problem. At the same time, molecular simulations in the condensed phase often yield large amounts of data that are extremely challenging to interpret and understand using only chemical imagination. In this section, we will highlight efforts that have been made in recent years to overcome the sampling problem, with specific examples associated with water chemistry. Afterward, we will present recent developments in data-mining techniques that are providing new insights into the chemical physics and physical chemistry of aqueous systems.

\subsubsection{The sampling problem}

In the last two decades, numerous methods have been developed to deal with the sampling problem. The reader is referred to other detailed reviews in the topic\cite{barduccibonomoparrinello2011,vallsonparrinello2016,pietrucci_strategies_2017,pietrucci_novel_2018,Bussi2020}. 
Here, we briefly overview some of the main applications in this area that have been instrumental in driving our understanding of various aqueous-related phenomena. 
Specifically, the equilibrium between water and its constituent ions (proton and hydroxide), and more generally acids and their conjugated bases, requires fluctuations that are rather slow and cannot be achieved by brute force simulations. Through CPMD simulations,  Trout, Parrinello, and Sprik computed the free energy associated with water dissociation using the Blue-Moon ensemble technique\cite{trout_dissociation_1998,troutparrinello1999,sprik_computation_2000}. 
The principle idea was to constrain the system along a pre-defined reaction coordinate, typically chosen by chemical intuition, with an external bias potential. Although these early studies were limited by short simulation times, they demonstrated that most of the free energy of ionization arises from breaking the O-H covalent bond.  This observation is essentially reproduced using more accurate DFT functionals such as SCAN\cite{borguet2020}.  These constrained simulations, however, while effective at extracting thermodynamic quantities, do not say anything about dynamical mechanisms.

\begin{figure}[H]
        \includegraphics[width=\textwidth]{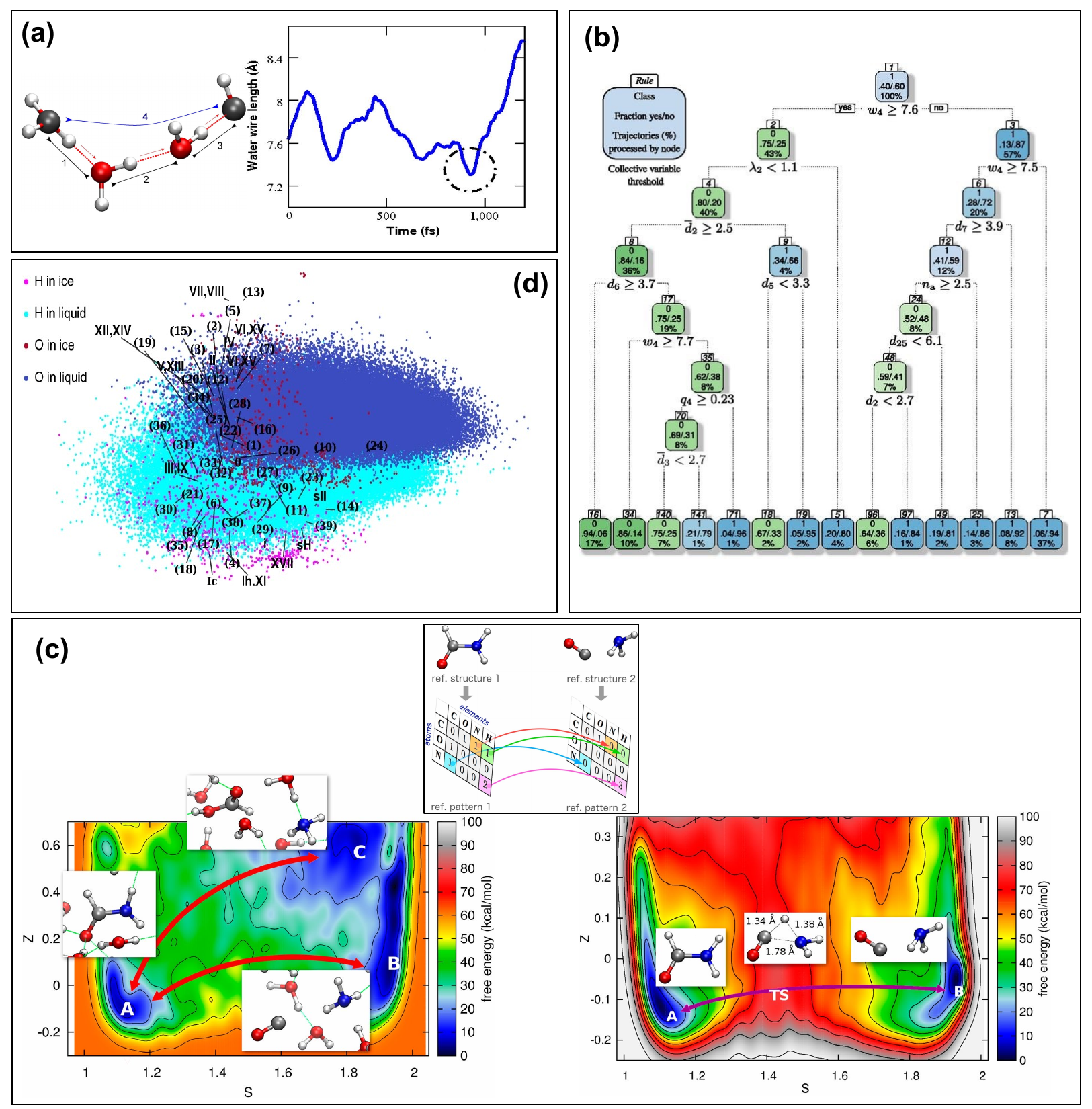}
		\caption{Panel (a) shows a water wire connecting the hydronium and hydroxide ion and an example of the water wire compression that leads to concerted proton transfer. This figure was reproduced from Reference~\citenum{HassanaliPrakashEshetParrinello2011}. Panel (b) shows the decision tree obtained by using a machine learning procedure on ionization trajectories generated from AIMD simulations. As seen the parameter $w_{4}$ corresponds to the water wire distance which is identified as the most important parameter. Others include the tetrahedrality and the local hydrogen bond patterns involving the number of accepting/donating hydrogen bonds of water molecules along the wire. This figure is reproduced from Reference~\citenum{vanErp2018}. Panel (c) (top panel) shows an illustration of the graph chemical network used to quantify different molecules. The bottom panels compare the free energy surface associated with the pre-biotic chemistry of formamide in the gas phase (left) and in solution (right panel). This figure is reproduced from Reference~\citenum{pietruccisaitta2015}. Panel (d) shows a two-dimensional scatter plot of SOAP descriptors of liquid water using principal component analysis. Also shown are the projections of the points associated with different phases of Ice. This figure is reproduced from Reference~\citenum{monserrat2020liquid}. }
		\label{FIG6}
\end{figure}

Chandler and Parrinello combined Transition Path Sampling\cite{Bolhuischandlergeissler2002} together with CPMD simulations to study the pathways associated with water ionization\cite{geisslerparrinello2001}. By generating reactive trajectories, they suggested that the breaking of the O-H covalent bond involves large electric field fluctuations followed by the reorganization of the hydrogen-bond network that separates the hydronium and hydroxide ion. Several other studies have used metadynamics simulations to investigate acid-base equilibria, for example, in acetic and carbonic acid\cite{parkparrinello2006,cunyhassanali2014}. These studies have shown that, due to the collective nature of the fluctuations of the hydrogen-bond network, coordinates on both short and intermediate-length scales are required.  The time-reversed process of the ionization, namely the recombination of the hydronium and hydroxide ions, was revisited by Hassanali and Parrinello\cite{HassanaliPrakashEshetParrinello2011}. The AIMD simulations confirmed the importance of medium-range correlations in the water network, as revealed by fluctuations of picosecond compressions of water wires that connect the two ions as illustrated in Figure \ref{FIG6}a. Though to a lesser extent, these water wires compressions are also observed during proton transfer events in crystalline ice phases\cite{Cassone_JPCB2014}. In addition, the amphiphilic nature of these ions\cite{vothpaesani2009,HassanaliGibertiSossoParrinello2014} implies that there are rather specific changes in the solvation patterns required on the donating/accepting hydrogen-bond directions of the ions.

In the last decade, machine learning and information science techniques have also been integrated into atomistic simulations to automatically discover relevant reaction coordinates, minimize chemical bias, and, subsequently, characterize the free energy landscapes of complex systems\cite{glielmo2021unsupervised}. In addition, identifying and constructing reaction coordinates 
involves a dimensionality reduction which can lead to uncontrolled information loss\cite{glielmo2022ranking}. A recent example that attempts to overcome this problem for aqueous chemistry is the work by van Erp and co-workers\cite{vanErp2018} where they revisit the ionization of water using a regression tree analysis, pictorially shown in Figure \ref{FIG6}b. This approach allows the determination of the relative weights of different variables (starting from 138 collective variables) and a more rigorous distinction between the reactive and non-reactive trajectories in the ionization event. Besides the role of the vibrations of the proton wire shown in Figure \ref{FIG6}a, which the decision tree ranks as one of the most important, van Erp and co-workers also demonstrate a non-insignificant role of the variables such as the number of accepting hydrogen bonds along the wire and tetrahedrality.

Chemical reactions involving proton transfer, as is the case of water ionization, are particularly challenging due to the fact that since protons move through the hydrogen-bond network via the Grotthuss mechanism, their identity always changes\cite{Agmon1995,ChemRev3}. Thus, a good reaction coordinate must automatically account for this feature. We highlight a relatively recent work by Pietrucci and co-authors that makes a significant stride in dealing with this problem, in which they develop chemical-topology-based coordinates\cite{prlpietrucciandreoni2011,pietruccisaitta2015}. 
For a more detailed review of the methodological aspects and developments of graph-based approaches for studying chemical reactions, the reader is directed to other papers\cite{ismail2022}. The idea is that chemical species in a reactant and product state can be defined by their bonding topology without a need to describe bonding interactions between specific atoms. This is depicted in the top panel of Figure 6c which illustrates the transformation from formamide to ammonia and carbon-monoxide. These collective variables have been interfaced with techniques such as metadynamics and have been used to explore chemical reactions relevant to problems associated with the origin of life\cite{pietruccisaitta2015,cassone_chemcomm2018}. A key finding of these simulations is the importance of solvent reorganization which changes the thermodynamics and dynamics of chemical reactions quite significantly. Specifically, the two bottom panels of Figure 6c show free energy surfaces coming from metadynamics simulations -- formamide chemistry in the gas phase (left) versus in solution (right) resulting in completely different pathways, intermediates, and products.

\subsubsection{Mining simulation data}

Another very active and growing area of research currently is the use of unsupervised learning approaches to characterize and understand patterns that arise in molecular simulations without prior imposition of knowledge often introduced by chemical bias\cite{glielmo2021unsupervised}. In this context, a wide range of local atomic descriptors are now being used to encode information about atomic environments because they preserve important symmetries. A particularly popular one, that is used in the study of liquid water for example, is the smooth overlap of atomic positions (SOAP) which essentially expands the atomic neighbor density of a chemical species onto a basis of radial basis functions and spherical harmonics\cite{bartok2010gaussian,debartokceriotti2016}. These types of atomic descriptors are high-dimensional vectors encoding details of the local environment. In order to extract useful and interpretable information from them, one needs to perform some form of dimensionality reduction and subsequently project along the relevant degrees of freedom using clustering. Figure 6d illustrates an example of this approach by 
Cheng and co-workers: SOAP descriptors of bulk liquid water are built and later used for a principal component analysis\cite{monserrat2020liquid}. These authors also try to relate the fluctuations in liquid water to milestone structures involving different phases of ice. In particular, they demonstrate that one can think about fluctuations in liquid water as transiently forming different local structures resembling phases of ice. Pavan and co-workers have also demonstrated similar ideas, again by taking advantage of the generality of the SOAP descriptors, comparing similarity measures of different empirical potentials of liquid water to phases of ice\cite{capelli_ephemeral_2022}.

\begin{figure}[H]
        \includegraphics[width=\textwidth]{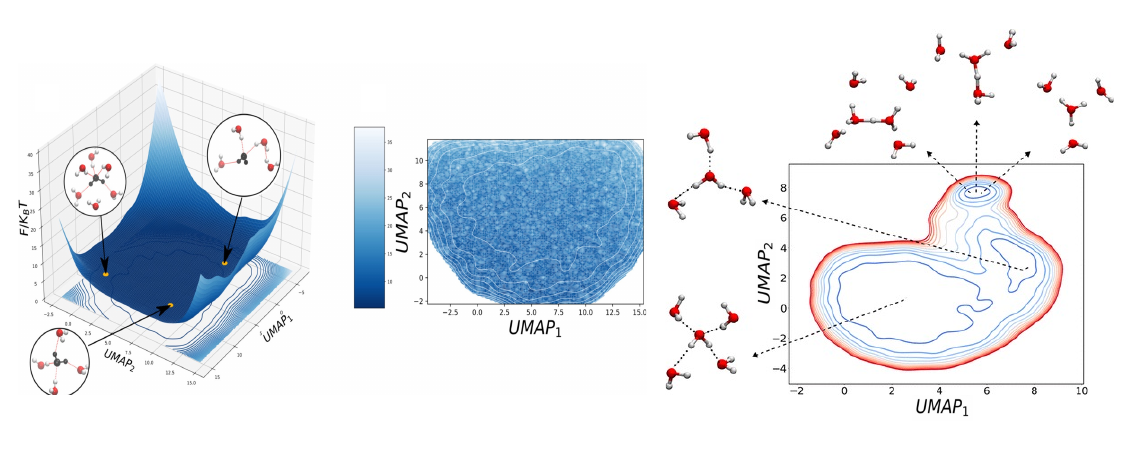}
		\caption{The 2-projections using UMAP of the SOAP variables of liquid water at room temperature on the left, and of 2M HCl on the right. This figure is reproduced from Reference~\citenum{hassanali2023}.}
		\label{FIG7}
\end{figure}

Recently, our group has also taken important steps in this direction in an effort to understand the thermodynamic landscape of liquid water and how it changes around model hydrophobic polymers, amino acids, peptide groups, and at interfaces\cite{ansari2018,ansari2019spontaneously,azizi2022model,azizi2023solvation,jong2018data,laiodonkorhassanali2023}. Specifically, using the SOAP descriptors we have investigated the number of independent degrees of freedom (often referred to as the intrinsic dimension (ID) of a data set) needed to characterize the fluctuations in liquid water at room temperature and, thereafter, we extracted the high dimensional free energies\cite{offei2022high}. One important observation is that even at the local structure level, for example, between the first and second solvation shells, the ID of the system is quite large. This implies that a correct description of the underlying thermodynamics and dynamics of the system requires looking at many orthogonal degrees of freedom concertedly. Contrary to current descriptions of water in terms of two-state liquid, the left panel of Figure 7 shows that liquid water at room temperature is characterized by a single broad minimum. Furthermore, we do not find any evidence for two states or stable local structures corresponding to low-density (LDL) and high-density liquid (HDL). More recently, we have applied similar techniques to address the question regarding the structure of the proton in liquid water, typically discussed in the literature in terms of a competition between idealized limiting states, namely the Eigen and Zundel\cite{hassanali2023}. Our agnostic approach, instead, shows that the Eigen and Zundel are neither limiting nor stable thermodynamic states. Contrasting the two panels of Figure \ref{FIG7}, shows that the simulations with the excess proton (concentrated HCl to be specific) lead to the creation of two additional minima on the free energy landscape. The excess proton in water is best seen as a charged topological defect that strongly perturbs its local environment, leading to an enhancement of the concentration of neutral water defects.  



\subsection{Future Perspectives}

The goal of this review is to give a broad overview of the historical context of studying aqueous chemistry in solution, both from the perspective of the use of DFT-based AIMD simulations as well as modern empirical potentials with a focus on dissociative schemes. This background motivated our discussion on the development of machine-learning potentials and, more generally, data-driven approaches to both model aqueous systems with greater accuracy and sophistication as well as to drive the learning of complex phenomena in these systems. Where possible, we have made an attempt to elucidate critical challenges that have arisen in different types of problems relevant to the physical chemistry and chemical physics of solvation.

The growth of machine-learning approaches to study aqueous chemistry, in particular the dissociation of water, offers enormous potential for future applications. Specifically, how these models can be extended to tackle other problems involving salt solutions\cite{bakker2016}, organic molecules such as amino acids, proteins, and DNA\cite{laagehynes2017}, and, finally, processes out of equilibrium such as in external electric fields\cite{Cassone_JPCL2020,English_SciAdv2020,Saitta_PRL2012,Cassone_PCCP17,Shafiei_JCP19,Futera_JCP17,ContiNibali_JCP2023,Kuhne_PCCP2023,cassone2023electrofreezing}, remains an open question. The challenge here involves both the generation of accurate data sets for training as well as assessing the transferability of existing models. For example, the exchange of protons in solution is a critical step underlying isotope fractionation of biomolecules in water, which is thought to be catalyzed by negatively charged hydroxide ions\cite{perssonhalle2015}. However, the underlying mechanisms by which this happens still remain uncharted territory. 

In the context of externally driven aqueous solutions, large external electric fields, for example, have been shown to enhance autoprotolysis in solution \cite{Cassone_JPCL2020,Saitta_PRL2012,Cassone_PCCP17}. It is clear that the breadth and depth of computer simulations that have been achieved with standard methods such as AIMD will play a critical role in helping push forward the development and application of machine-learning-based approaches, as recently reported in the literature \cite{FIREANN}. Another very interesting and challenging area that will certainly benefit from data-driven approaches is excited-state chemistry in solution, which presents many other methodological issues at the moment that are typically absent in the ground-state such as the \emph{near-sightedness} of the electronic degrees of freedom. The reader is referred to the following review and references therein for more details\cite{dral_molecular_2021}.

Besides the use of ML-based potentials to simulate aqueous chemistry with higher accuracy and on larger system sizes, another active area that we believe will continue to grow is the use of both supervised and unsupervised learning to unravel the complexity of chemical processes in aqueous solutions. While these approaches allow an agnostic inference of models based on the underlying structure of the data, chemical and physical interpretability is often a critical missing link. Therefore, attempts to provide physical interpretability, for example, to atomic descriptors used in neural networks (NN), is an area that requires more attention\cite{laiodonkorhassanali2023}. This understanding has important implications on how to translate physics-based interaction models into the structure of NNs. 

While the focus of this review has been on motivating data-driven potentials relevant for first-principles simulations, there are also other examples of ML being used to improve coarse-grained potentials.  The mW model developed by the Molinero group\cite{mw_2009} despite being a coarse-grained model, outperforms many of the other empirical potentials such as SPC and TIP in predicting thermodynamic properties such as surface-tension. While this model does not capture realistically the dynamics of water, it has been successful at studying phenomena such ice-nucleation and thermodynamics of water under confinement\cite{nature_ice, mw_LL, mw_LL_conf, mw_nanoice, mw_hydrosurface,mw_vp, mw_bubbles}.  More recently, it has also been interfaced to combining it with ions, molecules and polymers to tackle more complex aqueous solutions as well as systems involving interfaces such as membranes\cite{mw_clathrates, mw_ions, mw_vp_electrolyte, mw_membrane, mw_antifreeze}. ML approaches have the potential to also enable coarse-grained models such as mW. To highlight one specific case, the coarse-grained ML potentials developed by 
Sankaranarayanan and co-workers have devised an analytical force-field for monoatomic water based on the Pauling bond-order concept, containing up to 4-body terms.\cite{cg_ML} 
In particular, they have used MD trajectories and experimental data to train their NN for the optimal parameters to feed their MB potential, named ML-BOP, and also the mW model. The possibilty to combine ML approaches with coarse-grained potentials like mW has potential to open up other interesting applications of aqueous solutions.

\color{black}
\begin{acknowledgement}

GDM, DB, CE, MM and AH thank the European Commission for funding on the ERC Grant HyBOP 101043272. KA thanks the Center of International Science and Technology Cooperation (CISTC) of the Iranian Vice-Presidency of Science and Technology for the Connect+ grant.

\end{acknowledgement}

\clearpage


\bibliography{used_refs}

\providecommand{\latin}[1]{#1}
\makeatletter
\providecommand{\doi}
  {\begingroup\let\do\@makeother\dospecials
  \catcode`\{=1 \catcode`\}=2 \doi@aux}
\providecommand{\doi@aux}[1]{\endgroup\texttt{#1}}
\makeatother
\providecommand*\mcitethebibliography{\thebibliography}
\csname @ifundefined\endcsname{endmcitethebibliography}
  {\let\endmcitethebibliography\endthebibliography}{}
\begin{mcitethebibliography}{273}
\providecommand*\natexlab[1]{#1}
\providecommand*\mciteSetBstSublistMode[1]{}
\providecommand*\mciteSetBstMaxWidthForm[2]{}
\providecommand*\mciteBstWouldAddEndPuncttrue
  {\def\EndOfBibitem{\unskip.}}
\providecommand*\mciteBstWouldAddEndPunctfalse
  {\let\EndOfBibitem\relax}
\providecommand*\mciteSetBstMidEndSepPunct[3]{}
\providecommand*\mciteSetBstSublistLabelBeginEnd[3]{}
\providecommand*\EndOfBibitem{}
\mciteSetBstSublistMode{f}
\mciteSetBstMaxWidthForm{subitem}{(\alph{mcitesubitemcount})}
\mciteSetBstSublistLabelBeginEnd
  {\mcitemaxwidthsubitemform\space}
  {\relax}
  {\relax}

\bibitem[Ball(2008)]{ball2008water}
Ball,~P. Water as an active constituent in cell biology. \emph{Chemical
  reviews} \textbf{2008}, \emph{108}, 74--108\relax
\mciteBstWouldAddEndPuncttrue
\mciteSetBstMidEndSepPunct{\mcitedefaultmidpunct}
{\mcitedefaultendpunct}{\mcitedefaultseppunct}\relax
\EndOfBibitem
\bibitem[Kärkäs \latin{et~al.}(2014)Kärkäs, Verho, Johnston, and
  Åkermark]{chemrevphotosynthesis2014}
Kärkäs,~M.~D.; Verho,~O.; Johnston,~E.~V.; Åkermark,~B. Artificial
  Photosynthesis: Molecular Systems for Catalytic Water Oxidation.
  \emph{Chemical Reviews} \textbf{2014}, \emph{114}, 11863--12001, PMID:
  25354019\relax
\mciteBstWouldAddEndPuncttrue
\mciteSetBstMidEndSepPunct{\mcitedefaultmidpunct}
{\mcitedefaultendpunct}{\mcitedefaultseppunct}\relax
\EndOfBibitem
\bibitem[Bj\"{o}rneholm \latin{et~al.}(2016)Bj\"{o}rneholm, Hansen, Hodgson,
  and Liu]{chemrevinterfaces2016}
Bj\"{o}rneholm,~O.; Hansen,~M.~H.; Hodgson,~A.; Liu,~L. Water at Interfaces.
  \emph{Chemical Reviews} \textbf{2016}, \emph{116}, 7698--7726, PMID:
  27232062\relax
\mciteBstWouldAddEndPuncttrue
\mciteSetBstMidEndSepPunct{\mcitedefaultmidpunct}
{\mcitedefaultendpunct}{\mcitedefaultseppunct}\relax
\EndOfBibitem
\bibitem[Ma \latin{et~al.}(2022)Ma, Shi, Liu, Li, Cui, Tan, Zhao, and
  Wang]{HBN_watersplitting}
Ma,~X.; Shi,~Y.; Liu,~J.; Li,~X.; Cui,~X.; Tan,~S.; Zhao,~J.; Wang,~B.
  Hydrogen-Bond Network Promotes Water Splitting on the TiO2 Surface.
  \emph{Journal of the American Chemical Society} \textbf{2022}, \emph{144},
  13565--13573, PMID: 35852138\relax
\mciteBstWouldAddEndPuncttrue
\mciteSetBstMidEndSepPunct{\mcitedefaultmidpunct}
{\mcitedefaultendpunct}{\mcitedefaultseppunct}\relax
\EndOfBibitem
\bibitem[Stillinger and Rahman(1974)Stillinger, and
  Rahman]{stillinger1974improved}
Stillinger,~F.~H.; Rahman,~A. Improved simulation of liquid water by molecular
  dynamics. \emph{The Journal of Chemical Physics} \textbf{1974}, \emph{60},
  1545--1557\relax
\mciteBstWouldAddEndPuncttrue
\mciteSetBstMidEndSepPunct{\mcitedefaultmidpunct}
{\mcitedefaultendpunct}{\mcitedefaultseppunct}\relax
\EndOfBibitem
\bibitem[Chatterjee \latin{et~al.}(2008)Chatterjee, Debenedetti, Stillinger,
  and Lynden-Bell]{tet4}
Chatterjee,~S.; Debenedetti,~P.~G.; Stillinger,~F.~H.; Lynden-Bell,~R.~M. A
  computational investigation of thermodynamics, structure, dynamics and
  solvation behavior in modified water models. \emph{The Journal of chemical
  physics} \textbf{2008}, \emph{128}, 124511\relax
\mciteBstWouldAddEndPuncttrue
\mciteSetBstMidEndSepPunct{\mcitedefaultmidpunct}
{\mcitedefaultendpunct}{\mcitedefaultseppunct}\relax
\EndOfBibitem
\bibitem[Stillinger and Rahman(1972)Stillinger, and
  Rahman]{stillinger1972molecular}
Stillinger,~F.~H.; Rahman,~A. Molecular dynamics study of temperature effects
  on water structure and kinetics. \emph{The Journal of chemical physics}
  \textbf{1972}, \emph{57}, 1281--1292\relax
\mciteBstWouldAddEndPuncttrue
\mciteSetBstMidEndSepPunct{\mcitedefaultmidpunct}
{\mcitedefaultendpunct}{\mcitedefaultseppunct}\relax
\EndOfBibitem
\bibitem[Rahman and Stillinger(1973)Rahman, and Stillinger]{rahman1973hydrogen}
Rahman,~A.; Stillinger,~F.~H. Hydrogen-bond patterns in liquid water.
  \emph{Journal of the American Chemical Society} \textbf{1973}, \emph{95},
  7943--7948\relax
\mciteBstWouldAddEndPuncttrue
\mciteSetBstMidEndSepPunct{\mcitedefaultmidpunct}
{\mcitedefaultendpunct}{\mcitedefaultseppunct}\relax
\EndOfBibitem
\bibitem[Gallo \latin{et~al.}(2016)Gallo, Amann-Winkel, Angell, Anisimov,
  Caupin, Chakravarty, Lascaris, Loerting, Panagiotopoulos, Russo,
  \latin{et~al.} others]{gallo2016water}
Gallo,~P.; Amann-Winkel,~K.; Angell,~C.~A.; Anisimov,~M.~A.; Caupin,~F.;
  Chakravarty,~C.; Lascaris,~E.; Loerting,~T.; Panagiotopoulos,~A.~Z.;
  Russo,~J., \latin{et~al.}  Water: A tale of two liquids. \emph{Chemical
  reviews} \textbf{2016}, \emph{116}, 7463--7500\relax
\mciteBstWouldAddEndPuncttrue
\mciteSetBstMidEndSepPunct{\mcitedefaultmidpunct}
{\mcitedefaultendpunct}{\mcitedefaultseppunct}\relax
\EndOfBibitem
\bibitem[Pettersson(2018)]{pettersson2018two}
Pettersson,~L.~G. A Two-State Picture of Water and the Funnel of Life.
  International Conference Physics of Liquid Matter: Modern Problems. 2018; pp
  3--39\relax
\mciteBstWouldAddEndPuncttrue
\mciteSetBstMidEndSepPunct{\mcitedefaultmidpunct}
{\mcitedefaultendpunct}{\mcitedefaultseppunct}\relax
\EndOfBibitem
\bibitem[Agmon \latin{et~al.}(2016)Agmon, Bakker, Campen, Henchman, Pohl, Roke,
  Th\"amer, and Hassanali]{ChemRev3}
Agmon,~N.; Bakker,~H.~J.; Campen,~R.~K.; Henchman,~R.~H.; Pohl,~P.; Roke,~S.;
  Th\"amer,~M.; Hassanali,~A. Protons and Hydroxide Ions in Aqueous Systems.
  \emph{Chemical Reviews} \textbf{2016}, \emph{116}, 7642--7672\relax
\mciteBstWouldAddEndPuncttrue
\mciteSetBstMidEndSepPunct{\mcitedefaultmidpunct}
{\mcitedefaultendpunct}{\mcitedefaultseppunct}\relax
\EndOfBibitem
\bibitem[Marx \latin{et~al.}(2010)Marx, Chandra, and
  Tuckerman]{marx2010aqueous}
Marx,~D.; Chandra,~A.; Tuckerman,~M.~E. Aqueous basic solutions: Hydroxide
  solvation, structural diffusion, and comparison to the hydrated proton.
  \emph{Chemical reviews} \textbf{2010}, \emph{110}, 2174--2216\relax
\mciteBstWouldAddEndPuncttrue
\mciteSetBstMidEndSepPunct{\mcitedefaultmidpunct}
{\mcitedefaultendpunct}{\mcitedefaultseppunct}\relax
\EndOfBibitem
\bibitem[Hassanali \latin{et~al.}(2014)Hassanali, Cuny, Verdolino, and
  Parrinello]{rscparrinello2014}
Hassanali,~A.~A.; Cuny,~J.; Verdolino,~V.; Parrinello,~M. Aqueous solutions:
  state of the art in \emph{ab initio} molecular dynamics. \emph{Philosophical
  Transactions of the Royal Society A: Mathematical, Physical and Engineering
  Sciences} \textbf{2014}, \emph{372}, 20120482\relax
\mciteBstWouldAddEndPuncttrue
\mciteSetBstMidEndSepPunct{\mcitedefaultmidpunct}
{\mcitedefaultendpunct}{\mcitedefaultseppunct}\relax
\EndOfBibitem
\bibitem[Dykstra(1993)]{Dykstra1993}
Dykstra,~C.~E. Electrostatic interaction potentials in molecular force fields.
  \emph{Chemical Reviews} \textbf{1993}, \emph{93}, 2339--2353\relax
\mciteBstWouldAddEndPuncttrue
\mciteSetBstMidEndSepPunct{\mcitedefaultmidpunct}
{\mcitedefaultendpunct}{\mcitedefaultseppunct}\relax
\EndOfBibitem
\bibitem[Iftimie \latin{et~al.}(2005)Iftimie, Minary, and
  Tuckerman]{tuckerman2005}
Iftimie,~R.; Minary,~P.; Tuckerman,~M.~E. \emph{Ab initio} molecular dynamics:
  Concepts, recent developments, and future trends. \emph{Proceedings of the
  National Academy of Sciences} \textbf{2005}, \emph{102}, 6654--6659\relax
\mciteBstWouldAddEndPuncttrue
\mciteSetBstMidEndSepPunct{\mcitedefaultmidpunct}
{\mcitedefaultendpunct}{\mcitedefaultseppunct}\relax
\EndOfBibitem
\bibitem[Scheraga \latin{et~al.}(2007)Scheraga, Khalili, and
  Liwo]{scheraga2007}
Scheraga,~H.~A.; Khalili,~M.; Liwo,~A. Protein-Folding Dynamics: Overview of
  Molecular Simulation Techniques. \emph{Annual Review of Physical Chemistry}
  \textbf{2007}, \emph{58}, 57--83, PMID: 17034338\relax
\mciteBstWouldAddEndPuncttrue
\mciteSetBstMidEndSepPunct{\mcitedefaultmidpunct}
{\mcitedefaultendpunct}{\mcitedefaultseppunct}\relax
\EndOfBibitem
\bibitem[Jungwirth and Tobias(2006)Jungwirth, and Tobias]{jungwirthtobias2006}
Jungwirth,~P.; Tobias,~D.~J. Specific Ion Effects at the Air/Water Interface.
  \emph{Chemical Reviews} \textbf{2006}, \emph{106}, 1259--1281, PMID:
  16608180\relax
\mciteBstWouldAddEndPuncttrue
\mciteSetBstMidEndSepPunct{\mcitedefaultmidpunct}
{\mcitedefaultendpunct}{\mcitedefaultseppunct}\relax
\EndOfBibitem
\bibitem[Laage \latin{et~al.}(2017)Laage, Elsaesser, and Hynes]{laagehynes2017}
Laage,~D.; Elsaesser,~T.; Hynes,~J.~T. Water Dynamics in the Hydration Shells
  of Biomolecules. \emph{Chemical Reviews} \textbf{2017}, \emph{117},
  10694--10725, PMID: 28248491\relax
\mciteBstWouldAddEndPuncttrue
\mciteSetBstMidEndSepPunct{\mcitedefaultmidpunct}
{\mcitedefaultendpunct}{\mcitedefaultseppunct}\relax
\EndOfBibitem
\bibitem[Gillan \latin{et~al.}(2016)Gillan, Alfè, and
  Michaelides]{gillanalfemichaelides2016}
Gillan,~M.~J.; Alfè,~D.; Michaelides,~A. Perspective: How good is DFT for
  water? \emph{The Journal of Chemical Physics} \textbf{2016}, \emph{144},
  130901\relax
\mciteBstWouldAddEndPuncttrue
\mciteSetBstMidEndSepPunct{\mcitedefaultmidpunct}
{\mcitedefaultendpunct}{\mcitedefaultseppunct}\relax
\EndOfBibitem
\bibitem[Zen \latin{et~al.}(2015)Zen, Luo, Mazzola, Guidoni, and
  Sorella]{zen_ab_2015}
Zen,~A.; Luo,~Y.; Mazzola,~G.; Guidoni,~L.; Sorella,~S. Ab initio molecular
  dynamics simulation of liquid water by quantum {Monte} {Carlo}. \emph{The
  Journal of Chemical Physics} \textbf{2015}, \emph{142}, 144111, \_eprint:
  https://pubs.aip.org/aip/jcp/article-pdf/doi/10.1063/1.4917171/13525553/144111\_1\_online.pdf\relax
\mciteBstWouldAddEndPuncttrue
\mciteSetBstMidEndSepPunct{\mcitedefaultmidpunct}
{\mcitedefaultendpunct}{\mcitedefaultseppunct}\relax
\EndOfBibitem
\bibitem[Del~Ben \latin{et~al.}(2013)Del~Ben, Schönherr, Hutter, and
  VandeVondele]{joostmp2}
Del~Ben,~M.; Schönherr,~M.; Hutter,~J.; VandeVondele,~J. Bulk Liquid Water at
  Ambient Temperature and Pressure from MP2 Theory. \emph{The Journal of
  Physical Chemistry Letters} \textbf{2013}, \emph{4}, 3753--3759\relax
\mciteBstWouldAddEndPuncttrue
\mciteSetBstMidEndSepPunct{\mcitedefaultmidpunct}
{\mcitedefaultendpunct}{\mcitedefaultseppunct}\relax
\EndOfBibitem
\bibitem[Reddy \latin{et~al.}(2016)Reddy, Straight, Bajaj, Huy~Pham, Riera,
  Moberg, Morales, Knight, G{\"o}tz, and Paesani]{reddy2016accuracy}
Reddy,~S.~K.; Straight,~S.~C.; Bajaj,~P.; Huy~Pham,~C.; Riera,~M.;
  Moberg,~D.~R.; Morales,~M.~A.; Knight,~C.; G{\"o}tz,~A.~W.; Paesani,~F. On
  the accuracy of the MB-pol many-body potential for water: Interaction
  energies, vibrational frequencies, and classical thermodynamic and dynamical
  properties from clusters to liquid water and ice. \emph{The Journal of
  chemical physics} \textbf{2016}, \emph{145}, 194504\relax
\mciteBstWouldAddEndPuncttrue
\mciteSetBstMidEndSepPunct{\mcitedefaultmidpunct}
{\mcitedefaultendpunct}{\mcitedefaultseppunct}\relax
\EndOfBibitem
\bibitem[Lambros and Paesani(2020)Lambros, and Paesani]{lambros_how_2020}
Lambros,~E.; Paesani,~F. How good are polarizable and flexible models for
  water: Insights from a many-body perspective. \emph{The Journal of Chemical
  Physics} \textbf{2020}, \emph{153}\relax
\mciteBstWouldAddEndPuncttrue
\mciteSetBstMidEndSepPunct{\mcitedefaultmidpunct}
{\mcitedefaultendpunct}{\mcitedefaultseppunct}\relax
\EndOfBibitem
\bibitem[Bonomi \latin{et~al.}(2009)Bonomi, Branduardi, Bussi, Camilloni,
  Provasi, Raiteri, Donadio, Marinelli, Pietrucci, Broglia, \latin{et~al.}
  others]{bonomi2009plumed}
Bonomi,~M.; Branduardi,~D.; Bussi,~G.; Camilloni,~C.; Provasi,~D.; Raiteri,~P.;
  Donadio,~D.; Marinelli,~F.; Pietrucci,~F.; Broglia,~R.~A., \latin{et~al.}
  PLUMED: A portable plugin for free-energy calculations with molecular
  dynamics. \emph{Computer Physics Communications} \textbf{2009}, \emph{180},
  1961--1972\relax
\mciteBstWouldAddEndPuncttrue
\mciteSetBstMidEndSepPunct{\mcitedefaultmidpunct}
{\mcitedefaultendpunct}{\mcitedefaultseppunct}\relax
\EndOfBibitem
\bibitem[Barducci \latin{et~al.}(2011)Barducci, Bonomi, and
  Parrinello]{barduccibonomoparrinello2011}
Barducci,~A.; Bonomi,~M.; Parrinello,~M. Metadynamics. \emph{WIREs
  Computational Molecular Science} \textbf{2011}, \emph{1}, 826--843\relax
\mciteBstWouldAddEndPuncttrue
\mciteSetBstMidEndSepPunct{\mcitedefaultmidpunct}
{\mcitedefaultendpunct}{\mcitedefaultseppunct}\relax
\EndOfBibitem
\bibitem[Geissler \latin{et~al.}(2001)Geissler, Dellago, Chandler, Hutter, and
  Parrinello]{Geissler2001}
Geissler,~P.~L.; Dellago,~C.; Chandler,~D.; Hutter,~J.; Parrinello,~M.
  Autoionization in Liquid Water. \emph{Science} \textbf{2001}, \emph{291},
  2121--2124\relax
\mciteBstWouldAddEndPuncttrue
\mciteSetBstMidEndSepPunct{\mcitedefaultmidpunct}
{\mcitedefaultendpunct}{\mcitedefaultseppunct}\relax
\EndOfBibitem
\bibitem[Sirkin \latin{et~al.}(2018)Sirkin, Hassanali, and
  Scherlis]{sirkin2018}
Sirkin,~Y. A.~P.; Hassanali,~A.; Scherlis,~D.~A. One-Dimensional Confinement
  Inhibits Water Dissociation in Carbon Nanotubes. \emph{The Journal of
  Physical Chemistry Letters} \textbf{2018}, \emph{9}, 5029--5033, PMID:
  30113846\relax
\mciteBstWouldAddEndPuncttrue
\mciteSetBstMidEndSepPunct{\mcitedefaultmidpunct}
{\mcitedefaultendpunct}{\mcitedefaultseppunct}\relax
\EndOfBibitem
\bibitem[Di~Pino \latin{et~al.}(2023)Di~Pino, Perez~Sirkin, Morzan, Sánchez,
  Hassanali, and Scherlis]{DiPino_angewandte2023}
Di~Pino,~S.; Perez~Sirkin,~Y.~A.; Morzan,~U.~N.; Sánchez,~V.~M.;
  Hassanali,~A.; Scherlis,~D.~A. Water Self-Dissociation is Insensitive to
  Nanoscale Environments. \emph{Angewandte Chemie International Edition}
  \textbf{2023}, \emph{62}, e202306526\relax
\mciteBstWouldAddEndPuncttrue
\mciteSetBstMidEndSepPunct{\mcitedefaultmidpunct}
{\mcitedefaultendpunct}{\mcitedefaultseppunct}\relax
\EndOfBibitem
\bibitem[Debenedetti \latin{et~al.}(2020)Debenedetti, Sciortino, and
  Zerze]{debenedettiscience2020}
Debenedetti,~P.~G.; Sciortino,~F.; Zerze,~G.~H. Second critical point in two
  realistic models of water. \emph{Science} \textbf{2020}, \emph{369},
  289--292\relax
\mciteBstWouldAddEndPuncttrue
\mciteSetBstMidEndSepPunct{\mcitedefaultmidpunct}
{\mcitedefaultendpunct}{\mcitedefaultseppunct}\relax
\EndOfBibitem
\bibitem[Warshel(2002)]{Warshel2002}
Warshel,~A. Molecular Dynamics Simulations of Biological Reactions.
  \emph{Accounts of Chemical Research} \textbf{2002}, \emph{35}, 385--395,
  PMID: 12069623\relax
\mciteBstWouldAddEndPuncttrue
\mciteSetBstMidEndSepPunct{\mcitedefaultmidpunct}
{\mcitedefaultendpunct}{\mcitedefaultseppunct}\relax
\EndOfBibitem
\bibitem[Valsson \latin{et~al.}(2016)Valsson, Tiwary, and
  Parrinello]{vallsonparrinello2016}
Valsson,~O.; Tiwary,~P.; Parrinello,~M. Enhancing Important Fluctuations: Rare
  Events and Metadynamics from a Conceptual Viewpoint. \emph{Annual Review of
  Physical Chemistry} \textbf{2016}, \emph{67}, 159--184, PMID: 26980304\relax
\mciteBstWouldAddEndPuncttrue
\mciteSetBstMidEndSepPunct{\mcitedefaultmidpunct}
{\mcitedefaultendpunct}{\mcitedefaultseppunct}\relax
\EndOfBibitem
\bibitem[Parr and Yang(1995)Parr, and Yang]{parryang1995}
Parr,~R.~G.; Yang,~W. Density-Functional Theory of the Electronic Structure of
  Molecules. \emph{Annual Review of Physical Chemistry} \textbf{1995},
  \emph{46}, 701--728, PMID: 24341393\relax
\mciteBstWouldAddEndPuncttrue
\mciteSetBstMidEndSepPunct{\mcitedefaultmidpunct}
{\mcitedefaultendpunct}{\mcitedefaultseppunct}\relax
\EndOfBibitem
\bibitem[Kohn \latin{et~al.}(1996)Kohn, Becke, and Parr]{kohnparr1996}
Kohn,~W.; Becke,~A.~D.; Parr,~R.~G. Density Functional Theory of Electronic
  Structure. \emph{The Journal of Physical Chemistry} \textbf{1996},
  \emph{100}, 12974--12980\relax
\mciteBstWouldAddEndPuncttrue
\mciteSetBstMidEndSepPunct{\mcitedefaultmidpunct}
{\mcitedefaultendpunct}{\mcitedefaultseppunct}\relax
\EndOfBibitem
\bibitem[Iftimie \latin{et~al.}(2005)Iftimie, Minary, and
  Tuckerman]{tuckermaniftimieminary2005}
Iftimie,~R.; Minary,~P.; Tuckerman,~M.~E. <i>Ab initio</i> molecular dynamics:
  Concepts, recent developments, and future trends. \emph{Proceedings of the
  National Academy of Sciences} \textbf{2005}, \emph{102}, 6654--6659\relax
\mciteBstWouldAddEndPuncttrue
\mciteSetBstMidEndSepPunct{\mcitedefaultmidpunct}
{\mcitedefaultendpunct}{\mcitedefaultseppunct}\relax
\EndOfBibitem
\bibitem[Marx and Hutter(2000)Marx, and Hutter]{marx2000modern}
Marx,~D.; Hutter,~J. In \emph{Modern Methods and Algorithms of Quantum
  Chemistry}; Grotendorst,~J., Ed.; Forschungszentrum, J{\"u}lich, Germany,
  John von Neumann-Institut f{\"u}r Computing: J{\"u}lich, Germany, 2000;
  Vol.~1; pp 301--449\relax
\mciteBstWouldAddEndPuncttrue
\mciteSetBstMidEndSepPunct{\mcitedefaultmidpunct}
{\mcitedefaultendpunct}{\mcitedefaultseppunct}\relax
\EndOfBibitem
\bibitem[Tuckerman \latin{et~al.}(1995)Tuckerman, Laasonen, Sprik, and
  Parrinello]{TuckermanLaasonenSprikParrinello1995}
Tuckerman,~M.; Laasonen,~K.; Sprik,~M.; Parrinello,~M. Ab Initio Molecular
  Dynamics Simulation of the Solvation and Transport of {H3O}+ and {OH}- Ions
  in Water. \emph{The Journal of Physical Chemistry} \textbf{1995}, \emph{99},
  5749--5752\relax
\mciteBstWouldAddEndPuncttrue
\mciteSetBstMidEndSepPunct{\mcitedefaultmidpunct}
{\mcitedefaultendpunct}{\mcitedefaultseppunct}\relax
\EndOfBibitem
\bibitem[Tuckerman \latin{et~al.}(1995)Tuckerman, Laasonen, Sprik, and
  Parrinello]{TuckermanLaasonenSprikParrinello1995JCP}
Tuckerman,~M.; Laasonen,~K.; Sprik,~M.; Parrinello,~M. Ab initio molecular
  dynamics simulation of the solvation and transport of hydronium and hydroxyl
  ions in water. \emph{The Journal of Chemical Physics} \textbf{1995},
  \emph{103}, 150--161\relax
\mciteBstWouldAddEndPuncttrue
\mciteSetBstMidEndSepPunct{\mcitedefaultmidpunct}
{\mcitedefaultendpunct}{\mcitedefaultseppunct}\relax
\EndOfBibitem
\bibitem[Tuckerman \latin{et~al.}(1997)Tuckerman, Marx, Klein, and
  Parrinello]{TuckermanMarxKleinParrinello1997}
Tuckerman,~M.~E.; Marx,~D.; Klein,~M.~L.; Parrinello,~M. On the quantum nature
  of the shared proton in hydrogen bonds. \emph{Science} \textbf{1997},
  \emph{275}, 817\relax
\mciteBstWouldAddEndPuncttrue
\mciteSetBstMidEndSepPunct{\mcitedefaultmidpunct}
{\mcitedefaultendpunct}{\mcitedefaultseppunct}\relax
\EndOfBibitem
\bibitem[Marx \latin{et~al.}(1999)Marx, Tuckerman, Hutter, and
  Parrinello]{MarxTuckermanHutterParrinello1999}
Marx,~D.; Tuckerman,~M.~E.; Hutter,~J.; Parrinello,~M. The nature of the
  hydrated excess proton in water. \emph{Nature} \textbf{1999}, \emph{397},
  601--604\relax
\mciteBstWouldAddEndPuncttrue
\mciteSetBstMidEndSepPunct{\mcitedefaultmidpunct}
{\mcitedefaultendpunct}{\mcitedefaultseppunct}\relax
\EndOfBibitem
\bibitem[Agmon(1995)]{Agmon1995}
Agmon,~N. The {G}rotthuss mechanism. \emph{Chemical Physics Letters}
  \textbf{1995}, \emph{244}, 456 -- 462\relax
\mciteBstWouldAddEndPuncttrue
\mciteSetBstMidEndSepPunct{\mcitedefaultmidpunct}
{\mcitedefaultendpunct}{\mcitedefaultseppunct}\relax
\EndOfBibitem
\bibitem[Carpenter \latin{et~al.}(2018)Carpenter, Fournier, Lewis, and
  Tokmakoff]{williamtokmakoff2018}
Carpenter,~W.~B.; Fournier,~J.~A.; Lewis,~N. H.~C.; Tokmakoff,~A. Picosecond
  Proton Transfer Kinetics in Water Revealed with Ultrafast IR Spectroscopy.
  \emph{The Journal of Physical Chemistry B} \textbf{2018}, \emph{122},
  2792--2802, PMID: 29452488\relax
\mciteBstWouldAddEndPuncttrue
\mciteSetBstMidEndSepPunct{\mcitedefaultmidpunct}
{\mcitedefaultendpunct}{\mcitedefaultseppunct}\relax
\EndOfBibitem
\bibitem[Fournier \latin{et~al.}(2018)Fournier, Carpenter, Lewis, and
  Tokmakoff]{fournierbroadband_2018}
Fournier,~J.~A.; Carpenter,~W.~B.; Lewis,~N. H.~C.; Tokmakoff,~A. Broadband
  {2D} {IR} spectroscopy reveals dominant asymmetric {H5O2}+ proton hydration
  structures in acid solutions. \emph{Nature Chemistry} \textbf{2018},
  \emph{10}, 932--937\relax
\mciteBstWouldAddEndPuncttrue
\mciteSetBstMidEndSepPunct{\mcitedefaultmidpunct}
{\mcitedefaultendpunct}{\mcitedefaultseppunct}\relax
\EndOfBibitem
\bibitem[Daly \latin{et~al.}(2017)Daly, Streacker, Sun, Pattenaude, Hassanali,
  Petersen, Corcelli, and Ben-Amotz]{daly2017}
Daly,~C. A.~J.; Streacker,~L.~M.; Sun,~Y.; Pattenaude,~S.~R.; Hassanali,~A.~A.;
  Petersen,~P.~B.; Corcelli,~S.~A.; Ben-Amotz,~D. Decomposition of the
  Experimental Raman and Infrared Spectra of Acidic Water into Proton, Special
  Pair, and Counterion Contributions. \emph{The Journal of Physical Chemistry
  Letters} \textbf{2017}, \emph{8}, 5246--5252, PMID: 28976760\relax
\mciteBstWouldAddEndPuncttrue
\mciteSetBstMidEndSepPunct{\mcitedefaultmidpunct}
{\mcitedefaultendpunct}{\mcitedefaultseppunct}\relax
\EndOfBibitem
\bibitem[Kozari \latin{et~al.}(2021)Kozari, Sigalov, Pines, Fingerhut, and
  Pines]{kozari2021}
Kozari,~E.; Sigalov,~M.; Pines,~D.; Fingerhut,~B.~P.; Pines,~E. Infrared and
  NMR Spectroscopic Fingerprints of the Asymmetric H7+O3 Complex in Solution.
  \emph{ChemPhysChem} \textbf{2021}, \emph{22}, 716--725\relax
\mciteBstWouldAddEndPuncttrue
\mciteSetBstMidEndSepPunct{\mcitedefaultmidpunct}
{\mcitedefaultendpunct}{\mcitedefaultseppunct}\relax
\EndOfBibitem
\bibitem[Grimme \latin{et~al.}(2010)Grimme, Antony, Ehrlich, and
  Krieg]{Grimme2010}
Grimme,~S.; Antony,~J.; Ehrlich,~S.; Krieg,~H. {A consistent and accurate ab
  initio parametrization of density functional dispersion correction (DFT-D)
  for the 94 elements H-Pu}. \emph{The Journal of Chemical Physics}
  \textbf{2010}, \emph{132}, 154104\relax
\mciteBstWouldAddEndPuncttrue
\mciteSetBstMidEndSepPunct{\mcitedefaultmidpunct}
{\mcitedefaultendpunct}{\mcitedefaultseppunct}\relax
\EndOfBibitem
\bibitem[Skinner \latin{et~al.}(2013)Skinner, Huang, Schlesinger, Pettersson,
  Nilsson, and Benmore]{skinner_jcp13}
Skinner,~L.~B.; Huang,~C.; Schlesinger,~D.; Pettersson,~L. G.~M.; Nilsson,~A.;
  Benmore,~C.~J. {Benchmark oxygen-oxygen pair-distribution function of ambient
  water from x-ray diffraction measurements with a wide Q-range}. \emph{The
  Journal of Chemical Physics} \textbf{2013}, \emph{138}, 074506\relax
\mciteBstWouldAddEndPuncttrue
\mciteSetBstMidEndSepPunct{\mcitedefaultmidpunct}
{\mcitedefaultendpunct}{\mcitedefaultseppunct}\relax
\EndOfBibitem
\bibitem[Soper(2007)]{Soper_2007}
Soper,~A.~K. Joint structure refinement of x-ray and neutron diffraction data
  on disordered materials: application to liquid water. \emph{Journal of
  Physics: Condensed Matter} \textbf{2007}, \emph{19}, 335206\relax
\mciteBstWouldAddEndPuncttrue
\mciteSetBstMidEndSepPunct{\mcitedefaultmidpunct}
{\mcitedefaultendpunct}{\mcitedefaultseppunct}\relax
\EndOfBibitem
\bibitem[VandeVondele \latin{et~al.}(2004)VandeVondele, Mohamed, Krack, Hutter,
  Sprik, and Parrinello]{vondeleparrinello2004}
VandeVondele,~J.; Mohamed,~F.; Krack,~M.; Hutter,~J.; Sprik,~M.; Parrinello,~M.
  {The influence of temperature and density functional models in ab initio
  molecular dynamics simulation of liquid water}. \emph{The Journal of Chemical
  Physics} \textbf{2004}, \emph{122}, 014515\relax
\mciteBstWouldAddEndPuncttrue
\mciteSetBstMidEndSepPunct{\mcitedefaultmidpunct}
{\mcitedefaultendpunct}{\mcitedefaultseppunct}\relax
\EndOfBibitem
\bibitem[Kuo \latin{et~al.}(2004)Kuo, Mundy, McGrath, Siepmann, VandeVondele,
  Sprik, Hutter, Chen, Klein, Mohamed, Krack, and Parrinello]{KuoMundy2004}
Kuo,~I.-F.~W.; Mundy,~C.~J.; McGrath,~M.~J.; Siepmann,~J.~I.; VandeVondele,~J.;
  Sprik,~M.; Hutter,~J.; Chen,~B.; Klein,~M.~L.; Mohamed,~F.; Krack,~M.;
  Parrinello,~M. Liquid Water from First Principles: Investigation of Different
  Sampling Approaches. \emph{The Journal of Physical Chemistry B}
  \textbf{2004}, \emph{108}, 12990--12998\relax
\mciteBstWouldAddEndPuncttrue
\mciteSetBstMidEndSepPunct{\mcitedefaultmidpunct}
{\mcitedefaultendpunct}{\mcitedefaultseppunct}\relax
\EndOfBibitem
\bibitem[K{\"u}hne \latin{et~al.}(2009)K{\"u}hne, Krack, and
  Parrinello]{kuhneparrinello2009}
K{\"u}hne,~T.~D.; Krack,~M.; Parrinello,~M. Static and dynamical properties of
  liquid water from first principles by a novel Car- Parrinello-like approach.
  \emph{Journal of chemical theory and computation} \textbf{2009}, \emph{5},
  235--241\relax
\mciteBstWouldAddEndPuncttrue
\mciteSetBstMidEndSepPunct{\mcitedefaultmidpunct}
{\mcitedefaultendpunct}{\mcitedefaultseppunct}\relax
\EndOfBibitem
\bibitem[Chen \latin{et~al.}(2017)Chen, Ko, Remsing, Andrade, Santra, Sun,
  Selloni, Car, Klein, Perdew, and Wu]{mohan2017}
Chen,~M.; Ko,~H.-Y.; Remsing,~R.~C.; Andrade,~M. F.~C.; Santra,~B.; Sun,~Z.;
  Selloni,~A.; Car,~R.; Klein,~M.~L.; Perdew,~J.~P.; Wu,~X. Ab initio theory
  and modeling of water. \emph{Proceedings of the National Academy of Sciences}
  \textbf{2017}, \emph{114}, 10846--10851\relax
\mciteBstWouldAddEndPuncttrue
\mciteSetBstMidEndSepPunct{\mcitedefaultmidpunct}
{\mcitedefaultendpunct}{\mcitedefaultseppunct}\relax
\EndOfBibitem
\bibitem[Wang \latin{et~al.}(2011)Wang, Román-Pérez, Soler, Artacho, and
  Fernández-Serra]{serra2011}
Wang,~J.; Román-Pérez,~G.; Soler,~J.~M.; Artacho,~E.; Fernández-Serra,~M.-V.
  {Density, structure, and dynamics of water: The effect of van der Waals
  interactions}. \emph{The Journal of Chemical Physics} \textbf{2011},
  \emph{134}, 024516\relax
\mciteBstWouldAddEndPuncttrue
\mciteSetBstMidEndSepPunct{\mcitedefaultmidpunct}
{\mcitedefaultendpunct}{\mcitedefaultseppunct}\relax
\EndOfBibitem
\bibitem[Lin \latin{et~al.}(2012)Lin, Seitsonen, Tavernelli, and
  Rothlisberger]{chuntavernelliroth2012}
Lin,~I.-C.; Seitsonen,~A.~P.; Tavernelli,~I.; Rothlisberger,~U. Structure and
  Dynamics of Liquid Water from ab Initio Molecular Dynamics—Comparison of
  BLYP, PBE, and revPBE Density Functionals with and without van der Waals
  Corrections. \emph{Journal of Chemical Theory and Computation} \textbf{2012},
  \emph{8}, 3902--3910, PMID: 26593030\relax
\mciteBstWouldAddEndPuncttrue
\mciteSetBstMidEndSepPunct{\mcitedefaultmidpunct}
{\mcitedefaultendpunct}{\mcitedefaultseppunct}\relax
\EndOfBibitem
\bibitem[DiStasio \latin{et~al.}(2014)DiStasio, Santra, Li, Wu, and
  Car]{distasio2014}
DiStasio,~J.,~Robert~A.; Santra,~B.; Li,~Z.; Wu,~X.; Car,~R. {The individual
  and collective effects of exact exchange and dispersion interactions on the
  ab initio structure of liquid water}. \emph{The Journal of Chemical Physics}
  \textbf{2014}, \emph{141}, 084502\relax
\mciteBstWouldAddEndPuncttrue
\mciteSetBstMidEndSepPunct{\mcitedefaultmidpunct}
{\mcitedefaultendpunct}{\mcitedefaultseppunct}\relax
\EndOfBibitem
\bibitem[Morrone and Car(2008)Morrone, and Car]{morronecar2008}
Morrone,~J.~A.; Car,~R. Nuclear Quantum Effects in Water. \emph{Phys. Rev.
  Lett.} \textbf{2008}, \emph{101}, 017801\relax
\mciteBstWouldAddEndPuncttrue
\mciteSetBstMidEndSepPunct{\mcitedefaultmidpunct}
{\mcitedefaultendpunct}{\mcitedefaultseppunct}\relax
\EndOfBibitem
\bibitem[Ceriotti \latin{et~al.}(2011)Ceriotti, Manolopoulos, and
  Parrinello]{ceriotti2011accelerating}
Ceriotti,~M.; Manolopoulos,~D.~E.; Parrinello,~M. Accelerating the convergence
  of path integral dynamics with a generalized Langevin equation.
  \emph{{J}.~{C}hem.\ {P}hys.} \textbf{2011}, \emph{134}, 084104\relax
\mciteBstWouldAddEndPuncttrue
\mciteSetBstMidEndSepPunct{\mcitedefaultmidpunct}
{\mcitedefaultendpunct}{\mcitedefaultseppunct}\relax
\EndOfBibitem
\bibitem[Ceriotti \latin{et~al.}(2016)Ceriotti, Fang, Kusalik, McKenzie,
  Michaelides, Morales, and Markland]{ceriottimarkland2016}
Ceriotti,~M.; Fang,~W.; Kusalik,~P.~G.; McKenzie,~R.~H.; Michaelides,~A.;
  Morales,~M.~A.; Markland,~T.~E. Nuclear Quantum Effects in Water and Aqueous
  Systems: Experiment, Theory, and Current Challenges. \emph{Chemical Reviews}
  \textbf{2016}, \emph{116}, 7529--7550, PMID: 27049513\relax
\mciteBstWouldAddEndPuncttrue
\mciteSetBstMidEndSepPunct{\mcitedefaultmidpunct}
{\mcitedefaultendpunct}{\mcitedefaultseppunct}\relax
\EndOfBibitem
\bibitem[Markland and Ceriotti(2018)Markland, and Ceriotti]{Markland2018}
Markland,~T.~E.; Ceriotti,~M. Nuclear quantum effects enter the mainstream.
  \emph{Nature Reviews Chemistry} \textbf{2018}, \emph{2}, 0109\relax
\mciteBstWouldAddEndPuncttrue
\mciteSetBstMidEndSepPunct{\mcitedefaultmidpunct}
{\mcitedefaultendpunct}{\mcitedefaultseppunct}\relax
\EndOfBibitem
\bibitem[Law and Hassanali(2015)Law, and Hassanali]{law2015role}
Law,~Y.~K.; Hassanali,~A.~A. Role of quantum vibrations on the structural,
  electronic, and optical properties of 9-methylguanine. \emph{{J}.~{P}hys.
  {C}hem. {A}} \textbf{2015}, \emph{119}, 10816--10827\relax
\mciteBstWouldAddEndPuncttrue
\mciteSetBstMidEndSepPunct{\mcitedefaultmidpunct}
{\mcitedefaultendpunct}{\mcitedefaultseppunct}\relax
\EndOfBibitem
\bibitem[Law and Hassanali(2018)Law, and Hassanali]{lawhassanali2018}
Law,~Y.~K.; Hassanali,~A.~A. The importance of nuclear quantum effects in
  spectral line broadening of optical spectra and electrostatic properties in
  aromatic chromophores. \emph{The Journal of Chemical Physics} \textbf{2018},
  \emph{148}, 102331\relax
\mciteBstWouldAddEndPuncttrue
\mciteSetBstMidEndSepPunct{\mcitedefaultmidpunct}
{\mcitedefaultendpunct}{\mcitedefaultseppunct}\relax
\EndOfBibitem
\bibitem[Sappati \latin{et~al.}(2016)Sappati, Hassanali, Gebauer, and
  Ghosh]{sappati2016nuclear}
Sappati,~S.; Hassanali,~A.; Gebauer,~R.; Ghosh,~P. Nuclear quantum effects in a
  HIV/cancer inhibitor: The case of ellipticine. \emph{{J}.~{C}hem.\ {P}hys.}
  \textbf{2016}, \emph{145}, 205102\relax
\mciteBstWouldAddEndPuncttrue
\mciteSetBstMidEndSepPunct{\mcitedefaultmidpunct}
{\mcitedefaultendpunct}{\mcitedefaultseppunct}\relax
\EndOfBibitem
\bibitem[Rossi \latin{et~al.}(2016)Rossi, Gasparotto, and
  Ceriotti]{rossi2016anharmonic}
Rossi,~M.; Gasparotto,~P.; Ceriotti,~M. Anharmonic and quantum fluctuations in
  molecular crystals: A first-principles study of the stability of paracetamol.
  \emph{{P}hys.\ {R}ev.\ {L}ett.} \textbf{2016}, \emph{117}, 115702\relax
\mciteBstWouldAddEndPuncttrue
\mciteSetBstMidEndSepPunct{\mcitedefaultmidpunct}
{\mcitedefaultendpunct}{\mcitedefaultseppunct}\relax
\EndOfBibitem
\bibitem[Chen \latin{et~al.}(2016)Chen, Ambrosio, Miceli, and
  Pasquarello]{alfredo2016}
Chen,~W.; Ambrosio,~F.; Miceli,~G.; Pasquarello,~A. Ab initio Electronic
  Structure of Liquid Water. \emph{Phys. Rev. Lett.} \textbf{2016}, \emph{117},
  186401\relax
\mciteBstWouldAddEndPuncttrue
\mciteSetBstMidEndSepPunct{\mcitedefaultmidpunct}
{\mcitedefaultendpunct}{\mcitedefaultseppunct}\relax
\EndOfBibitem
\bibitem[Litman \latin{et~al.}(2019)Litman, Richardson, Kumagai, and
  Rossi]{litman2019elucidating}
Litman,~Y.; Richardson,~J.~O.; Kumagai,~T.; Rossi,~M. Elucidating the nuclear
  quantum dynamics of intramolecular double hydrogen transfer in porphycene.
  \emph{{J}.~{A}m.\ {C}hem.\ {S}oc.} \textbf{2019}, \emph{141},
  2526--2534\relax
\mciteBstWouldAddEndPuncttrue
\mciteSetBstMidEndSepPunct{\mcitedefaultmidpunct}
{\mcitedefaultendpunct}{\mcitedefaultseppunct}\relax
\EndOfBibitem
\bibitem[Markland and Ceriotti(2018)Markland, and
  Ceriotti]{Ceriotti_Nature2018}
Markland,~T.~E.; Ceriotti,~M. Nuclear quantum effects enter the mainstream.
  \emph{Nature Reviews Chemistry} \textbf{2018}, \emph{2}, 0109\relax
\mciteBstWouldAddEndPuncttrue
\mciteSetBstMidEndSepPunct{\mcitedefaultmidpunct}
{\mcitedefaultendpunct}{\mcitedefaultseppunct}\relax
\EndOfBibitem
\bibitem[Cassone(2020)]{Cassone_JPCL2020}
Cassone,~G. Nuclear Quantum Effects Largely Influence Molecular Dissociation
  and Proton Transfer in Liquid Water under an Electric Field. \emph{The
  Journal of Physical Chemistry Letters} \textbf{2020}, \emph{11}, 8983--8988,
  PMID: 33035059\relax
\mciteBstWouldAddEndPuncttrue
\mciteSetBstMidEndSepPunct{\mcitedefaultmidpunct}
{\mcitedefaultendpunct}{\mcitedefaultseppunct}\relax
\EndOfBibitem
\bibitem[Thomsen and Shiga(2021)Thomsen, and Shiga]{Shiga_JCP2021}
Thomsen,~B.; Shiga,~M. {Ab initio study of nuclear quantum effects on sub- and
  supercritical water}. \emph{The Journal of Chemical Physics} \textbf{2021},
  \emph{155}, 194107\relax
\mciteBstWouldAddEndPuncttrue
\mciteSetBstMidEndSepPunct{\mcitedefaultmidpunct}
{\mcitedefaultendpunct}{\mcitedefaultseppunct}\relax
\EndOfBibitem
\bibitem[Ceriotti \latin{et~al.}(2013)Ceriotti, Cuny, Parrinello, and
  Manolopoulos]{CeriottiCunyParrinelloManolopoulos2013}
Ceriotti,~M.; Cuny,~J.; Parrinello,~M.; Manolopoulos,~D.~E. Nuclear quantum
  effects and hydrogen bond fluctuations in water. \emph{Proceedings of the
  National Academy of Sciences} \textbf{2013}, \relax
\mciteBstWouldAddEndPunctfalse
\mciteSetBstMidEndSepPunct{\mcitedefaultmidpunct}
{}{\mcitedefaultseppunct}\relax
\EndOfBibitem
\bibitem[Marsalek and Markland(2017)Marsalek, and
  Markland]{marselekmarkland2017}
Marsalek,~O.; Markland,~T.~E. Quantum Dynamics and Spectroscopy of Ab Initio
  Liquid Water: The Interplay of Nuclear and Electronic Quantum Effects.
  \emph{The Journal of Physical Chemistry Letters} \textbf{2017}, \emph{8},
  1545--1551, PMID: 28296422\relax
\mciteBstWouldAddEndPuncttrue
\mciteSetBstMidEndSepPunct{\mcitedefaultmidpunct}
{\mcitedefaultendpunct}{\mcitedefaultseppunct}\relax
\EndOfBibitem
\bibitem[Tse \latin{et~al.}(2015)Tse, Knight, and Voth]{tse_analysis_2015}
Tse,~Y.-L.~S.; Knight,~C.; Voth,~G.~A. An analysis of hydrated proton diffusion
  in ab initio molecular dynamics. \emph{The Journal of Chemical Physics}
  \textbf{2015}, \emph{142}, 014104, \_eprint:
  https://pubs.aip.org/aip/jcp/article-pdf/doi/10.1063/1.4905077/13607981/014104\_1\_online.pdf\relax
\mciteBstWouldAddEndPuncttrue
\mciteSetBstMidEndSepPunct{\mcitedefaultmidpunct}
{\mcitedefaultendpunct}{\mcitedefaultseppunct}\relax
\EndOfBibitem
\bibitem[Chen \latin{et~al.}(2018)Chen, Zheng, Santra, Ko, DiStasio~Jr, Klein,
  Car, and Wu]{Chen2018}
Chen,~M.; Zheng,~L.; Santra,~B.; Ko,~H.-Y.; DiStasio~Jr,~R.~A.; Klein,~M.~L.;
  Car,~R.; Wu,~X. Hydroxide diffuses slower than hydronium in water because its
  solvated structure inhibits correlated proton transfer. \emph{Nature
  Chemistry} \textbf{2018}, \emph{10}, 413–419\relax
\mciteBstWouldAddEndPuncttrue
\mciteSetBstMidEndSepPunct{\mcitedefaultmidpunct}
{\mcitedefaultendpunct}{\mcitedefaultseppunct}\relax
\EndOfBibitem
\bibitem[Hassanali \latin{et~al.}(2013)Hassanali, Giberti, Cuny, Kühne, and
  Parrinello]{HassanaliGibertiCunyKuhneParrinello2013}
Hassanali,~A.; Giberti,~F.; Cuny,~J.; Kühne,~T.~D.; Parrinello,~M. Proton
  transfer through the water gossamer. \emph{Proceedings of the National
  Academy of Sciences} \textbf{2013}, 13723\relax
\mciteBstWouldAddEndPuncttrue
\mciteSetBstMidEndSepPunct{\mcitedefaultmidpunct}
{\mcitedefaultendpunct}{\mcitedefaultseppunct}\relax
\EndOfBibitem
\bibitem[Wang \latin{et~al.}(2010)Wang, Jenness, Al-Saidi, and
  Jordan]{wang2010assessment}
Wang,~F.-F.; Jenness,~G.; Al-Saidi,~W.; Jordan,~K. Assessment of the
  performance of common density functional methods for describing the
  interaction energies of (H$_2$O)$_6$ clusters. \emph{The Journal of chemical
  physics} \textbf{2010}, \emph{132}\relax
\mciteBstWouldAddEndPuncttrue
\mciteSetBstMidEndSepPunct{\mcitedefaultmidpunct}
{\mcitedefaultendpunct}{\mcitedefaultseppunct}\relax
\EndOfBibitem
\bibitem[Howard \latin{et~al.}(2015)Howard, Enyard, and
  Tschumper]{howard2015assessing}
Howard,~J.~C.; Enyard,~J.~D.; Tschumper,~G.~S. Assessing the accuracy of some
  popular DFT methods for computing harmonic vibrational frequencies of water
  clusters. \emph{The Journal of chemical physics} \textbf{2015},
  \emph{143}\relax
\mciteBstWouldAddEndPuncttrue
\mciteSetBstMidEndSepPunct{\mcitedefaultmidpunct}
{\mcitedefaultendpunct}{\mcitedefaultseppunct}\relax
\EndOfBibitem
\bibitem[Babin \latin{et~al.}(2013)Babin, Leforestier, and
  Paesani]{babin_v_2013_b}
Babin,~V.; Leforestier,~C.; Paesani,~F. {Development of a ``First Principles"
  Water Potential with Flexible Monomers: Dimer Potential Energy Surface, VRT
  Spectrum, and Second Virial Coefficient}. \emph{J. Chem. Theory Comput.}
  \textbf{2013}, \emph{9}, 5395--5403\relax
\mciteBstWouldAddEndPuncttrue
\mciteSetBstMidEndSepPunct{\mcitedefaultmidpunct}
{\mcitedefaultendpunct}{\mcitedefaultseppunct}\relax
\EndOfBibitem
\bibitem[Babin \latin{et~al.}(2014)Babin, Medders, and Paesani]{babin_v_2014}
Babin,~V.; Medders,~G.~R.; Paesani,~F. {Development of a ``First Principles''
  Water Potential with Flexible Monomers. II: Trimer Potential Energy Surface,
  Third Virial Coefficient, and Small Clusters}. \emph{J. Chem. Theory Comput.}
  \textbf{2014}, \emph{10}, 1599--1607\relax
\mciteBstWouldAddEndPuncttrue
\mciteSetBstMidEndSepPunct{\mcitedefaultmidpunct}
{\mcitedefaultendpunct}{\mcitedefaultseppunct}\relax
\EndOfBibitem
\bibitem[Medders \latin{et~al.}(2014)Medders, Babin, and
  Paesani]{medders_g_2014}
Medders,~G.~R.; Babin,~V.; Paesani,~F. {Development of a ``First-Principles"
  Water Potential with Flexible Monomers. III. Liquid Phase Properties}.
  \emph{J. Chem. Theory Comput.} \textbf{2014}, \emph{10}, 2906--2910\relax
\mciteBstWouldAddEndPuncttrue
\mciteSetBstMidEndSepPunct{\mcitedefaultmidpunct}
{\mcitedefaultendpunct}{\mcitedefaultseppunct}\relax
\EndOfBibitem
\bibitem[Paesani(2016)]{paesani_f_2016}
Paesani,~F. {Getting the Right Answers for the Right Reasons: Toward Predictive
  Molecular Simulations of Water with Many-Body Potential Energy Functions}.
  \emph{Acc. Chem. Res.} \textbf{2016}, \emph{49}, 1844--1851\relax
\mciteBstWouldAddEndPuncttrue
\mciteSetBstMidEndSepPunct{\mcitedefaultmidpunct}
{\mcitedefaultendpunct}{\mcitedefaultseppunct}\relax
\EndOfBibitem
\bibitem[Stone(2013)]{stones_book}
Stone,~A. \emph{The Theory of Intermolecular Forces}; OUP Oxford, 2013\relax
\mciteBstWouldAddEndPuncttrue
\mciteSetBstMidEndSepPunct{\mcitedefaultmidpunct}
{\mcitedefaultendpunct}{\mcitedefaultseppunct}\relax
\EndOfBibitem
\bibitem[Jensen(2017)]{jensen2017introduction}
Jensen,~F. \emph{Introduction to computational chemistry}; John wiley \& sons,
  2017\relax
\mciteBstWouldAddEndPuncttrue
\mciteSetBstMidEndSepPunct{\mcitedefaultmidpunct}
{\mcitedefaultendpunct}{\mcitedefaultseppunct}\relax
\EndOfBibitem
\bibitem[Cisneros \latin{et~al.}(2016)Cisneros, Wikfeldt, Ojam\"{a}e, Lu, Xu,
  Torabifard, Bart{\'o}k, Cs{\'a}nyi, Molinero, and
  Paesani]{cisneros2016modeling}
Cisneros,~G.~A.; Wikfeldt,~K.~T.; Ojam\"{a}e,~L.; Lu,~J.; Xu,~Y.;
  Torabifard,~H.; Bart{\'o}k,~A.~P.; Cs{\'a}nyi,~G.; Molinero,~V.; Paesani,~F.
  Modeling molecular interactions in water: From pairwise to many-body
  potential energy functions. \emph{Chemical reviews} \textbf{2016},
  \emph{116}, 7501--7528\relax
\mciteBstWouldAddEndPuncttrue
\mciteSetBstMidEndSepPunct{\mcitedefaultmidpunct}
{\mcitedefaultendpunct}{\mcitedefaultseppunct}\relax
\EndOfBibitem
\bibitem[Umeyama and Morokuma(1977)Umeyama, and Morokuma]{umeyama1977origin}
Umeyama,~H.; Morokuma,~K. The origin of hydrogen bonding. An energy
  decomposition study. \emph{Journal of the American Chemical Society}
  \textbf{1977}, \emph{99}, 1316--1332\relax
\mciteBstWouldAddEndPuncttrue
\mciteSetBstMidEndSepPunct{\mcitedefaultmidpunct}
{\mcitedefaultendpunct}{\mcitedefaultseppunct}\relax
\EndOfBibitem
\bibitem[Chen and Gordon(1996)Chen, and Gordon]{chen1996energy}
Chen,~W.; Gordon,~M.~S. Energy decomposition analyses for many-body interaction
  and applications to water complexes. \emph{The Journal of Physical Chemistry}
  \textbf{1996}, \emph{100}, 14316--14328\relax
\mciteBstWouldAddEndPuncttrue
\mciteSetBstMidEndSepPunct{\mcitedefaultmidpunct}
{\mcitedefaultendpunct}{\mcitedefaultseppunct}\relax
\EndOfBibitem
\bibitem[Mo \latin{et~al.}(2000)Mo, Gao, and Peyerimhoff]{mo2000energy}
Mo,~Y.; Gao,~J.; Peyerimhoff,~S.~D. Energy decomposition analysis of
  intermolecular interactions using a block-localized wave function approach.
  \emph{The Journal of Chemical Physics} \textbf{2000}, \emph{112},
  5530--5538\relax
\mciteBstWouldAddEndPuncttrue
\mciteSetBstMidEndSepPunct{\mcitedefaultmidpunct}
{\mcitedefaultendpunct}{\mcitedefaultseppunct}\relax
\EndOfBibitem
\bibitem[Glendening(2005)]{glendening2005natural}
Glendening,~E.~D. Natural energy decomposition analysis: Extension to density
  functional methods and analysis of cooperative effects in water clusters.
  \emph{The Journal of Physical Chemistry A} \textbf{2005}, \emph{109},
  11936--11940\relax
\mciteBstWouldAddEndPuncttrue
\mciteSetBstMidEndSepPunct{\mcitedefaultmidpunct}
{\mcitedefaultendpunct}{\mcitedefaultseppunct}\relax
\EndOfBibitem
\bibitem[van Duijneveldt-van~de Rijdt \latin{et~al.}(2003)van
  Duijneveldt-van~de Rijdt, Mooij, and van Duijneveldt]{van2003testing}
van Duijneveldt-van~de Rijdt,~J.; Mooij,~W.; van Duijneveldt,~F. Testing the
  quality of some recent water--water potentials. \emph{Physical Chemistry
  Chemical Physics} \textbf{2003}, \emph{5}, 1169--1180\relax
\mciteBstWouldAddEndPuncttrue
\mciteSetBstMidEndSepPunct{\mcitedefaultmidpunct}
{\mcitedefaultendpunct}{\mcitedefaultseppunct}\relax
\EndOfBibitem
\bibitem[Khaliullin \latin{et~al.}(2007)Khaliullin, Cobar, Lochan, Bell, and
  Head-Gordon]{khaliullin2007unravelling}
Khaliullin,~R.~Z.; Cobar,~E.~A.; Lochan,~R.~C.; Bell,~A.~T.; Head-Gordon,~M.
  Unravelling the origin of intermolecular interactions using absolutely
  localized molecular orbitals. \emph{The Journal of Physical Chemistry A}
  \textbf{2007}, \emph{111}, 8753--8765\relax
\mciteBstWouldAddEndPuncttrue
\mciteSetBstMidEndSepPunct{\mcitedefaultmidpunct}
{\mcitedefaultendpunct}{\mcitedefaultseppunct}\relax
\EndOfBibitem
\bibitem[Khaliullin \latin{et~al.}(2009)Khaliullin, Bell, and
  Head-Gordon]{khaliullin2009electron}
Khaliullin,~R.~Z.; Bell,~A.~T.; Head-Gordon,~M. Electron donation in the
  water--water hydrogen bond. \emph{Chemistry--A European Journal}
  \textbf{2009}, \emph{15}, 851--855\relax
\mciteBstWouldAddEndPuncttrue
\mciteSetBstMidEndSepPunct{\mcitedefaultmidpunct}
{\mcitedefaultendpunct}{\mcitedefaultseppunct}\relax
\EndOfBibitem
\bibitem[Khaliullin and K{\"u}hne(2013)Khaliullin, and
  K{\"u}hne]{khaliullin2013microscopic}
Khaliullin,~R.~Z.; K{\"u}hne,~T.~D. Microscopic properties of liquid water from
  combined ab initio molecular dynamics and energy decomposition studies.
  \emph{Physical Chemistry Chemical Physics} \textbf{2013}, \emph{15},
  15746--15766\relax
\mciteBstWouldAddEndPuncttrue
\mciteSetBstMidEndSepPunct{\mcitedefaultmidpunct}
{\mcitedefaultendpunct}{\mcitedefaultseppunct}\relax
\EndOfBibitem
\bibitem[K\"{u}hne and Khaliullin(2014)K\"{u}hne, and
  Khaliullin]{kuhne2014nature}
K\"{u}hne,~T.~D.; Khaliullin,~R.~Z. Nature of the asymmetry in the
  hydrogen-bond networks of hexagonal ice and liquid water. \emph{Journal of
  the American Chemical Society} \textbf{2014}, \emph{136}, 3395--3399\relax
\mciteBstWouldAddEndPuncttrue
\mciteSetBstMidEndSepPunct{\mcitedefaultmidpunct}
{\mcitedefaultendpunct}{\mcitedefaultseppunct}\relax
\EndOfBibitem
\bibitem[Mao \latin{et~al.}(2017)Mao, Horn, and Head-Gordon]{mao2017energy}
Mao,~Y.; Horn,~P.~R.; Head-Gordon,~M. Energy decomposition analysis in an
  adiabatic picture. \emph{Physical Chemistry Chemical Physics} \textbf{2017},
  \emph{19}, 5944--5958\relax
\mciteBstWouldAddEndPuncttrue
\mciteSetBstMidEndSepPunct{\mcitedefaultmidpunct}
{\mcitedefaultendpunct}{\mcitedefaultseppunct}\relax
\EndOfBibitem
\bibitem[Das \latin{et~al.}(2019)Das, Urban, Leven, Loipersberger, Aldossary,
  Head-Gordon, and Head-Gordon]{das2019development}
Das,~A.~K.; Urban,~L.; Leven,~I.; Loipersberger,~M.; Aldossary,~A.;
  Head-Gordon,~M.; Head-Gordon,~T. Development of an advanced force field for
  water using variational energy decomposition analysis. \emph{Journal of
  chemical theory and computation} \textbf{2019}, \emph{15}, 5001--5013\relax
\mciteBstWouldAddEndPuncttrue
\mciteSetBstMidEndSepPunct{\mcitedefaultmidpunct}
{\mcitedefaultendpunct}{\mcitedefaultseppunct}\relax
\EndOfBibitem
\bibitem[Egan \latin{et~al.}(2020)Egan, Bizzarro, Riera, and
  Paesani]{egan2020nature}
Egan,~C.~K.; Bizzarro,~B.~B.; Riera,~M.; Paesani,~F. Nature of alkali
  ion--water interactions: Insights from many-body representations and density
  functional theory. II. \emph{Journal of Chemical Theory and Computation}
  \textbf{2020}, \emph{16}, 3055--3072\relax
\mciteBstWouldAddEndPuncttrue
\mciteSetBstMidEndSepPunct{\mcitedefaultmidpunct}
{\mcitedefaultendpunct}{\mcitedefaultseppunct}\relax
\EndOfBibitem
\bibitem[Palos \latin{et~al.}(2022)Palos, Lambros, Swee, Hu, Dasgupta, and
  Paesani]{palos2022assessing}
Palos,~E.; Lambros,~E.; Swee,~S.; Hu,~J.; Dasgupta,~S.; Paesani,~F. Assessing
  the interplay between functional-driven and density-driven errors in DFT
  models of water. \emph{Journal of Chemical Theory and Computation}
  \textbf{2022}, \emph{18}, 3410--3426\relax
\mciteBstWouldAddEndPuncttrue
\mciteSetBstMidEndSepPunct{\mcitedefaultmidpunct}
{\mcitedefaultendpunct}{\mcitedefaultseppunct}\relax
\EndOfBibitem
\bibitem[Hankins \latin{et~al.}(1970)Hankins, Moskowitz, and
  Stillinger]{hankins1970water}
Hankins,~D.; Moskowitz,~J.; Stillinger,~F. Water molecule interactions.
  \emph{The Journal of Chemical Physics} \textbf{1970}, \emph{53},
  4544--4554\relax
\mciteBstWouldAddEndPuncttrue
\mciteSetBstMidEndSepPunct{\mcitedefaultmidpunct}
{\mcitedefaultendpunct}{\mcitedefaultseppunct}\relax
\EndOfBibitem
\bibitem[Mark and Nilsson(2001)Mark, and Nilsson]{mark2001structure}
Mark,~P.; Nilsson,~L. Structure and dynamics of the TIP3P, SPC, and SPC/E water
  models at 298 K. \emph{The Journal of Physical Chemistry A} \textbf{2001},
  \emph{105}, 9954--9960\relax
\mciteBstWouldAddEndPuncttrue
\mciteSetBstMidEndSepPunct{\mcitedefaultmidpunct}
{\mcitedefaultendpunct}{\mcitedefaultseppunct}\relax
\EndOfBibitem
\bibitem[Zielkiewicz(2005)]{zielkiewicz2005structural}
Zielkiewicz,~J. Structural properties of water: Comparison of the SPC, SPCE,
  TIP4P, and TIP5P models of water. \emph{The Journal of chemical physics}
  \textbf{2005}, \emph{123}\relax
\mciteBstWouldAddEndPuncttrue
\mciteSetBstMidEndSepPunct{\mcitedefaultmidpunct}
{\mcitedefaultendpunct}{\mcitedefaultseppunct}\relax
\EndOfBibitem
\bibitem[Berendsen \latin{et~al.}(1987)Berendsen, Grigera, and
  Straatsma]{berendsen1987missing}
Berendsen,~H.~J.; Grigera,~J.~R.; Straatsma,~T.~P. The missing term in
  effective pair potentials. \emph{Journal of Physical Chemistry}
  \textbf{1987}, \emph{91}, 6269--6271\relax
\mciteBstWouldAddEndPuncttrue
\mciteSetBstMidEndSepPunct{\mcitedefaultmidpunct}
{\mcitedefaultendpunct}{\mcitedefaultseppunct}\relax
\EndOfBibitem
\bibitem[Berweger \latin{et~al.}(1995)Berweger, van Gunsteren, and
  M{\"u}ller-Plathe]{berweger1995force}
Berweger,~C.~D.; van Gunsteren,~W.~F.; M{\"u}ller-Plathe,~F. Force field
  parametrization by weak coupling. Re-engineering SPC water. \emph{Chemical
  physics letters} \textbf{1995}, \emph{232}, 429--436\relax
\mciteBstWouldAddEndPuncttrue
\mciteSetBstMidEndSepPunct{\mcitedefaultmidpunct}
{\mcitedefaultendpunct}{\mcitedefaultseppunct}\relax
\EndOfBibitem
\bibitem[Jorgensen \latin{et~al.}(1983)Jorgensen, Chandrasekhar, Madura, Impey,
  and Klein]{jorgensen1983comparison}
Jorgensen,~W.~L.; Chandrasekhar,~J.; Madura,~J.~D.; Impey,~R.~W.; Klein,~M.~L.
  Comparison of simple potential functions for simulating liquid water.
  \emph{The Journal of chemical physics} \textbf{1983}, \emph{79},
  926--935\relax
\mciteBstWouldAddEndPuncttrue
\mciteSetBstMidEndSepPunct{\mcitedefaultmidpunct}
{\mcitedefaultendpunct}{\mcitedefaultseppunct}\relax
\EndOfBibitem
\bibitem[Mahoney and Jorgensen(2000)Mahoney, and Jorgensen]{mahoney2000five}
Mahoney,~M.~W.; Jorgensen,~W.~L. A five-site model for liquid water and the
  reproduction of the density anomaly by rigid, nonpolarizable potential
  functions. \emph{The Journal of chemical physics} \textbf{2000}, \emph{112},
  8910--8922\relax
\mciteBstWouldAddEndPuncttrue
\mciteSetBstMidEndSepPunct{\mcitedefaultmidpunct}
{\mcitedefaultendpunct}{\mcitedefaultseppunct}\relax
\EndOfBibitem
\bibitem[Horn \latin{et~al.}(2004)Horn, Swope, Pitera, Madura, Dick, Hura, and
  Head-Gordon]{horn2004development}
Horn,~H.~W.; Swope,~W.~C.; Pitera,~J.~W.; Madura,~J.~D.; Dick,~T.~J.;
  Hura,~G.~L.; Head-Gordon,~T. Development of an improved four-site water model
  for biomolecular simulations: TIP4P-Ew. \emph{The Journal of chemical
  physics} \textbf{2004}, \emph{120}, 9665--9678\relax
\mciteBstWouldAddEndPuncttrue
\mciteSetBstMidEndSepPunct{\mcitedefaultmidpunct}
{\mcitedefaultendpunct}{\mcitedefaultseppunct}\relax
\EndOfBibitem
\bibitem[Xantheas(1994)]{xantheas1994ab}
Xantheas,~S.~S. Ab initio studies of cyclic water clusters (H2O) n, n= 1--6.
  II. Analysis of many-body interactions. \emph{The Journal of chemical
  physics} \textbf{1994}, \emph{100}, 7523--7534\relax
\mciteBstWouldAddEndPuncttrue
\mciteSetBstMidEndSepPunct{\mcitedefaultmidpunct}
{\mcitedefaultendpunct}{\mcitedefaultseppunct}\relax
\EndOfBibitem
\bibitem[Cui \latin{et~al.}(2006)Cui, Liu, and Jordan]{cui2006theoretical}
Cui,~J.; Liu,~H.; Jordan,~K.~D. Theoretical Characterization of the (H2O) 21
  Cluster: Application of an n-body Decomposition Procedure. \emph{The Journal
  of Physical Chemistry B} \textbf{2006}, \emph{110}, 18872--18878\relax
\mciteBstWouldAddEndPuncttrue
\mciteSetBstMidEndSepPunct{\mcitedefaultmidpunct}
{\mcitedefaultendpunct}{\mcitedefaultseppunct}\relax
\EndOfBibitem
\bibitem[G{\'o}ra \latin{et~al.}(2011)G{\'o}ra, Podeszwa, Cencek, and
  Szalewicz]{gora2011interaction}
G{\'o}ra,~U.; Podeszwa,~R.; Cencek,~W.; Szalewicz,~K. Interaction energies of
  large clusters from many-body expansion. \emph{The Journal of chemical
  physics} \textbf{2011}, \emph{135}\relax
\mciteBstWouldAddEndPuncttrue
\mciteSetBstMidEndSepPunct{\mcitedefaultmidpunct}
{\mcitedefaultendpunct}{\mcitedefaultseppunct}\relax
\EndOfBibitem
\bibitem[Heindel and Xantheas(2020)Heindel, and Xantheas]{heindel2020many}
Heindel,~J.~P.; Xantheas,~S.~S. The many-body expansion for aqueous systems
  revisited: I. Water--water interactions. \emph{Journal of Chemical Theory and
  Computation} \textbf{2020}, \emph{16}, 6843--6855\relax
\mciteBstWouldAddEndPuncttrue
\mciteSetBstMidEndSepPunct{\mcitedefaultmidpunct}
{\mcitedefaultendpunct}{\mcitedefaultseppunct}\relax
\EndOfBibitem
\bibitem[Bukowski \latin{et~al.}(2007)Bukowski, Szalewicz, Groenenboom, and
  Van~der Avoird]{bukowski2007predictions}
Bukowski,~R.; Szalewicz,~K.; Groenenboom,~G.~C.; Van~der Avoird,~A. Predictions
  of the properties of water from first principles. \emph{Science}
  \textbf{2007}, \emph{315}, 1249--1252\relax
\mciteBstWouldAddEndPuncttrue
\mciteSetBstMidEndSepPunct{\mcitedefaultmidpunct}
{\mcitedefaultendpunct}{\mcitedefaultseppunct}\relax
\EndOfBibitem
\bibitem[Bukowski \latin{et~al.}(2008)Bukowski, Szalewicz, Groenenboom, and
  van~der Avoird]{bukowski2008polarizable}
Bukowski,~R.; Szalewicz,~K.; Groenenboom,~G.~C.; van~der Avoird,~A. Polarizable
  interaction potential for water from coupled cluster calculations. I.
  Analysis of dimer potential energy surface. \emph{The Journal of chemical
  physics} \textbf{2008}, \emph{128}\relax
\mciteBstWouldAddEndPuncttrue
\mciteSetBstMidEndSepPunct{\mcitedefaultmidpunct}
{\mcitedefaultendpunct}{\mcitedefaultseppunct}\relax
\EndOfBibitem
\bibitem[Cencek \latin{et~al.}(2008)Cencek, Szalewicz, Leforestier, van
  Harrevelt, and van~der Avoird]{cencek2008accurate}
Cencek,~W.; Szalewicz,~K.; Leforestier,~C.; van Harrevelt,~R.; van~der
  Avoird,~A. An accurate analytic representation of the water pair potential.
  \emph{Physical Chemistry Chemical Physics} \textbf{2008}, \emph{10},
  4716--4731\relax
\mciteBstWouldAddEndPuncttrue
\mciteSetBstMidEndSepPunct{\mcitedefaultmidpunct}
{\mcitedefaultendpunct}{\mcitedefaultseppunct}\relax
\EndOfBibitem
\bibitem[G{\'o}ra \latin{et~al.}(2014)G{\'o}ra, Cencek, Podeszwa, van~der
  Avoird, and Szalewicz]{gora2014predictions}
G{\'o}ra,~U.; Cencek,~W.; Podeszwa,~R.; van~der Avoird,~A.; Szalewicz,~K.
  Predictions for water clusters from a first-principles two-and three-body
  force field. \emph{The Journal of Chemical Physics} \textbf{2014},
  \emph{140}\relax
\mciteBstWouldAddEndPuncttrue
\mciteSetBstMidEndSepPunct{\mcitedefaultmidpunct}
{\mcitedefaultendpunct}{\mcitedefaultseppunct}\relax
\EndOfBibitem
\bibitem[Jankowski \latin{et~al.}(2015)Jankowski, Murdachaew, Bukowski,
  Akin-Ojo, Leforestier, and Szalewicz]{jankowski2015ab}
Jankowski,~P.; Murdachaew,~G.; Bukowski,~R.; Akin-Ojo,~O.; Leforestier,~C.;
  Szalewicz,~K. Ab initio water pair potential with flexible monomers.
  \emph{The Journal of Physical Chemistry A} \textbf{2015}, \emph{119},
  2940--2964\relax
\mciteBstWouldAddEndPuncttrue
\mciteSetBstMidEndSepPunct{\mcitedefaultmidpunct}
{\mcitedefaultendpunct}{\mcitedefaultseppunct}\relax
\EndOfBibitem
\bibitem[Huang \latin{et~al.}(2006)Huang, Braams, and Bowman]{huang2006ab}
Huang,~X.; Braams,~B.~J.; Bowman,~J.~M. Ab initio potential energy and dipole
  moment surfaces of (H2O) 2. \emph{The Journal of Physical Chemistry A}
  \textbf{2006}, \emph{110}, 445--451\relax
\mciteBstWouldAddEndPuncttrue
\mciteSetBstMidEndSepPunct{\mcitedefaultmidpunct}
{\mcitedefaultendpunct}{\mcitedefaultseppunct}\relax
\EndOfBibitem
\bibitem[Wang \latin{et~al.}(2009)Wang, Shepler, Braams, and
  Bowman]{wang2009full}
Wang,~Y.; Shepler,~B.~C.; Braams,~B.~J.; Bowman,~J.~M. Full-dimensional, ab
  initio potential energy and dipole moment surfaces for water. \emph{The
  Journal of chemical physics} \textbf{2009}, \emph{131}\relax
\mciteBstWouldAddEndPuncttrue
\mciteSetBstMidEndSepPunct{\mcitedefaultmidpunct}
{\mcitedefaultendpunct}{\mcitedefaultseppunct}\relax
\EndOfBibitem
\bibitem[Wang \latin{et~al.}(2011)Wang, Huang, Shepler, Braams, and
  Bowman]{wang2011flexible}
Wang,~Y.; Huang,~X.; Shepler,~B.~C.; Braams,~B.~J.; Bowman,~J.~M. Flexible, ab
  initio potential, and dipole moment surfaces for water. I. Tests and
  applications for clusters up to the 22-mer. \emph{The Journal of chemical
  physics} \textbf{2011}, \emph{134}\relax
\mciteBstWouldAddEndPuncttrue
\mciteSetBstMidEndSepPunct{\mcitedefaultmidpunct}
{\mcitedefaultendpunct}{\mcitedefaultseppunct}\relax
\EndOfBibitem
\bibitem[Babin \latin{et~al.}(2012)Babin, Medders, and
  Paesani]{babin2012toward}
Babin,~V.; Medders,~G.~R.; Paesani,~F. Toward a universal water model: First
  principles simulations from the dimer to the liquid phase. \emph{The journal
  of physical chemistry letters} \textbf{2012}, \emph{3}, 3765--3769\relax
\mciteBstWouldAddEndPuncttrue
\mciteSetBstMidEndSepPunct{\mcitedefaultmidpunct}
{\mcitedefaultendpunct}{\mcitedefaultseppunct}\relax
\EndOfBibitem
\bibitem[Purvis~III and Bartlett(1982)Purvis~III, and Bartlett]{purvis_gd_1982}
Purvis~III,~G.~D.; Bartlett,~R.~J. {A Full Coupled-Cluster Singles and Doubles
  Model: The Inclusion of Disconnected Triples}. \emph{J. Chem. Phys.}
  \textbf{1982}, \emph{76}, 1910--1918\relax
\mciteBstWouldAddEndPuncttrue
\mciteSetBstMidEndSepPunct{\mcitedefaultmidpunct}
{\mcitedefaultendpunct}{\mcitedefaultseppunct}\relax
\EndOfBibitem
\bibitem[Raghavachari \latin{et~al.}(1989)Raghavachari, Trucks, Pople, and
  Head-Gordon]{raghavachari_k_1989}
Raghavachari,~K.; Trucks,~G.~W.; Pople,~J.~A.; Head-Gordon,~M. {A Fifth-Order
  Perturbation Comparison of Electron Correlation Theories}. \emph{Chem. Phys.
  Lett} \textbf{1989}, \emph{157}, 479--483\relax
\mciteBstWouldAddEndPuncttrue
\mciteSetBstMidEndSepPunct{\mcitedefaultmidpunct}
{\mcitedefaultendpunct}{\mcitedefaultseppunct}\relax
\EndOfBibitem
\bibitem[Adler \latin{et~al.}(2007)Adler, Knizia, and Werner]{adler_tb_2007}
Adler,~T.~B.; Knizia,~G.; Werner,~H.~J. {A Simple and Efficient CCSD(T)-F12
  Approximation}. \emph{J. Chem. Phys.} \textbf{2007}, \emph{127}, 221106\relax
\mciteBstWouldAddEndPuncttrue
\mciteSetBstMidEndSepPunct{\mcitedefaultmidpunct}
{\mcitedefaultendpunct}{\mcitedefaultseppunct}\relax
\EndOfBibitem
\bibitem[Knizia \latin{et~al.}(2009)Knizia, Adler, and Werner]{knizia_g_2009}
Knizia,~G.; Adler,~T.~B.; Werner,~H.~J. {Simplified CCSD(T)-F12 Methods: Theory
  and Benchmarks}. \emph{J. Chem. Phys.} \textbf{2009}, \emph{130},
  054104\relax
\mciteBstWouldAddEndPuncttrue
\mciteSetBstMidEndSepPunct{\mcitedefaultmidpunct}
{\mcitedefaultendpunct}{\mcitedefaultseppunct}\relax
\EndOfBibitem
\bibitem[Sun \latin{et~al.}(2018)Sun, Zheng, Chen, Klein, Paesani, and
  Wu]{sun2018electron}
Sun,~Z.; Zheng,~L.; Chen,~M.; Klein,~M.~L.; Paesani,~F.; Wu,~X. Electron-hole
  theory of the effect of quantum nuclei on the X-ray absorption spectra of
  liquid water. \emph{Physical Review Letters} \textbf{2018}, \emph{121},
  137401\relax
\mciteBstWouldAddEndPuncttrue
\mciteSetBstMidEndSepPunct{\mcitedefaultmidpunct}
{\mcitedefaultendpunct}{\mcitedefaultseppunct}\relax
\EndOfBibitem
\bibitem[Gartner~III \latin{et~al.}(2022)Gartner~III, Hunter, Lambros, Caruso,
  Riera, Medders, Panagiotopoulos, Debenedetti, and
  Paesani]{gartner2022anomalies}
Gartner~III,~T.~E.; Hunter,~K.~M.; Lambros,~E.; Caruso,~A.; Riera,~M.;
  Medders,~G.~R.; Panagiotopoulos,~A.~Z.; Debenedetti,~P.~G.; Paesani,~F.
  Anomalies and local structure of liquid water from boiling to the supercooled
  regime as predicted by the many-body MB-pol model. \emph{The Journal of
  Physical Chemistry Letters} \textbf{2022}, \emph{13}, 3652--3658\relax
\mciteBstWouldAddEndPuncttrue
\mciteSetBstMidEndSepPunct{\mcitedefaultmidpunct}
{\mcitedefaultendpunct}{\mcitedefaultseppunct}\relax
\EndOfBibitem
\bibitem[Zhuang \latin{et~al.}(2022)Zhuang, Riera, Zhou, Deary, and
  Paesani]{zhuang2022hydration}
Zhuang,~D.; Riera,~M.; Zhou,~R.; Deary,~A.; Paesani,~F. Hydration Structure of
  Na+ and K+ Ions in Solution Predicted by Data-Driven Many-Body Potentials.
  \emph{The Journal of Physical Chemistry B} \textbf{2022}, \emph{126},
  9349--9360\relax
\mciteBstWouldAddEndPuncttrue
\mciteSetBstMidEndSepPunct{\mcitedefaultmidpunct}
{\mcitedefaultendpunct}{\mcitedefaultseppunct}\relax
\EndOfBibitem
\bibitem[Nguyen \latin{et~al.}(2018)Nguyen, Sz{\'e}kely, Imbalzano, Behler,
  Cs{\'a}nyi, Ceriotti, G{\"o}tz, and Paesani]{nguyen2018comparison}
Nguyen,~T.~T.; Sz{\'e}kely,~E.; Imbalzano,~G.; Behler,~J.; Cs{\'a}nyi,~G.;
  Ceriotti,~M.; G{\"o}tz,~A.~W.; Paesani,~F. Comparison of permutationally
  invariant polynomials, neural networks, and Gaussian approximation potentials
  in representing water interactions through many-body expansions. \emph{The
  Journal of chemical physics} \textbf{2018}, \emph{148}\relax
\mciteBstWouldAddEndPuncttrue
\mciteSetBstMidEndSepPunct{\mcitedefaultmidpunct}
{\mcitedefaultendpunct}{\mcitedefaultseppunct}\relax
\EndOfBibitem
\bibitem[Zhu \latin{et~al.}(2023)Zhu, Riera, Bull-Vulpe, and
  Paesani]{zhu2023mb}
Zhu,~X.; Riera,~M.; Bull-Vulpe,~E.~F.; Paesani,~F. MB-pol (2023): Sub-chemical
  Accuracy for Water Simulations from the Gas to the Liquid Phase.
  \emph{Journal of Chemical Theory and Computation} \textbf{2023}, \relax
\mciteBstWouldAddEndPunctfalse
\mciteSetBstMidEndSepPunct{\mcitedefaultmidpunct}
{}{\mcitedefaultseppunct}\relax
\EndOfBibitem
\bibitem[Bajaj \latin{et~al.}(2016)Bajaj, Gotz, and Paesani]{bajaj2016toward}
Bajaj,~P.; Gotz,~A.~W.; Paesani,~F. Toward chemical accuracy in the description
  of ion--water interactions through many-body representations. I.
  Halide--water dimer potential energy surfaces. \emph{Journal of chemical
  theory and computation} \textbf{2016}, \emph{12}, 2698--2705\relax
\mciteBstWouldAddEndPuncttrue
\mciteSetBstMidEndSepPunct{\mcitedefaultmidpunct}
{\mcitedefaultendpunct}{\mcitedefaultseppunct}\relax
\EndOfBibitem
\bibitem[Riera \latin{et~al.}(2017)Riera, Mardirossian, Bajaj, G{\"o}tz, and
  Paesani]{riera_m_2017}
Riera,~M.; Mardirossian,~N.; Bajaj,~P.; G{\"o}tz,~A.~W.; Paesani,~F. {Toward
  Chemical Accuracy in the Description of Ion$-$Water Interactions Through
  Many-Body Representations. Alkali-Water Dimer Potential Energy Surfaces}.
  \emph{J. Chem. Phys.} \textbf{2017}, \emph{147}, 161715\relax
\mciteBstWouldAddEndPuncttrue
\mciteSetBstMidEndSepPunct{\mcitedefaultmidpunct}
{\mcitedefaultendpunct}{\mcitedefaultseppunct}\relax
\EndOfBibitem
\bibitem[Paesani \latin{et~al.}(2019)Paesani, Bajaj, and
  Riera]{paesani2019chemical}
Paesani,~F.; Bajaj,~P.; Riera,~M. Chemical accuracy in modeling halide ion
  hydration from many-body representations. \emph{Advances in Physics: X}
  \textbf{2019}, \emph{4}, 1631212\relax
\mciteBstWouldAddEndPuncttrue
\mciteSetBstMidEndSepPunct{\mcitedefaultmidpunct}
{\mcitedefaultendpunct}{\mcitedefaultseppunct}\relax
\EndOfBibitem
\bibitem[Bizzarro \latin{et~al.}(2019)Bizzarro, Egan, and
  Paesani]{bizzarro2019nature}
Bizzarro,~B.~B.; Egan,~C.~K.; Paesani,~F. Nature of halide--water interactions:
  Insights from many-body representations and density functional theory.
  \emph{Journal of Chemical Theory and Computation} \textbf{2019}, \emph{15},
  2983--2995\relax
\mciteBstWouldAddEndPuncttrue
\mciteSetBstMidEndSepPunct{\mcitedefaultmidpunct}
{\mcitedefaultendpunct}{\mcitedefaultseppunct}\relax
\EndOfBibitem
\bibitem[Zhuang \latin{et~al.}(2019)Zhuang, Riera, Schenter, Fulton, and
  Paesani]{zhuang2019many}
Zhuang,~D.; Riera,~M.; Schenter,~G.~K.; Fulton,~J.~L.; Paesani,~F. Many-body
  effects determine the local hydration structure of Cs+ in solution. \emph{The
  journal of physical chemistry letters} \textbf{2019}, \emph{10},
  406--412\relax
\mciteBstWouldAddEndPuncttrue
\mciteSetBstMidEndSepPunct{\mcitedefaultmidpunct}
{\mcitedefaultendpunct}{\mcitedefaultseppunct}\relax
\EndOfBibitem
\bibitem[Caruso and Paesani(2021)Caruso, and Paesani]{caruso2021data}
Caruso,~A.; Paesani,~F. Data-driven many-body models enable a quantitative
  description of chloride hydration from clusters to bulk. \emph{The Journal of
  Chemical Physics} \textbf{2021}, \emph{155}\relax
\mciteBstWouldAddEndPuncttrue
\mciteSetBstMidEndSepPunct{\mcitedefaultmidpunct}
{\mcitedefaultendpunct}{\mcitedefaultseppunct}\relax
\EndOfBibitem
\bibitem[Caruso \latin{et~al.}(2022)Caruso, Zhu, Fulton, and
  Paesani]{caruso2022accurate}
Caruso,~A.; Zhu,~X.; Fulton,~J.~L.; Paesani,~F. Accurate modeling of bromide
  and iodide hydration with data-driven many-body potentials. \emph{The Journal
  of Physical Chemistry B} \textbf{2022}, \emph{126}, 8266--8278\relax
\mciteBstWouldAddEndPuncttrue
\mciteSetBstMidEndSepPunct{\mcitedefaultmidpunct}
{\mcitedefaultendpunct}{\mcitedefaultseppunct}\relax
\EndOfBibitem
\bibitem[Wernet \latin{et~al.}(2004)Wernet, Nordlund, Bergmann, Cavalleri,
  Odelius, Ogasawara, Naslund, Hirsch, Ojamae, Glatzel, \latin{et~al.}
  others]{wernet2004structure}
Wernet,~P.; Nordlund,~D.; Bergmann,~U.; Cavalleri,~M.; Odelius,~M.;
  Ogasawara,~H.; Naslund,~L.~A.; Hirsch,~T.; Ojamae,~L.; Glatzel,~P.,
  \latin{et~al.}  The structure of the first coordination shell in liquid
  water. \emph{Science} \textbf{2004}, \emph{304}, 995--999\relax
\mciteBstWouldAddEndPuncttrue
\mciteSetBstMidEndSepPunct{\mcitedefaultmidpunct}
{\mcitedefaultendpunct}{\mcitedefaultseppunct}\relax
\EndOfBibitem
\bibitem[Fransson \latin{et~al.}(2016)Fransson, Harada, Kosugi, Besley, Winter,
  Rehr, Pettersson, and Nilsson]{fransson2016x}
Fransson,~T.; Harada,~Y.; Kosugi,~N.; Besley,~N.~A.; Winter,~B.; Rehr,~J.~J.;
  Pettersson,~L.~G.; Nilsson,~A. X-ray and electron spectroscopy of water.
  \emph{Chemical reviews} \textbf{2016}, \emph{116}, 7551--7569\relax
\mciteBstWouldAddEndPuncttrue
\mciteSetBstMidEndSepPunct{\mcitedefaultmidpunct}
{\mcitedefaultendpunct}{\mcitedefaultseppunct}\relax
\EndOfBibitem
\bibitem[Kell(1967)]{kell1967precise}
Kell,~G. Precise representation of volume properties of water at one
  atmosphere. \emph{Journal of Chemical and Engineering data} \textbf{1967},
  \emph{12}, 66--69\relax
\mciteBstWouldAddEndPuncttrue
\mciteSetBstMidEndSepPunct{\mcitedefaultmidpunct}
{\mcitedefaultendpunct}{\mcitedefaultseppunct}\relax
\EndOfBibitem
\bibitem[Skinner \latin{et~al.}(2014)Skinner, Benmore, Neuefeind, and
  Parise]{skinner2014structure}
Skinner,~L.~B.; Benmore,~C.; Neuefeind,~J.~C.; Parise,~J.~B. The structure of
  water around the compressibility minimum. \emph{The Journal of chemical
  physics} \textbf{2014}, \emph{141}\relax
\mciteBstWouldAddEndPuncttrue
\mciteSetBstMidEndSepPunct{\mcitedefaultmidpunct}
{\mcitedefaultendpunct}{\mcitedefaultseppunct}\relax
\EndOfBibitem
\bibitem[Gillan \latin{et~al.}(2016)Gillan, Alfè, and
  Michaelides]{Gillan_JCP2016}
Gillan,~M.~J.; Alfè,~D.; Michaelides,~A. {Perspective: How good is DFT for
  water?} \emph{The Journal of Chemical Physics} \textbf{2016}, \emph{144},
  130901\relax
\mciteBstWouldAddEndPuncttrue
\mciteSetBstMidEndSepPunct{\mcitedefaultmidpunct}
{\mcitedefaultendpunct}{\mcitedefaultseppunct}\relax
\EndOfBibitem
\bibitem[Mahoney and Jorgensen(2001)Mahoney, and Jorgensen]{mahoney2001quantum}
Mahoney,~M.~W.; Jorgensen,~W.~L. Quantum, intramolecular flexibility, and
  polarizability effects on the reproduction of the density anomaly of liquid
  water by simple potential functions. \emph{The Journal of Chemical Physics}
  \textbf{2001}, \emph{115}, 10758--10768\relax
\mciteBstWouldAddEndPuncttrue
\mciteSetBstMidEndSepPunct{\mcitedefaultmidpunct}
{\mcitedefaultendpunct}{\mcitedefaultseppunct}\relax
\EndOfBibitem
\bibitem[Kim \latin{et~al.}(2017)Kim, Sp{\"a}h, Pathak, Perakis, Mariedahl,
  Amann-Winkel, Sellberg, Lee, Kim, Park, \latin{et~al.} others]{kim2017maxima}
Kim,~K.~H.; Sp{\"a}h,~A.; Pathak,~H.; Perakis,~F.; Mariedahl,~D.;
  Amann-Winkel,~K.; Sellberg,~J.~A.; Lee,~J.~H.; Kim,~S.; Park,~J.,
  \latin{et~al.}  Maxima in the thermodynamic response and correlation
  functions of deeply supercooled water. \emph{Science} \textbf{2017},
  \emph{358}, 1589--1593\relax
\mciteBstWouldAddEndPuncttrue
\mciteSetBstMidEndSepPunct{\mcitedefaultmidpunct}
{\mcitedefaultendpunct}{\mcitedefaultseppunct}\relax
\EndOfBibitem
\bibitem[Medders and Paesani(2015)Medders, and Paesani]{medders2015infrared}
Medders,~G.~R.; Paesani,~F. Infrared and Raman spectroscopy of liquid water
  through “first-principles” many-body molecular dynamics. \emph{Journal of
  Chemical Theory and Computation} \textbf{2015}, \emph{11}, 1145--1154\relax
\mciteBstWouldAddEndPuncttrue
\mciteSetBstMidEndSepPunct{\mcitedefaultmidpunct}
{\mcitedefaultendpunct}{\mcitedefaultseppunct}\relax
\EndOfBibitem
\bibitem[Medders and Paesani(2015)Medders, and Paesani]{medders2015interplay}
Medders,~G.~R.; Paesani,~F. On the interplay of the potential energy and dipole
  moment surfaces in controlling the infrared activity of liquid water.
  \emph{The Journal of Chemical Physics} \textbf{2015}, \emph{142}\relax
\mciteBstWouldAddEndPuncttrue
\mciteSetBstMidEndSepPunct{\mcitedefaultmidpunct}
{\mcitedefaultendpunct}{\mcitedefaultseppunct}\relax
\EndOfBibitem
\bibitem[Medders and Paesani(2016)Medders, and Paesani]{medders2016dissecting}
Medders,~G.~R.; Paesani,~F. Dissecting the molecular structure of the air/water
  interface from quantum simulations of the sum-frequency generation spectrum.
  \emph{Journal of the American Chemical Society} \textbf{2016}, \emph{138},
  3912--3919\relax
\mciteBstWouldAddEndPuncttrue
\mciteSetBstMidEndSepPunct{\mcitedefaultmidpunct}
{\mcitedefaultendpunct}{\mcitedefaultseppunct}\relax
\EndOfBibitem
\bibitem[Riera \latin{et~al.}(2020)Riera, Hirales, Ghosh, and
  Paesani]{riera2020data}
Riera,~M.; Hirales,~A.; Ghosh,~R.; Paesani,~F. Data-driven many-body models
  with chemical accuracy for CH4/H2O mixtures. \emph{The Journal of Physical
  Chemistry B} \textbf{2020}, \emph{124}, 11207--11221\relax
\mciteBstWouldAddEndPuncttrue
\mciteSetBstMidEndSepPunct{\mcitedefaultmidpunct}
{\mcitedefaultendpunct}{\mcitedefaultseppunct}\relax
\EndOfBibitem
\bibitem[Robinson \latin{et~al.}(2022)Robinson, Ghosh, Egan, Riera, Knight,
  Paesani, and Hassanali]{robinson2022behavior}
Robinson,~V.~N.; Ghosh,~R.; Egan,~C.~K.; Riera,~M.; Knight,~C.; Paesani,~F.;
  Hassanali,~A. The behavior of methane--water mixtures under elevated
  pressures from simulations using many-body potentials. \emph{The Journal of
  Chemical Physics} \textbf{2022}, \emph{156}\relax
\mciteBstWouldAddEndPuncttrue
\mciteSetBstMidEndSepPunct{\mcitedefaultmidpunct}
{\mcitedefaultendpunct}{\mcitedefaultseppunct}\relax
\EndOfBibitem
\bibitem[Pruteanu \latin{et~al.}(2017)Pruteanu, Ackland, Poon, and
  Loveday]{pruteanu2017immiscible}
Pruteanu,~C.~G.; Ackland,~G.~J.; Poon,~W.~C.; Loveday,~J.~S. When immiscible
  becomes miscible—Methane in water at high pressures. \emph{Science
  advances} \textbf{2017}, \emph{3}, e1700240\relax
\mciteBstWouldAddEndPuncttrue
\mciteSetBstMidEndSepPunct{\mcitedefaultmidpunct}
{\mcitedefaultendpunct}{\mcitedefaultseppunct}\relax
\EndOfBibitem
\bibitem[Pruteanu \latin{et~al.}(2020)Pruteanu, Naden~Robinson, Ansari,
  Hassanali, Scandolo, and Loveday]{pruteanu2020squeezing}
Pruteanu,~C.~G.; Naden~Robinson,~V.; Ansari,~N.; Hassanali,~A.; Scandolo,~S.;
  Loveday,~J.~S. Squeezing oil into water under pressure: Inverting the
  hydrophobic effect. \emph{The Journal of Physical Chemistry Letters}
  \textbf{2020}, \emph{11}, 4826--4833\relax
\mciteBstWouldAddEndPuncttrue
\mciteSetBstMidEndSepPunct{\mcitedefaultmidpunct}
{\mcitedefaultendpunct}{\mcitedefaultseppunct}\relax
\EndOfBibitem
\bibitem[Eigen and De~Maeyer(1958)Eigen, and De~Maeyer]{eigen1958self}
Eigen,~M.; De~Maeyer,~L. Self-dissociation and protonic charge transport in
  water and ice. \emph{Proceedings of the Royal Society of London. Series A.
  Mathematical and Physical Sciences} \textbf{1958}, \emph{247}, 505--533\relax
\mciteBstWouldAddEndPuncttrue
\mciteSetBstMidEndSepPunct{\mcitedefaultmidpunct}
{\mcitedefaultendpunct}{\mcitedefaultseppunct}\relax
\EndOfBibitem
\bibitem[Eigen(1964)]{eigen1964proton}
Eigen,~M. Proton transfer, acid-base catalysis, and enzymatic hydrolysis. I.
  Elementary processes. \emph{Angewandte Chemie International Edition in
  English} \textbf{1964}, \emph{3}, 1--499\relax
\mciteBstWouldAddEndPuncttrue
\mciteSetBstMidEndSepPunct{\mcitedefaultmidpunct}
{\mcitedefaultendpunct}{\mcitedefaultseppunct}\relax
\EndOfBibitem
\bibitem[Thamer \latin{et~al.}(2015)Thamer, De~Marco, Ramasesha, Mandal, and
  Tokmakoff]{tokmakoff2015}
Thamer,~M.; De~Marco,~L.; Ramasesha,~K.; Mandal,~A.; Tokmakoff,~A. Ultrafast 2d
  Ir Spectroscopy of the Excess Proton in Liquid Water. \emph{Science}
  \textbf{2015}, \emph{350}, 78\relax
\mciteBstWouldAddEndPuncttrue
\mciteSetBstMidEndSepPunct{\mcitedefaultmidpunct}
{\mcitedefaultendpunct}{\mcitedefaultseppunct}\relax
\EndOfBibitem
\bibitem[Stillinger and Weber(1981)Stillinger, and
  Weber]{stillinger_computer_1981}
Stillinger,~F.~H.; Weber,~T.~A. Computer simulation of proton hydration
  dynamics. \emph{Chemical Physics Letters} \textbf{1981}, \emph{79},
  259--260\relax
\mciteBstWouldAddEndPuncttrue
\mciteSetBstMidEndSepPunct{\mcitedefaultmidpunct}
{\mcitedefaultendpunct}{\mcitedefaultseppunct}\relax
\EndOfBibitem
\bibitem[Stillinger and Weber(1982)Stillinger, and
  Weber]{stillinger_polarization_1982}
Stillinger,~F.~H.; Weber,~T.~A. Polarization model study of isotope effects in
  the gas phase hydronium–hydroxide neutralization reaction. \emph{The
  Journal of Chemical Physics} \textbf{1982}, \emph{76}, 4028--4036, \_eprint:
  https://pubs.aip.org/aip/jcp/article-pdf/76/8/4028/7365084/4028\_1\_online.pdf\relax
\mciteBstWouldAddEndPuncttrue
\mciteSetBstMidEndSepPunct{\mcitedefaultmidpunct}
{\mcitedefaultendpunct}{\mcitedefaultseppunct}\relax
\EndOfBibitem
\bibitem[Stillinger and David(2008)Stillinger, and
  David]{stillinger_polarization_2008}
Stillinger,~F.~H.; David,~C.~W. Polarization model for water and its ionic
  dissociation products. \emph{The Journal of Chemical Physics} \textbf{2008},
  \emph{69}, 1473--1484, \_eprint:
  https://pubs.aip.org/aip/jcp/article-pdf/69/4/1473/11141983/1473\_1\_online.pdf\relax
\mciteBstWouldAddEndPuncttrue
\mciteSetBstMidEndSepPunct{\mcitedefaultmidpunct}
{\mcitedefaultendpunct}{\mcitedefaultseppunct}\relax
\EndOfBibitem
\bibitem[Halley \latin{et~al.}(1993)Halley, Rustad, and
  Rahman]{halley_polarizable_1993}
Halley,~J.~W.; Rustad,~J.~R.; Rahman,~A. A polarizable, dissociating molecular
  dynamics model for liquid water. \emph{The Journal of Chemical Physics}
  \textbf{1993}, \emph{98}, 4110--4119, \_eprint:
  https://pubs.aip.org/aip/jcp/article-pdf/98/5/4110/14805771/4110\_1\_online.pdf\relax
\mciteBstWouldAddEndPuncttrue
\mciteSetBstMidEndSepPunct{\mcitedefaultmidpunct}
{\mcitedefaultendpunct}{\mcitedefaultseppunct}\relax
\EndOfBibitem
\bibitem[Ojamäe \latin{et~al.}(1998)Ojamäe, Shavitt, and
  Singer]{larssinger1998}
Ojamäe,~L.; Shavitt,~I.; Singer,~S.~J. {Potential models for simulations of
  the solvated proton in water}. \emph{The Journal of Chemical Physics}
  \textbf{1998}, \emph{109}, 5547--5564\relax
\mciteBstWouldAddEndPuncttrue
\mciteSetBstMidEndSepPunct{\mcitedefaultmidpunct}
{\mcitedefaultendpunct}{\mcitedefaultseppunct}\relax
\EndOfBibitem
\bibitem[Lee and Rasaiah(2010)Lee, and Rasaiah]{LeeRasiah2010}
Lee,~S.~H.; Rasaiah,~J.~C. Local dynamics and structure of the solvated
  hydroxide ion in water. \emph{Molecular Simulation} \textbf{2010}, \emph{36},
  69--73\relax
\mciteBstWouldAddEndPuncttrue
\mciteSetBstMidEndSepPunct{\mcitedefaultmidpunct}
{\mcitedefaultendpunct}{\mcitedefaultseppunct}\relax
\EndOfBibitem
\bibitem[Lee and Rasaiah(2011)Lee, and Rasaiah]{lee_proton_2011}
Lee,~S.~H.; Rasaiah,~J.~C. Proton transfer and the mobilities of the H+ and OH-
  ions from studies of a dissociating model for water. \emph{The Journal of
  Chemical Physics} \textbf{2011}, \emph{135}, 124505, \_eprint:
  https://pubs.aip.org/aip/jcp/article-pdf/doi/10.1063/1.3632990/15443072/124505\_1\_online.pdf\relax
\mciteBstWouldAddEndPuncttrue
\mciteSetBstMidEndSepPunct{\mcitedefaultmidpunct}
{\mcitedefaultendpunct}{\mcitedefaultseppunct}\relax
\EndOfBibitem
\bibitem[Lee and Rasaiah(2013)Lee, and Rasaiah]{lee_note_2013}
Lee,~S.~H.; Rasaiah,~J.~C. Note: Recombination of H+ and OH- ions along water
  wires. \emph{The Journal of Chemical Physics} \textbf{2013}, \emph{139},
  036102, \_eprint:
  https://pubs.aip.org/aip/jcp/article-pdf/doi/10.1063/1.4811294/15464423/036102\_1\_online.pdf\relax
\mciteBstWouldAddEndPuncttrue
\mciteSetBstMidEndSepPunct{\mcitedefaultmidpunct}
{\mcitedefaultendpunct}{\mcitedefaultseppunct}\relax
\EndOfBibitem
\bibitem[Mahadevan and Garofalini(2007)Mahadevan, and
  Garofalini]{mahadevangarofalini2007}
Mahadevan,~T.~S.; Garofalini,~S.~H. Dissociative Water Potential for Molecular
  Dynamics Simulations. \emph{The Journal of Physical Chemistry B}
  \textbf{2007}, \emph{111}, 8919--8927, PMID: 17604393\relax
\mciteBstWouldAddEndPuncttrue
\mciteSetBstMidEndSepPunct{\mcitedefaultmidpunct}
{\mcitedefaultendpunct}{\mcitedefaultseppunct}\relax
\EndOfBibitem
\bibitem[Garofalini and Lentz(2023)Garofalini, and Lentz]{Garofalini_JPCB2023}
Garofalini,~S.~H.; Lentz,~J. Subpicosecond Molecular Rearrangements Affect
  Local Electric Fields and Auto-Dissociation in Water. \emph{The Journal of
  Physical Chemistry B} \textbf{2023}, \emph{127}, 3392--3401, PMID:
  37036747\relax
\mciteBstWouldAddEndPuncttrue
\mciteSetBstMidEndSepPunct{\mcitedefaultmidpunct}
{\mcitedefaultendpunct}{\mcitedefaultseppunct}\relax
\EndOfBibitem
\bibitem[Wiedemair and Hofer(2017)Wiedemair, and Hofer]{wiedamair2017}
Wiedemair,~M.~J.; Hofer,~T.~S. Towards a dissociative SPC-like water model –
  probing the impact of intramolecular Coulombic contributions. \emph{Phys.
  Chem. Chem. Phys.} \textbf{2017}, \emph{19}, 31910--31920\relax
\mciteBstWouldAddEndPuncttrue
\mciteSetBstMidEndSepPunct{\mcitedefaultmidpunct}
{\mcitedefaultendpunct}{\mcitedefaultseppunct}\relax
\EndOfBibitem
\bibitem[Pavese \latin{et~al.}(1997)Pavese, Chawla, Lu, Lobaugh, and
  Voth]{PaveseChawlaLuLobaughVoth1997}
Pavese,~M.; Chawla,~S.; Lu,~D.; Lobaugh,~J.; Voth,~G.~A. Quantum effects and
  the excess proton in water. \emph{The Journal of Chemical Physics}
  \textbf{1997}, \emph{107}, 7428--7432\relax
\mciteBstWouldAddEndPuncttrue
\mciteSetBstMidEndSepPunct{\mcitedefaultmidpunct}
{\mcitedefaultendpunct}{\mcitedefaultseppunct}\relax
\EndOfBibitem
\bibitem[Pavese and Voth(1998)Pavese, and Voth]{PaveseVoth1998}
Pavese,~M.; Voth,~G.~A. Quantum and classical simulations of an excess proton
  in water. \emph{Berichte der Bunsengesellschaft für physikalische Chemie}
  \textbf{1998}, \emph{102}, 527--532\relax
\mciteBstWouldAddEndPuncttrue
\mciteSetBstMidEndSepPunct{\mcitedefaultmidpunct}
{\mcitedefaultendpunct}{\mcitedefaultseppunct}\relax
\EndOfBibitem
\bibitem[Lapid \latin{et~al.}(2005)Lapid, Agmon, Petersen, and
  Voth]{LapidAgmonPetersenVoth2005}
Lapid,~H.; Agmon,~N.; Petersen,~M.~K.; Voth,~G.~A. A bond-order analysis of the
  mechanism for hydrated proton mobility in liquid water. \emph{The Journal of
  Chemical Physics} \textbf{2005}, \emph{122}, 014506\relax
\mciteBstWouldAddEndPuncttrue
\mciteSetBstMidEndSepPunct{\mcitedefaultmidpunct}
{\mcitedefaultendpunct}{\mcitedefaultseppunct}\relax
\EndOfBibitem
\bibitem[Knight and Voth(2012)Knight, and Voth]{knight2012curious}
Knight,~C.; Voth,~G.~A. The curious case of the hydrated proton. \emph{Accounts
  of chemical research} \textbf{2012}, \emph{45}, 101--109\relax
\mciteBstWouldAddEndPuncttrue
\mciteSetBstMidEndSepPunct{\mcitedefaultmidpunct}
{\mcitedefaultendpunct}{\mcitedefaultseppunct}\relax
\EndOfBibitem
\bibitem[Wu \latin{et~al.}(2008)Wu, Chen, Wang, Paesani, and
  Voth]{WuChenWangPaesaniVoth2008}
Wu,~Y.; Chen,~H.; Wang,~F.; Paesani,~F.; Voth,~G.~A. An Improved Multistate
  Empirical Valence Bond Model for Aqueous Proton Solvation and Transport.
  \emph{The Journal of Physical Chemistry B} \textbf{2008}, \emph{112},
  467--482, PMID: 17999484\relax
\mciteBstWouldAddEndPuncttrue
\mciteSetBstMidEndSepPunct{\mcitedefaultmidpunct}
{\mcitedefaultendpunct}{\mcitedefaultseppunct}\relax
\EndOfBibitem
\bibitem[Markovitch \latin{et~al.}(2008)Markovitch, Chen, Izvekov, Paesani,
  Voth, and Agmon]{markovitch2008special}
Markovitch,~O.; Chen,~H.; Izvekov,~S.; Paesani,~F.; Voth,~G.~A.; Agmon,~N.
  Special pair dance and partner selection: Elementary steps in proton
  transport in liquid water. \emph{The Journal of Physical Chemistry B}
  \textbf{2008}, \emph{112}, 9456--9466\relax
\mciteBstWouldAddEndPuncttrue
\mciteSetBstMidEndSepPunct{\mcitedefaultmidpunct}
{\mcitedefaultendpunct}{\mcitedefaultseppunct}\relax
\EndOfBibitem
\bibitem[Marx \latin{et~al.}(2010)Marx, Chandra, and
  Tuckerman]{MarxChandraTuckerman2010}
Marx,~D.; Chandra,~A.; Tuckerman,~M.~E. Aqueous {B}asic {S}olutions:
  {H}ydroxide {S}olvation, {S}tructural {D}iffusion, and {C}omparison to the
  {H}ydrated {P}roton. \emph{Chemical Reviews} \textbf{2010}, \emph{110},
  2174--2216\relax
\mciteBstWouldAddEndPuncttrue
\mciteSetBstMidEndSepPunct{\mcitedefaultmidpunct}
{\mcitedefaultendpunct}{\mcitedefaultseppunct}\relax
\EndOfBibitem
\bibitem[Beattie \latin{et~al.}(2009)Beattie, Djerdjev, and
  Warr]{BeattieDjerdjevWarr2009}
Beattie,~J.~K.; Djerdjev,~A.~M.; Warr,~G.~G. The surface of neat water is
  basic. \emph{Faraday discussions} \textbf{2009}, \emph{141}, 31--39\relax
\mciteBstWouldAddEndPuncttrue
\mciteSetBstMidEndSepPunct{\mcitedefaultmidpunct}
{\mcitedefaultendpunct}{\mcitedefaultseppunct}\relax
\EndOfBibitem
\bibitem[Poli \latin{et~al.}(2020)Poli, Jong, and Hassanali]{poli2020charge}
Poli,~E.; Jong,~K.~H.; Hassanali,~A. Charge transfer as a ubiquitous mechanism
  in determining the negative charge at hydrophobic interfaces. \emph{Nature
  communications} \textbf{2020}, \emph{11}, 1--13\relax
\mciteBstWouldAddEndPuncttrue
\mciteSetBstMidEndSepPunct{\mcitedefaultmidpunct}
{\mcitedefaultendpunct}{\mcitedefaultseppunct}\relax
\EndOfBibitem
\bibitem[Tse \latin{et~al.}(2015)Tse, Chen, Lindberg, Kumar, and
  Voth]{TseLindbergKumarVoth2015}
Tse,~Y.-L.~S.; Chen,~C.; Lindberg,~G.~E.; Kumar,~R.; Voth,~G.~A. Propensity of
  Hydrated Excess Protons and Hydroxide Anions for the Air--Water Interface.
  \emph{Journal of the American Chemical Society} \textbf{2015}, \emph{137},
  12610--12616\relax
\mciteBstWouldAddEndPuncttrue
\mciteSetBstMidEndSepPunct{\mcitedefaultmidpunct}
{\mcitedefaultendpunct}{\mcitedefaultseppunct}\relax
\EndOfBibitem
\bibitem[Li \latin{et~al.}(2020)Li, Li, Wang, and Voth]{voth2020}
Li,~Z.; Li,~C.; Wang,~Z.; Voth,~G.~A. What Coordinate Best Describes the
  Affinity of the Hydrated Excess Proton for the Air–Water Interface?
  \emph{The Journal of Physical Chemistry B} \textbf{2020}, \emph{124},
  5039--5046, PMID: 32426982\relax
\mciteBstWouldAddEndPuncttrue
\mciteSetBstMidEndSepPunct{\mcitedefaultmidpunct}
{\mcitedefaultendpunct}{\mcitedefaultseppunct}\relax
\EndOfBibitem
\bibitem[Mundy \latin{et~al.}(2009)Mundy, Kuo, Tuckerman, Lee, and
  Tobias]{Mundy20092}
Mundy,~C.~J.; Kuo,~I.-F.~W.; Tuckerman,~M.~E.; Lee,~H.-S.; Tobias,~D.~J.
  Hydroxide anion at the air–water interface. \emph{Chemical Physics Letters}
  \textbf{2009}, \emph{481}, 2 -- 8\relax
\mciteBstWouldAddEndPuncttrue
\mciteSetBstMidEndSepPunct{\mcitedefaultmidpunct}
{\mcitedefaultendpunct}{\mcitedefaultseppunct}\relax
\EndOfBibitem
\bibitem[Baer \latin{et~al.}(2014)Baer, Kuo, Tobias, and Mundy]{baermundy2014}
Baer,~M.~D.; Kuo,~I.-F.~W.; Tobias,~D.~J.; Mundy,~C.~J. Toward a Unified
  Picture of the Water Self-Ions at the Air–Water Interface: A Density
  Functional Theory Perspective. \emph{The Journal of Physical Chemistry B}
  \textbf{2014}, \emph{118}, 8364--8372, PMID: 24762096\relax
\mciteBstWouldAddEndPuncttrue
\mciteSetBstMidEndSepPunct{\mcitedefaultmidpunct}
{\mcitedefaultendpunct}{\mcitedefaultseppunct}\relax
\EndOfBibitem
\bibitem[Rashid \latin{et~al.}(2023)Rashid, Rahman, Acter, and
  Uddin]{Rashid_PCCP23}
Rashid,~M. A.~M.; Rahman,~M.; Acter,~T.; Uddin,~N. Identify the Acidic or Basic
  Behavior of Surface Water: A QM/MM-MD Study. \emph{Phys. Chem. Chem. Phys.}
  \textbf{2023}, --\relax
\mciteBstWouldAddEndPuncttrue
\mciteSetBstMidEndSepPunct{\mcitedefaultmidpunct}
{\mcitedefaultendpunct}{\mcitedefaultseppunct}\relax
\EndOfBibitem
\bibitem[Bai and Herzfeld(2016)Bai, and Herzfeld]{baihertzfeld2016}
Bai,~C.; Herzfeld,~J. Surface Propensities of the Self-Ions of Water. \emph{ACS
  Central Science} \textbf{2016}, \emph{2}, 225--231, PMID: 27163053\relax
\mciteBstWouldAddEndPuncttrue
\mciteSetBstMidEndSepPunct{\mcitedefaultmidpunct}
{\mcitedefaultendpunct}{\mcitedefaultseppunct}\relax
\EndOfBibitem
\bibitem[Bai and Herzfeld(2017)Bai, and Herzfeld]{baihertzfeld2017}
Bai,~C.; Herzfeld,~J. Special Pairs Are Decisive in the Autoionization and
  Recombination of Water. \emph{The Journal of Physical Chemistry B}
  \textbf{2017}, \emph{121}, 4213--4219, PMID: 28381087\relax
\mciteBstWouldAddEndPuncttrue
\mciteSetBstMidEndSepPunct{\mcitedefaultmidpunct}
{\mcitedefaultendpunct}{\mcitedefaultseppunct}\relax
\EndOfBibitem
\bibitem[Mundy \latin{et~al.}(2009)Mundy, Kuo, Tuckerman, Lee, and
  Tobias]{MundyKuoTuckermanLeeTobias2009}
Mundy,~C.~J.; Kuo,~I.-F.~W.; Tuckerman,~M.~E.; Lee,~H.-S.; Tobias,~D.~J.
  Hydroxide anion at the air--water interface. \emph{Chemical Physics Letters}
  \textbf{2009}, \emph{481}, 2--8\relax
\mciteBstWouldAddEndPuncttrue
\mciteSetBstMidEndSepPunct{\mcitedefaultmidpunct}
{\mcitedefaultendpunct}{\mcitedefaultseppunct}\relax
\EndOfBibitem
\bibitem[Mamatkulov \latin{et~al.}(2017)Mamatkulov, Allolio, Netz, and
  Bonthuis]{jansjavkatnetz2017}
Mamatkulov,~S.~I.; Allolio,~C.; Netz,~R.~R.; Bonthuis,~D.~J.
  Orientation-Induced Adsorption of Hydrated Protons at the Air–Water
  Interface. \emph{Angewandte Chemie International Edition} \textbf{2017},
  \emph{56}, 15846--15851\relax
\mciteBstWouldAddEndPuncttrue
\mciteSetBstMidEndSepPunct{\mcitedefaultmidpunct}
{\mcitedefaultendpunct}{\mcitedefaultseppunct}\relax
\EndOfBibitem
\bibitem[van Duin \latin{et~al.}(2001)van Duin, Dasgupta, Lorant, and
  Goddard]{reaxff_2001}
van Duin,~A. C.~T.; Dasgupta,~S.; Lorant,~F.; Goddard,~W.~A. ReaxFF: A Reactive
  Force Field for Hydrocarbons. \emph{The Journal of Physical Chemistry A}
  \textbf{2001}, \emph{105}, 9396--9409\relax
\mciteBstWouldAddEndPuncttrue
\mciteSetBstMidEndSepPunct{\mcitedefaultmidpunct}
{\mcitedefaultendpunct}{\mcitedefaultseppunct}\relax
\EndOfBibitem
\bibitem[Chenoweth \latin{et~al.}(2008)Chenoweth, van Duin, and
  Goddard]{reaxff_2008}
Chenoweth,~K.; van Duin,~A. C.~T.; Goddard,~W.~A. ReaxFF Reactive Force Field
  for Molecular Dynamics Simulations of Hydrocarbon Oxidation. \emph{The
  Journal of Physical Chemistry A} \textbf{2008}, \emph{112}, 1040--1053, PMID:
  18197648\relax
\mciteBstWouldAddEndPuncttrue
\mciteSetBstMidEndSepPunct{\mcitedefaultmidpunct}
{\mcitedefaultendpunct}{\mcitedefaultseppunct}\relax
\EndOfBibitem
\bibitem[Senftle \latin{et~al.}(2016)Senftle, Hong, Islam, Kylasa, Zheng, Shin,
  Junkermeier, Engel-Herbert, Janik, Aktulga, Verstraelen, Grama, and van
  Duin]{reaxff_nature_review}
Senftle,~T.~P.; Hong,~S.; Islam,~M.~M.; Kylasa,~S.~B.; Zheng,~Y.; Shin,~Y.~K.;
  Junkermeier,~C.; Engel-Herbert,~R.; Janik,~M.~J.; Aktulga,~H.~M.;
  Verstraelen,~T.; Grama,~A.; van Duin,~A. C.~T. The ReaxFF reactive
  force-field: development, applications and future directions. \emph{npj
  Computational Materials} \textbf{2016}, \emph{2}, 15011\relax
\mciteBstWouldAddEndPuncttrue
\mciteSetBstMidEndSepPunct{\mcitedefaultmidpunct}
{\mcitedefaultendpunct}{\mcitedefaultseppunct}\relax
\EndOfBibitem
\bibitem[van Duin \latin{et~al.}(2013)van Duin, Zou, Joshi, Bryantsev, and
  Goddard]{vanDuin2013}
van Duin,~A. C.~T.; Zou,~C.; Joshi,~K.; Bryantsev,~V.; Goddard,~W.~A.
  \emph{Catalysis Series}; Royal Society of Chemistry, 2013\relax
\mciteBstWouldAddEndPuncttrue
\mciteSetBstMidEndSepPunct{\mcitedefaultmidpunct}
{\mcitedefaultendpunct}{\mcitedefaultseppunct}\relax
\EndOfBibitem
\bibitem[Asthagiri and Janik(2013)Asthagiri, and Janik]{reaxff_diffusion_book}
Asthagiri,~A.; Janik,~M.~J. \emph{{Computational Catalysis}}; The Royal Society
  of Chemistry, 2013\relax
\mciteBstWouldAddEndPuncttrue
\mciteSetBstMidEndSepPunct{\mcitedefaultmidpunct}
{\mcitedefaultendpunct}{\mcitedefaultseppunct}\relax
\EndOfBibitem
\bibitem[Zhang and van Duin(2017)Zhang, and van Duin]{reaxff2}
Zhang,~W.; van Duin,~A. C.~T. Second-Generation ReaxFF Water Force Field:
  Improvements in the Description of Water Density and OH-Anion Diffusion.
  \emph{The Journal of Physical Chemistry B} \textbf{2017}, \emph{121},
  6021--6032, PMID: 28570806\relax
\mciteBstWouldAddEndPuncttrue
\mciteSetBstMidEndSepPunct{\mcitedefaultmidpunct}
{\mcitedefaultendpunct}{\mcitedefaultseppunct}\relax
\EndOfBibitem
\bibitem[Raju \latin{et~al.}(2013)Raju, Kim, van Duin, and
  Fichthorn]{reaxff_solid_dissoc}
Raju,~M.; Kim,~S.-Y.; van Duin,~A. C.~T.; Fichthorn,~K.~A. ReaxFF Reactive
  Force Field Study of the Dissociation of Water on Titania Surfaces. \emph{The
  Journal of Physical Chemistry C} \textbf{2013}, \emph{117},
  10558--10572\relax
\mciteBstWouldAddEndPuncttrue
\mciteSetBstMidEndSepPunct{\mcitedefaultmidpunct}
{\mcitedefaultendpunct}{\mcitedefaultseppunct}\relax
\EndOfBibitem
\bibitem[Hornik \latin{et~al.}(1989)Hornik, Stinchcombe, and White]{HORNIK1989}
Hornik,~K.; Stinchcombe,~M.; White,~H. Multilayer feedforward networks are
  universal approximators. \emph{Neural Networks} \textbf{1989}, \emph{2},
  359--366\relax
\mciteBstWouldAddEndPuncttrue
\mciteSetBstMidEndSepPunct{\mcitedefaultmidpunct}
{\mcitedefaultendpunct}{\mcitedefaultseppunct}\relax
\EndOfBibitem
\bibitem[Behler and Parrinello(2007)Behler, and
  Parrinello]{behler2007generalized}
Behler,~J.; Parrinello,~M. Generalized neural-network representation of
  high-dimensional potential-energy surfaces. \emph{Physical review letters}
  \textbf{2007}, \emph{98}, 146401\relax
\mciteBstWouldAddEndPuncttrue
\mciteSetBstMidEndSepPunct{\mcitedefaultmidpunct}
{\mcitedefaultendpunct}{\mcitedefaultseppunct}\relax
\EndOfBibitem
\bibitem[Gastegger and Marquetand(2020)Gastegger, and
  Marquetand]{gastegger2020molecular}
Gastegger,~M.; Marquetand,~P. \emph{Molecular dynamics with neural network
  potentials}; Springer, 2020\relax
\mciteBstWouldAddEndPuncttrue
\mciteSetBstMidEndSepPunct{\mcitedefaultmidpunct}
{\mcitedefaultendpunct}{\mcitedefaultseppunct}\relax
\EndOfBibitem
\bibitem[No\'{e} \latin{et~al.}(2020)No\'{e}, Tkatchenko, M\"{u}ller, and
  Clementi]{Noe_2020}
No\'{e},~F.; Tkatchenko,~A.; M\"{u}ller,~K.-R.; Clementi,~C. Machine Learning
  for Molecular Simulation. \emph{Annual Review of Physical Chemistry}
  \textbf{2020}, \emph{71}, 361--390, PMID: 32092281\relax
\mciteBstWouldAddEndPuncttrue
\mciteSetBstMidEndSepPunct{\mcitedefaultmidpunct}
{\mcitedefaultendpunct}{\mcitedefaultseppunct}\relax
\EndOfBibitem
\bibitem[Kocer \latin{et~al.}(2022)Kocer, Ko, and Behler]{Behler_2022}
Kocer,~E.; Ko,~T.~W.; Behler,~J. Neural Network Potentials: A Concise Overview
  of Methods. \emph{Annual Review of Physical Chemistry} \textbf{2022},
  \emph{73}, 163--186, PMID: 34982580\relax
\mciteBstWouldAddEndPuncttrue
\mciteSetBstMidEndSepPunct{\mcitedefaultmidpunct}
{\mcitedefaultendpunct}{\mcitedefaultseppunct}\relax
\EndOfBibitem
\bibitem[Morawietz \latin{et~al.}(2012)Morawietz, Sharma, and
  Behler]{morawietzbehler2012}
Morawietz,~T.; Sharma,~V.; Behler,~J. {A neural network potential-energy
  surface for the water dimer based on environment-dependent atomic energies
  and charges}. \emph{The Journal of Chemical Physics} \textbf{2012},
  \emph{136}, 064103\relax
\mciteBstWouldAddEndPuncttrue
\mciteSetBstMidEndSepPunct{\mcitedefaultmidpunct}
{\mcitedefaultendpunct}{\mcitedefaultseppunct}\relax
\EndOfBibitem
\bibitem[Morawietz and Behler(2013)Morawietz, and Behler]{morawietzbehler2013}
Morawietz,~T.; Behler,~J. A Density-Functional Theory-Based Neural Network
  Potential for Water Clusters Including van der Waals Corrections. \emph{The
  Journal of Physical Chemistry A} \textbf{2013}, \emph{117}, 7356--7366, PMID:
  23557541\relax
\mciteBstWouldAddEndPuncttrue
\mciteSetBstMidEndSepPunct{\mcitedefaultmidpunct}
{\mcitedefaultendpunct}{\mcitedefaultseppunct}\relax
\EndOfBibitem
\bibitem[Morawietz \latin{et~al.}(2016)Morawietz, Singraber, Dellago, and
  Behler]{morawietzbehler2016}
Morawietz,~T.; Singraber,~A.; Dellago,~C.; Behler,~J. How van der Waals
  interactions determine the unique properties of water. \emph{Proceedings of
  the National Academy of Sciences} \textbf{2016}, \emph{113}, 8368--8373\relax
\mciteBstWouldAddEndPuncttrue
\mciteSetBstMidEndSepPunct{\mcitedefaultmidpunct}
{\mcitedefaultendpunct}{\mcitedefaultseppunct}\relax
\EndOfBibitem
\bibitem[Cheng \latin{et~al.}(2019)Cheng, Engel, Behler, Dellago, and
  Ceriotti]{chengceriottibehler2019}
Cheng,~B.; Engel,~E.~A.; Behler,~J.; Dellago,~C.; Ceriotti,~M. Ab initio
  thermodynamics of liquid and solid water. \emph{Proceedings of the National
  Academy of Sciences} \textbf{2019}, \emph{116}, 1110--1115\relax
\mciteBstWouldAddEndPuncttrue
\mciteSetBstMidEndSepPunct{\mcitedefaultmidpunct}
{\mcitedefaultendpunct}{\mcitedefaultseppunct}\relax
\EndOfBibitem
\bibitem[Zhang \latin{et~al.}(2018)Zhang, Han, Wang, Car, and E]{DeepNN2018Car}
Zhang,~L.; Han,~J.; Wang,~H.; Car,~R.; E,~W. Deep Potential Molecular Dynamics:
  A Scalable Model with the Accuracy of Quantum Mechanics. \emph{Phys. Rev.
  Lett.} \textbf{2018}, \emph{120}, 143001\relax
\mciteBstWouldAddEndPuncttrue
\mciteSetBstMidEndSepPunct{\mcitedefaultmidpunct}
{\mcitedefaultendpunct}{\mcitedefaultseppunct}\relax
\EndOfBibitem
\bibitem[Wang \latin{et~al.}(2018)Wang, Zhang, Han, Wang, Li, Zhu, Zhang, and
  Zhang]{wang2018deepmd}
Wang,~H.; Zhang,~L.; Han,~J.; Wang,~W.; Li,~T.; Zhu,~T.; Zhang,~J.;
  Zhang,~L.-W. DeePMD-kit: A deep learning package for many-body potential
  energy representation and molecular dynamics. \emph{Computer Physics
  Communications} \textbf{2018}, \emph{228}, 178--184\relax
\mciteBstWouldAddEndPuncttrue
\mciteSetBstMidEndSepPunct{\mcitedefaultmidpunct}
{\mcitedefaultendpunct}{\mcitedefaultseppunct}\relax
\EndOfBibitem
\bibitem[Zhang \latin{et~al.}(2021)Zhang, Zhang, and Wang]{zhang2021deepmd}
Zhang,~L.; Zhang,~L.-W.; Wang,~H. DP-GEN: A concurrent learning platform for
  the generation of reliable deep learning based potential energy models.
  \emph{Computer Physics Communications} \textbf{2021}, \emph{260},
  107788\relax
\mciteBstWouldAddEndPuncttrue
\mciteSetBstMidEndSepPunct{\mcitedefaultmidpunct}
{\mcitedefaultendpunct}{\mcitedefaultseppunct}\relax
\EndOfBibitem
\bibitem[Han \latin{et~al.}(2018)Han, Zhang, Car, and E]{han2018deep}
Han,~J.; Zhang,~L.; Car,~R.; E,~W. Deep potential: A general representation of
  a many-body potential energy surface. \emph{Communications in Computational
  Physics} \textbf{2018}, \emph{23}, 629--639\relax
\mciteBstWouldAddEndPuncttrue
\mciteSetBstMidEndSepPunct{\mcitedefaultmidpunct}
{\mcitedefaultendpunct}{\mcitedefaultseppunct}\relax
\EndOfBibitem
\bibitem[Zhang \latin{et~al.}(2020)Zhang, Chen, Wu, Wang, E, and
  Car]{zhangcar2020}
Zhang,~L.; Chen,~M.; Wu,~X.; Wang,~H.; E,~W.; Car,~R. Deep neural network for
  the dielectric response of insulators. \emph{Phys. Rev. B} \textbf{2020},
  \emph{102}, 041121\relax
\mciteBstWouldAddEndPuncttrue
\mciteSetBstMidEndSepPunct{\mcitedefaultmidpunct}
{\mcitedefaultendpunct}{\mcitedefaultseppunct}\relax
\EndOfBibitem
\bibitem[Sommers \latin{et~al.}(2020)Sommers, Calegari~Andrade, Zhang, Wang,
  and Car]{pccprobertocar2020}
Sommers,~G.~M.; Calegari~Andrade,~M.~F.; Zhang,~L.; Wang,~H.; Car,~R. Raman
  spectrum and polarizability of liquid water from deep neural networks.
  \emph{Phys. Chem. Chem. Phys.} \textbf{2020}, \emph{22}, 10592--10602\relax
\mciteBstWouldAddEndPuncttrue
\mciteSetBstMidEndSepPunct{\mcitedefaultmidpunct}
{\mcitedefaultendpunct}{\mcitedefaultseppunct}\relax
\EndOfBibitem
\bibitem[Zhang \latin{et~al.}(2021)Zhang, Wang, Car, and
  E]{deepmd_water_phase_diagram}
Zhang,~L.; Wang,~H.; Car,~R.; E,~W. Phase Diagram of a Deep Potential Water
  Model. \emph{Phys. Rev. Lett.} \textbf{2021}, \emph{126}, 236001\relax
\mciteBstWouldAddEndPuncttrue
\mciteSetBstMidEndSepPunct{\mcitedefaultmidpunct}
{\mcitedefaultendpunct}{\mcitedefaultseppunct}\relax
\EndOfBibitem
\bibitem[Gartner \latin{et~al.}(2020)Gartner, Zhang, Piaggi, Car,
  Panagiotopoulos, and Debenedetti]{gartnerLLCP2020}
Gartner,~T.~E.; Zhang,~L.; Piaggi,~P.~M.; Car,~R.; Panagiotopoulos,~A.~Z.;
  Debenedetti,~P.~G. Signatures of a liquid–liquid transition in an ab initio
  deep neural network model for water. \emph{Proceedings of the National
  Academy of Sciences} \textbf{2020}, \emph{117}, 26040--26046\relax
\mciteBstWouldAddEndPuncttrue
\mciteSetBstMidEndSepPunct{\mcitedefaultmidpunct}
{\mcitedefaultendpunct}{\mcitedefaultseppunct}\relax
\EndOfBibitem
\bibitem[Bore and Paesani(2023)Bore, and Paesani]{Bore2023}
Bore,~S.~L.; Paesani,~F. Realistic phase diagram of water from ``first
  principles'' data-driven quantum simulations. \emph{Nature Communications}
  \textbf{2023}, \emph{14}, 3349\relax
\mciteBstWouldAddEndPuncttrue
\mciteSetBstMidEndSepPunct{\mcitedefaultmidpunct}
{\mcitedefaultendpunct}{\mcitedefaultseppunct}\relax
\EndOfBibitem
\bibitem[Malosso \latin{et~al.}(2022)Malosso, Zhang, Car, Baroni, and
  Tisi]{Malosso2022}
Malosso,~C.; Zhang,~L.; Car,~R.; Baroni,~S.; Tisi,~D. Viscosity in water from
  first-principles and deep-neural-network simulations. \emph{npj Computational
  Materials} \textbf{2022}, \emph{8}, 139\relax
\mciteBstWouldAddEndPuncttrue
\mciteSetBstMidEndSepPunct{\mcitedefaultmidpunct}
{\mcitedefaultendpunct}{\mcitedefaultseppunct}\relax
\EndOfBibitem
\bibitem[Tisi \latin{et~al.}(2021)Tisi, Zhang, Bertossa, Wang, Car, and
  Baroni]{Baroni2021}
Tisi,~D.; Zhang,~L.; Bertossa,~R.; Wang,~H.; Car,~R.; Baroni,~S. Heat transport
  in liquid water from first-principles and deep neural network simulations.
  \emph{Phys. Rev. B} \textbf{2021}, \emph{104}, 224202\relax
\mciteBstWouldAddEndPuncttrue
\mciteSetBstMidEndSepPunct{\mcitedefaultmidpunct}
{\mcitedefaultendpunct}{\mcitedefaultseppunct}\relax
\EndOfBibitem
\bibitem[Kondati~Natarajan \latin{et~al.}(2015)Kondati~Natarajan, Morawietz,
  and Behler]{natarajan2015}
Kondati~Natarajan,~S.; Morawietz,~T.; Behler,~J. Representing the
  potential-energy surface of protonated water clusters by high-dimensional
  neural network potentials. \emph{Phys. Chem. Chem. Phys.} \textbf{2015},
  \emph{17}, 8356--8371\relax
\mciteBstWouldAddEndPuncttrue
\mciteSetBstMidEndSepPunct{\mcitedefaultmidpunct}
{\mcitedefaultendpunct}{\mcitedefaultseppunct}\relax
\EndOfBibitem
\bibitem[Schran \latin{et~al.}(2020)Schran, Behler, and Marx]{schranmarx2020}
Schran,~C.; Behler,~J.; Marx,~D. Automated Fitting of Neural Network Potentials
  at Coupled Cluster Accuracy: Protonated Water Clusters as Testing Ground.
  \emph{Journal of Chemical Theory and Computation} \textbf{2020}, \emph{16},
  88--99, PMID: 31743025\relax
\mciteBstWouldAddEndPuncttrue
\mciteSetBstMidEndSepPunct{\mcitedefaultmidpunct}
{\mcitedefaultendpunct}{\mcitedefaultseppunct}\relax
\EndOfBibitem
\bibitem[Hellström and Behler(2016)Hellström, and Behler]{matti2016}
Hellström,~M.; Behler,~J. Concentration-Dependent Proton Transfer Mechanisms
  in Aqueous NaOH Solutions: From Acceptor-Driven to Donor-Driven and Back.
  \emph{The Journal of Physical Chemistry Letters} \textbf{2016}, \emph{7},
  3302--3306, PMID: 27504986\relax
\mciteBstWouldAddEndPuncttrue
\mciteSetBstMidEndSepPunct{\mcitedefaultmidpunct}
{\mcitedefaultendpunct}{\mcitedefaultseppunct}\relax
\EndOfBibitem
\bibitem[Hellström \latin{et~al.}(2018)Hellström, Ceriotti, and
  Behler]{ceriottibehler2018}
Hellström,~M.; Ceriotti,~M.; Behler,~J. Nuclear Quantum Effects in Sodium
  Hydroxide Solutions from Neural Network Molecular Dynamics Simulations.
  \emph{The Journal of Physical Chemistry B} \textbf{2018}, \emph{122},
  10158--10171, PMID: 30335385\relax
\mciteBstWouldAddEndPuncttrue
\mciteSetBstMidEndSepPunct{\mcitedefaultmidpunct}
{\mcitedefaultendpunct}{\mcitedefaultseppunct}\relax
\EndOfBibitem
\bibitem[Atsango \latin{et~al.}(2023)Atsango, Morawietz, Marsalek, and
  Markland]{atsango_developing_2023}
Atsango,~A.~O.; Morawietz,~T.; Marsalek,~O.; Markland,~T.~E. Developing
  machine-learned potentials to simultaneously capture the dynamics of excess
  protons and hydroxide ions in classical and path integral simulations.
  \emph{The Journal of Chemical Physics} \textbf{2023}, \emph{159}, 074101,
  \_eprint:
  https://pubs.aip.org/aip/jcp/article-pdf/doi/10.1063/5.0162066/18085072/074101\_1\_5.0162066.pdf\relax
\mciteBstWouldAddEndPuncttrue
\mciteSetBstMidEndSepPunct{\mcitedefaultmidpunct}
{\mcitedefaultendpunct}{\mcitedefaultseppunct}\relax
\EndOfBibitem
\bibitem[Andrade \latin{et~al.}(2023)Andrade, Car, and Selloni]{marcos2023}
Andrade,~M.~C.; Car,~R.; Selloni,~A. Probing the self-ionization of liquid
  water with ab initio deep potential molecular dynamics. \emph{Proceedings of
  the National Academy of Sciences} \textbf{2023}, \emph{120},
  e2302468120\relax
\mciteBstWouldAddEndPuncttrue
\mciteSetBstMidEndSepPunct{\mcitedefaultmidpunct}
{\mcitedefaultendpunct}{\mcitedefaultseppunct}\relax
\EndOfBibitem
\bibitem[Chmiela \latin{et~al.}(2017)Chmiela, Tkatchenko, Sauceda, Poltavsky,
  Schütt, and Müller]{gdml}
Chmiela,~S.; Tkatchenko,~A.; Sauceda,~H.~E.; Poltavsky,~I.; Schütt,~K.~T.;
  Müller,~K.-R. Machine learning of accurate energy-conserving molecular force
  fields. \emph{Science Advances} \textbf{2017}, \emph{3}, e1603015\relax
\mciteBstWouldAddEndPuncttrue
\mciteSetBstMidEndSepPunct{\mcitedefaultmidpunct}
{\mcitedefaultendpunct}{\mcitedefaultseppunct}\relax
\EndOfBibitem
\bibitem[Maldonado \latin{et~al.}(2023)Maldonado, Poltavsky, Vassilev-Galindo,
  Tkatchenko, and Keith]{mbGDML}
Maldonado,~A.~M.; Poltavsky,~I.; Vassilev-Galindo,~V.; Tkatchenko,~A.;
  Keith,~J.~A. Modeling molecular ensembles with gradient-domain machine
  learning force fields. \emph{Digital Discovery} \textbf{2023}, \emph{2},
  871--880\relax
\mciteBstWouldAddEndPuncttrue
\mciteSetBstMidEndSepPunct{\mcitedefaultmidpunct}
{\mcitedefaultendpunct}{\mcitedefaultseppunct}\relax
\EndOfBibitem
\bibitem[Pietrucci(2017)]{pietrucci_strategies_2017}
Pietrucci,~F. Strategies for the exploration of free energy landscapes: {Unity}
  in diversity and challenges ahead. \emph{Reviews in Physics} \textbf{2017},
  \emph{2}, 32--45\relax
\mciteBstWouldAddEndPuncttrue
\mciteSetBstMidEndSepPunct{\mcitedefaultmidpunct}
{\mcitedefaultendpunct}{\mcitedefaultseppunct}\relax
\EndOfBibitem
\bibitem[Pietrucci(2018)]{pietrucci_novel_2018}
Pietrucci,~F. In \emph{Handbook of {Materials} {Modeling} : {Methods}: {Theory}
  and {Modeling}}; Andreoni,~W., Yip,~S., Eds.; Springer International
  Publishing: Cham, 2018; pp 1--23\relax
\mciteBstWouldAddEndPuncttrue
\mciteSetBstMidEndSepPunct{\mcitedefaultmidpunct}
{\mcitedefaultendpunct}{\mcitedefaultseppunct}\relax
\EndOfBibitem
\bibitem[Bussi and Laio(2020)Bussi, and Laio]{Bussi2020}
Bussi,~G.; Laio,~A. Using metadynamics to explore complex free-energy
  landscapes. \emph{Nature Reviews Physics} \textbf{2020}, \emph{2},
  200--212\relax
\mciteBstWouldAddEndPuncttrue
\mciteSetBstMidEndSepPunct{\mcitedefaultmidpunct}
{\mcitedefaultendpunct}{\mcitedefaultseppunct}\relax
\EndOfBibitem
\bibitem[Trout and Parrinello(1998)Trout, and
  Parrinello]{trout_dissociation_1998}
Trout,~B.~L.; Parrinello,~M. The dissociation mechanism of {H2O} in water
  studied by first-principles molecular dynamics. \emph{Chemical Physics
  Letters} \textbf{1998}, \emph{288}, 343--347\relax
\mciteBstWouldAddEndPuncttrue
\mciteSetBstMidEndSepPunct{\mcitedefaultmidpunct}
{\mcitedefaultendpunct}{\mcitedefaultseppunct}\relax
\EndOfBibitem
\bibitem[Trout and Parrinello(1999)Trout, and Parrinello]{troutparrinello1999}
Trout,~B.~L.; Parrinello,~M. Analysis of the Dissociation of H2O in Water Using
  First-Principles Molecular Dynamics. \emph{The Journal of Physical Chemistry
  B} \textbf{1999}, \emph{103}, 7340--7345\relax
\mciteBstWouldAddEndPuncttrue
\mciteSetBstMidEndSepPunct{\mcitedefaultmidpunct}
{\mcitedefaultendpunct}{\mcitedefaultseppunct}\relax
\EndOfBibitem
\bibitem[Sprik(2000)]{sprik_computation_2000}
Sprik,~M. Computation of the {pK} of liquid water using coordination
  constraints. \emph{Chemical Physics} \textbf{2000}, \emph{258},
  139--150\relax
\mciteBstWouldAddEndPuncttrue
\mciteSetBstMidEndSepPunct{\mcitedefaultmidpunct}
{\mcitedefaultendpunct}{\mcitedefaultseppunct}\relax
\EndOfBibitem
\bibitem[Wang \latin{et~al.}(2020)Wang, Carnevale, Klein, and
  Borguet]{borguet2020}
Wang,~R.; Carnevale,~V.; Klein,~M.~L.; Borguet,~E. First-Principles Calculation
  of Water pKa Using the Newly Developed SCAN Functional. \emph{The Journal of
  Physical Chemistry Letters} \textbf{2020}, \emph{11}, 54--59, PMID:
  31834803\relax
\mciteBstWouldAddEndPuncttrue
\mciteSetBstMidEndSepPunct{\mcitedefaultmidpunct}
{\mcitedefaultendpunct}{\mcitedefaultseppunct}\relax
\EndOfBibitem
\bibitem[Bolhuis \latin{et~al.}(2002)Bolhuis, Chandler, Dellago, and
  Geissler]{Bolhuischandlergeissler2002}
Bolhuis,~P.~G.; Chandler,~D.; Dellago,~C.; Geissler,~P.~L. TRANSITION PATH
  SAMPLING: Throwing Ropes Over Rough Mountain Passes, in the Dark.
  \emph{Annual Review of Physical Chemistry} \textbf{2002}, \emph{53},
  291--318, PMID: 11972010\relax
\mciteBstWouldAddEndPuncttrue
\mciteSetBstMidEndSepPunct{\mcitedefaultmidpunct}
{\mcitedefaultendpunct}{\mcitedefaultseppunct}\relax
\EndOfBibitem
\bibitem[Geissler \latin{et~al.}(2001)Geissler, Dellago, Chandler, Hutter, and
  Parrinello]{geisslerparrinello2001}
Geissler,~P.~L.; Dellago,~C.; Chandler,~D.; Hutter,~J.; Parrinello,~M.
  Autoionization in Liquid Water. \emph{Science} \textbf{2001}, \emph{291},
  2121--2124\relax
\mciteBstWouldAddEndPuncttrue
\mciteSetBstMidEndSepPunct{\mcitedefaultmidpunct}
{\mcitedefaultendpunct}{\mcitedefaultseppunct}\relax
\EndOfBibitem
\bibitem[Park \latin{et~al.}(2006)Park, Laio, Iannuzzi, and
  Parrinello]{parkparrinello2006}
Park,~J.~M.; Laio,~A.; Iannuzzi,~M.; Parrinello,~M. Dissociation Mechanism of
  Acetic Acid in Water. \emph{Journal of the American Chemical Society}
  \textbf{2006}, \emph{128}, 11318--11319, PMID: 16939231\relax
\mciteBstWouldAddEndPuncttrue
\mciteSetBstMidEndSepPunct{\mcitedefaultmidpunct}
{\mcitedefaultendpunct}{\mcitedefaultseppunct}\relax
\EndOfBibitem
\bibitem[Cuny and Hassanali(2014)Cuny, and Hassanali]{cunyhassanali2014}
Cuny,~J.; Hassanali,~A.~A. Ab Initio Molecular Dynamics Study of the Mechanism
  of Proton Recombination with a Weak Base. \emph{The Journal of Physical
  Chemistry B} \textbf{2014}, \emph{118}, 13903--13912, PMID: 25415885\relax
\mciteBstWouldAddEndPuncttrue
\mciteSetBstMidEndSepPunct{\mcitedefaultmidpunct}
{\mcitedefaultendpunct}{\mcitedefaultseppunct}\relax
\EndOfBibitem
\bibitem[Hassanali \latin{et~al.}(2011)Hassanali, Prakash, Eshet, and
  Parrinello]{HassanaliPrakashEshetParrinello2011}
Hassanali,~A.; Prakash,~M.~K.; Eshet,~H.; Parrinello,~M. On the recombination
  of hydronium and hydroxide ions in water. \emph{Proceedings of the National
  Academy of Sciences} \textbf{2011}, \emph{108}, 20410--20415\relax
\mciteBstWouldAddEndPuncttrue
\mciteSetBstMidEndSepPunct{\mcitedefaultmidpunct}
{\mcitedefaultendpunct}{\mcitedefaultseppunct}\relax
\EndOfBibitem
\bibitem[Cassone \latin{et~al.}(2014)Cassone, Giaquinta, Saija, and
  Saitta]{Cassone_JPCB2014}
Cassone,~G.; Giaquinta,~P.~V.; Saija,~F.; Saitta,~A.~M. Proton Conduction in
  Water Ices under an Electric Field. \emph{The Journal of Physical Chemistry
  B} \textbf{2014}, \emph{118}, 4419--4424, PMID: 24689531\relax
\mciteBstWouldAddEndPuncttrue
\mciteSetBstMidEndSepPunct{\mcitedefaultmidpunct}
{\mcitedefaultendpunct}{\mcitedefaultseppunct}\relax
\EndOfBibitem
\bibitem[Iuchi \latin{et~al.}(2009)Iuchi, Chen, Paesani, and
  Voth]{vothpaesani2009}
Iuchi,~S.; Chen,~H.; Paesani,~F.; Voth,~G.~A. Hydrated Excess Proton at
  Water-Hydrophobic Interfaces. \emph{The Journal of Physical Chemistry B}
  \textbf{2009}, \emph{113}, 4017--4030\relax
\mciteBstWouldAddEndPuncttrue
\mciteSetBstMidEndSepPunct{\mcitedefaultmidpunct}
{\mcitedefaultendpunct}{\mcitedefaultseppunct}\relax
\EndOfBibitem
\bibitem[Hassanali \latin{et~al.}(2014)Hassanali, Giberti, Sosso, and
  Parrinello]{HassanaliGibertiSossoParrinello2014}
Hassanali,~A.~A.; Giberti,~F.; Sosso,~G.~C.; Parrinello,~M. The role of the
  umbrella inversion mode in proton diffusion. \emph{Chemical Physics Letters}
  \textbf{2014}, \emph{599}, 133 -- 138\relax
\mciteBstWouldAddEndPuncttrue
\mciteSetBstMidEndSepPunct{\mcitedefaultmidpunct}
{\mcitedefaultendpunct}{\mcitedefaultseppunct}\relax
\EndOfBibitem
\bibitem[Glielmo \latin{et~al.}(2021)Glielmo, Husic, Rodriguez, Clementi,
  No{\'e}, and Laio]{glielmo2021unsupervised}
Glielmo,~A.; Husic,~B.~E.; Rodriguez,~A.; Clementi,~C.; No{\'e},~F.; Laio,~A.
  Unsupervised Learning Methods for Molecular Simulation Data. \emph{Chemical
  Reviews} \textbf{2021}, \emph{121}, 9722--9758\relax
\mciteBstWouldAddEndPuncttrue
\mciteSetBstMidEndSepPunct{\mcitedefaultmidpunct}
{\mcitedefaultendpunct}{\mcitedefaultseppunct}\relax
\EndOfBibitem
\bibitem[Glielmo \latin{et~al.}(2022)Glielmo, Zeni, Cheng, Cs{\'a}nyi, and
  Laio]{glielmo2022ranking}
Glielmo,~A.; Zeni,~C.; Cheng,~B.; Cs{\'a}nyi,~G.; Laio,~A. Ranking the
  information content of distance measures. \emph{PNAS Nexus} \textbf{2022},
  \emph{1}, pgac039\relax
\mciteBstWouldAddEndPuncttrue
\mciteSetBstMidEndSepPunct{\mcitedefaultmidpunct}
{\mcitedefaultendpunct}{\mcitedefaultseppunct}\relax
\EndOfBibitem
\bibitem[Moqadam \latin{et~al.}(2018)Moqadam, Lervik, Riccardi, Venkatraman,
  Alsberg, and van Erp]{vanErp2018}
Moqadam,~M.; Lervik,~A.; Riccardi,~E.; Venkatraman,~V.; Alsberg,~B.~K.; van
  Erp,~T.~S. Local initiation conditions for water autoionization.
  \emph{Proceedings of the National Academy of Sciences} \textbf{2018},
  \emph{115}, E4569--E4576\relax
\mciteBstWouldAddEndPuncttrue
\mciteSetBstMidEndSepPunct{\mcitedefaultmidpunct}
{\mcitedefaultendpunct}{\mcitedefaultseppunct}\relax
\EndOfBibitem
\bibitem[Pietrucci and Andreoni(2011)Pietrucci, and
  Andreoni]{prlpietrucciandreoni2011}
Pietrucci,~F.; Andreoni,~W. Graph Theory Meets Ab Initio Molecular Dynamics:
  Atomic Structures and Transformations at the Nanoscale. \emph{Phys. Rev.
  Lett.} \textbf{2011}, \emph{107}, 085504\relax
\mciteBstWouldAddEndPuncttrue
\mciteSetBstMidEndSepPunct{\mcitedefaultmidpunct}
{\mcitedefaultendpunct}{\mcitedefaultseppunct}\relax
\EndOfBibitem
\bibitem[Pietrucci and Saitta(2015)Pietrucci, and Saitta]{pietruccisaitta2015}
Pietrucci,~F.; Saitta,~A.~M. Formamide reaction network in gas phase and
  solution via a unified theoretical approach: Toward a reconciliation of
  different prebiotic scenarios. \emph{Proceedings of the National Academy of
  Sciences} \textbf{2015}, \emph{112}, 15030--15035\relax
\mciteBstWouldAddEndPuncttrue
\mciteSetBstMidEndSepPunct{\mcitedefaultmidpunct}
{\mcitedefaultendpunct}{\mcitedefaultseppunct}\relax
\EndOfBibitem
\bibitem[Ismail \latin{et~al.}(2022)Ismail, Chantreau~Majerus, and
  Habershon]{ismail2022}
Ismail,~I.; Chantreau~Majerus,~R.; Habershon,~S. Graph-Driven Reaction
  Discovery: Progress, Challenges, and Future Opportunities. \emph{The Journal
  of Physical Chemistry A} \textbf{2022}, \emph{126}, 7051--7069, PMID:
  36190262\relax
\mciteBstWouldAddEndPuncttrue
\mciteSetBstMidEndSepPunct{\mcitedefaultmidpunct}
{\mcitedefaultendpunct}{\mcitedefaultseppunct}\relax
\EndOfBibitem
\bibitem[Cassone \latin{et~al.}(2018)Cassone, Sponer, Sponer, Pietrucci,
  Saitta, and Saija]{cassone_chemcomm2018}
Cassone,~G.; Sponer,~J.; Sponer,~J.~E.; Pietrucci,~F.; Saitta,~A.~M.; Saija,~F.
  Synthesis of (d)-erythrose from glycolaldehyde aqueous solutions under
  electric field. \emph{Chem. Commun.} \textbf{2018}, \emph{54},
  3211--3214\relax
\mciteBstWouldAddEndPuncttrue
\mciteSetBstMidEndSepPunct{\mcitedefaultmidpunct}
{\mcitedefaultendpunct}{\mcitedefaultseppunct}\relax
\EndOfBibitem
\bibitem[Bart{\'o}k \latin{et~al.}(2010)Bart{\'o}k, Payne, Kondor, and
  Cs{\'a}nyi]{bartok2010gaussian}
Bart{\'o}k,~A.~P.; Payne,~M.~C.; Kondor,~R.; Cs{\'a}nyi,~G. Gaussian
  approximation potentials: The accuracy of quantum mechanics, without the
  electrons. \emph{Physical review letters} \textbf{2010}, \emph{104},
  136403\relax
\mciteBstWouldAddEndPuncttrue
\mciteSetBstMidEndSepPunct{\mcitedefaultmidpunct}
{\mcitedefaultendpunct}{\mcitedefaultseppunct}\relax
\EndOfBibitem
\bibitem[De \latin{et~al.}(2016)De, Bart{\'o}k, Cs{\'a}nyi, and
  Ceriotti]{debartokceriotti2016}
De,~S.; Bart{\'o}k,~A.~P.; Cs{\'a}nyi,~G.; Ceriotti,~M. Comparing molecules and
  solids across structural and alchemical space. \emph{Physical Chemistry
  Chemical Physics} \textbf{2016}, \emph{18}, 13754--13769\relax
\mciteBstWouldAddEndPuncttrue
\mciteSetBstMidEndSepPunct{\mcitedefaultmidpunct}
{\mcitedefaultendpunct}{\mcitedefaultseppunct}\relax
\EndOfBibitem
\bibitem[Monserrat \latin{et~al.}(2020)Monserrat, Brandenburg, Engel, and
  Cheng]{monserrat2020liquid}
Monserrat,~B.; Brandenburg,~J.~G.; Engel,~E.~A.; Cheng,~B. Liquid water
  contains the building blocks of diverse ice phases. \emph{Nature
  communications} \textbf{2020}, \emph{11}, 1--8\relax
\mciteBstWouldAddEndPuncttrue
\mciteSetBstMidEndSepPunct{\mcitedefaultmidpunct}
{\mcitedefaultendpunct}{\mcitedefaultseppunct}\relax
\EndOfBibitem
\bibitem[Capelli \latin{et~al.}(2022)Capelli, Muniz-Miranda, and
  Pavan]{capelli_ephemeral_2022}
Capelli,~R.; Muniz-Miranda,~F.; Pavan,~G.~M. Ephemeral ice-like local
  environments in classical rigid models of liquid water. \emph{The Journal of
  Chemical Physics} \textbf{2022}, \emph{156}, 214503, \_eprint:
  https://pubs.aip.org/aip/jcp/article-pdf/doi/10.1063/5.0088599/16544848/214503\_1\_online.pdf\relax
\mciteBstWouldAddEndPuncttrue
\mciteSetBstMidEndSepPunct{\mcitedefaultmidpunct}
{\mcitedefaultendpunct}{\mcitedefaultseppunct}\relax
\EndOfBibitem
\bibitem[Ansari \latin{et~al.}(2018)Ansari, Dandekar, Caravati, Sosso, and
  Hassanali]{ansari2018}
Ansari,~N.; Dandekar,~R.; Caravati,~S.; Sosso,~G.; Hassanali,~A. High and low
  density patches in simulated liquid water. \emph{The Journal of Chemical
  Physics} \textbf{2018}, \emph{149}, 204507\relax
\mciteBstWouldAddEndPuncttrue
\mciteSetBstMidEndSepPunct{\mcitedefaultmidpunct}
{\mcitedefaultendpunct}{\mcitedefaultseppunct}\relax
\EndOfBibitem
\bibitem[Ansari \latin{et~al.}(2019)Ansari, Laio, and
  Hassanali]{ansari2019spontaneously}
Ansari,~N.; Laio,~A.; Hassanali,~A. Spontaneously forming dendritic voids in
  liquid water can host small polymers. \emph{The journal of physical chemistry
  letters} \textbf{2019}, \emph{10}, 5585--5591\relax
\mciteBstWouldAddEndPuncttrue
\mciteSetBstMidEndSepPunct{\mcitedefaultmidpunct}
{\mcitedefaultendpunct}{\mcitedefaultseppunct}\relax
\EndOfBibitem
\bibitem[Azizi \latin{et~al.}(2022)Azizi, Laio, and Hassanali]{azizi2022model}
Azizi,~K.; Laio,~A.; Hassanali,~A. Model Folded Hydrophobic Polymers Reside in
  Highly Branched Voids. \emph{The Journal of Physical Chemistry Letters}
  \textbf{2022}, \emph{13}, 183--189\relax
\mciteBstWouldAddEndPuncttrue
\mciteSetBstMidEndSepPunct{\mcitedefaultmidpunct}
{\mcitedefaultendpunct}{\mcitedefaultseppunct}\relax
\EndOfBibitem
\bibitem[Azizi \latin{et~al.}(2023)Azizi, Laio, and
  Hassanali]{azizi2023solvation}
Azizi,~K.; Laio,~A.; Hassanali,~A. Solvation thermodynamics from cavity shapes
  of amino acids. \emph{PNAS nexus} \textbf{2023}, \emph{2}, pgad239\relax
\mciteBstWouldAddEndPuncttrue
\mciteSetBstMidEndSepPunct{\mcitedefaultmidpunct}
{\mcitedefaultendpunct}{\mcitedefaultseppunct}\relax
\EndOfBibitem
\bibitem[Jong and Hassanali(2018)Jong, and Hassanali]{jong2018data}
Jong,~K.; Hassanali,~A.~A. A data science approach to understanding water
  networks around biomolecules: the case of tri-alanine in liquid water.
  \emph{The Journal of Physical Chemistry B} \textbf{2018}, \emph{122},
  7895--7906\relax
\mciteBstWouldAddEndPuncttrue
\mciteSetBstMidEndSepPunct{\mcitedefaultmidpunct}
{\mcitedefaultendpunct}{\mcitedefaultseppunct}\relax
\EndOfBibitem
\bibitem[Donkor \latin{et~al.}(0)Donkor, Laio, and
  Hassanali]{laiodonkorhassanali2023}
Donkor,~E.~D.; Laio,~A.; Hassanali,~A. Do Machine-Learning Atomic Descriptors
  and Order Parameters Tell the Same Story? The Case of Liquid Water.
  \emph{Journal of Chemical Theory and Computation} \textbf{0}, \emph{0}, null,
  PMID: 36920997\relax
\mciteBstWouldAddEndPuncttrue
\mciteSetBstMidEndSepPunct{\mcitedefaultmidpunct}
{\mcitedefaultendpunct}{\mcitedefaultseppunct}\relax
\EndOfBibitem
\bibitem[Offei-Danso \latin{et~al.}(2022)Offei-Danso, Hassanali, and
  Rodriguez]{offei2022high}
Offei-Danso,~A.; Hassanali,~A.; Rodriguez,~A. High-Dimensional Fluctuations in
  Liquid Water: Combining Chemical Intuition with Unsupervised Learning.
  \emph{Journal of Chemical Theory and Computation} \textbf{2022}, \emph{18},
  3136--3150\relax
\mciteBstWouldAddEndPuncttrue
\mciteSetBstMidEndSepPunct{\mcitedefaultmidpunct}
{\mcitedefaultendpunct}{\mcitedefaultseppunct}\relax
\EndOfBibitem
\bibitem[Di~Pino \latin{et~al.}(2023)Di~Pino, Donkor, Sánchez, Rodriguez,
  Cassone, Scherlis, and Hassanali]{hassanali2023}
Di~Pino,~S.; Donkor,~E.~D.; Sánchez,~V.~M.; Rodriguez,~A.; Cassone,~G.;
  Scherlis,~D.; Hassanali,~A. ZundEig: The Structure of the Proton in Liquid
  Water from Unsupervised Learning. \emph{The Journal of Physical Chemistry B}
  \textbf{2023}, \emph{127}, 9822--9832, PMID: 37930954\relax
\mciteBstWouldAddEndPuncttrue
\mciteSetBstMidEndSepPunct{\mcitedefaultmidpunct}
{\mcitedefaultendpunct}{\mcitedefaultseppunct}\relax
\EndOfBibitem
\bibitem[van~der Vegt \latin{et~al.}(2016)van~der Vegt, Haldrup, Roke, Zheng,
  Lund, and Bakker]{bakker2016}
van~der Vegt,~N. F.~A.; Haldrup,~K.; Roke,~S.; Zheng,~J.; Lund,~M.;
  Bakker,~H.~J. Water-Mediated Ion Pairing: Occurrence and Relevance.
  \emph{Chemical Reviews} \textbf{2016}, \emph{116}, 7626--7641, PMID:
  27153482\relax
\mciteBstWouldAddEndPuncttrue
\mciteSetBstMidEndSepPunct{\mcitedefaultmidpunct}
{\mcitedefaultendpunct}{\mcitedefaultseppunct}\relax
\EndOfBibitem
\bibitem[Futera \latin{et~al.}(2020)Futera, Tse, and
  English]{English_SciAdv2020}
Futera,~Z.; Tse,~J.~S.; English,~N.~J. Possibility of realizing superionic ice
  VII in external electric fields of planetary bodies. \emph{Science Advances}
  \textbf{2020}, \emph{6}, eaaz2915\relax
\mciteBstWouldAddEndPuncttrue
\mciteSetBstMidEndSepPunct{\mcitedefaultmidpunct}
{\mcitedefaultendpunct}{\mcitedefaultseppunct}\relax
\EndOfBibitem
\bibitem[Saitta \latin{et~al.}(2012)Saitta, Saija, and
  Giaquinta]{Saitta_PRL2012}
Saitta,~A.~M.; Saija,~F.; Giaquinta,~P.~V. Ab Initio Molecular Dynamics Study
  of Dissociation of Water under an Electric Field. \emph{Phys. Rev. Lett.}
  \textbf{2012}, \emph{108}, 207801\relax
\mciteBstWouldAddEndPuncttrue
\mciteSetBstMidEndSepPunct{\mcitedefaultmidpunct}
{\mcitedefaultendpunct}{\mcitedefaultseppunct}\relax
\EndOfBibitem
\bibitem[Cassone \latin{et~al.}(2017)Cassone, Creazzo, Giaquinta, Sponer, and
  Saija]{Cassone_PCCP17}
Cassone,~G.; Creazzo,~F.; Giaquinta,~P.~V.; Sponer,~J.; Saija,~F. Ionic
  diffusion and proton transfer in aqueous solutions of alkali metal salts.
  \emph{Phys. Chem. Chem. Phys.} \textbf{2017}, \emph{19}, 20420--20429\relax
\mciteBstWouldAddEndPuncttrue
\mciteSetBstMidEndSepPunct{\mcitedefaultmidpunct}
{\mcitedefaultendpunct}{\mcitedefaultseppunct}\relax
\EndOfBibitem
\bibitem[Shafiei \latin{et~al.}(2019)Shafiei, von Domaros, Bratko, and
  Luzar]{Shafiei_JCP19}
Shafiei,~M.; von Domaros,~M.; Bratko,~D.; Luzar,~A. Anisotropic structure and
  dynamics of water under static electric fields. \emph{The Journal of Chemical
  Physics} \textbf{2019}, \emph{150}, 074505\relax
\mciteBstWouldAddEndPuncttrue
\mciteSetBstMidEndSepPunct{\mcitedefaultmidpunct}
{\mcitedefaultendpunct}{\mcitedefaultseppunct}\relax
\EndOfBibitem
\bibitem[Futera and English(2017)Futera, and English]{Futera_JCP17}
Futera,~Z.; English,~N.~J. {Communication: Influence of external static and
  alternating electric fields on water from long-time non-equilibrium ab initio
  molecular dynamics}. \emph{The Journal of Chemical Physics} \textbf{2017},
  \emph{147}, 031102\relax
\mciteBstWouldAddEndPuncttrue
\mciteSetBstMidEndSepPunct{\mcitedefaultmidpunct}
{\mcitedefaultendpunct}{\mcitedefaultseppunct}\relax
\EndOfBibitem
\bibitem[Conti~Nibali \latin{et~al.}(2023)Conti~Nibali, Maiti, Saija, Heyden,
  and Cassone]{ContiNibali_JCP2023}
Conti~Nibali,~V.; Maiti,~S.; Saija,~F.; Heyden,~M.; Cassone,~G. {Electric-field
  induced entropic effects in liquid water}. \emph{The Journal of Chemical
  Physics} \textbf{2023}, \emph{158}, 184501\relax
\mciteBstWouldAddEndPuncttrue
\mciteSetBstMidEndSepPunct{\mcitedefaultmidpunct}
{\mcitedefaultendpunct}{\mcitedefaultseppunct}\relax
\EndOfBibitem
\bibitem[Ojha and Kühne(2023)Ojha, and Kühne]{Kuhne_PCCP2023}
Ojha,~D.; Kühne,~T.~D. Vibrational dynamics of liquid water in an external
  electric field. \emph{Phys. Chem. Chem. Phys.} \textbf{2023}, \emph{25},
  13442--13451\relax
\mciteBstWouldAddEndPuncttrue
\mciteSetBstMidEndSepPunct{\mcitedefaultmidpunct}
{\mcitedefaultendpunct}{\mcitedefaultseppunct}\relax
\EndOfBibitem
\bibitem[Cassone and Martelli(2024)Cassone, and
  Martelli]{cassone2023electrofreezing}
Cassone,~G.; Martelli,~F. Electrofreezing of liquid water at ambient
  conditions. \emph{Nature Communications} \textbf{2024}, \emph{15}, 1856\relax
\mciteBstWouldAddEndPuncttrue
\mciteSetBstMidEndSepPunct{\mcitedefaultmidpunct}
{\mcitedefaultendpunct}{\mcitedefaultseppunct}\relax
\EndOfBibitem
\bibitem[Persson and Halle(2015)Persson, and Halle]{perssonhalle2015}
Persson,~F.; Halle,~B. How amide hydrogens exchange in native proteins.
  \emph{Proceedings of the National Academy of Sciences} \textbf{2015},
  \emph{112}, 10383--10388\relax
\mciteBstWouldAddEndPuncttrue
\mciteSetBstMidEndSepPunct{\mcitedefaultmidpunct}
{\mcitedefaultendpunct}{\mcitedefaultseppunct}\relax
\EndOfBibitem
\bibitem[Zhang and Jiang(2023)Zhang, and Jiang]{FIREANN}
Zhang,~Y.; Jiang,~B. Universal machine learning for the response of atomistic
  systems to external fields. \emph{Nature Communications} \textbf{2023},
  \emph{14}, 6424\relax
\mciteBstWouldAddEndPuncttrue
\mciteSetBstMidEndSepPunct{\mcitedefaultmidpunct}
{\mcitedefaultendpunct}{\mcitedefaultseppunct}\relax
\EndOfBibitem
\bibitem[Dral and Barbatti(2021)Dral, and Barbatti]{dral_molecular_2021}
Dral,~P.~O.; Barbatti,~M. Molecular excited states through a machine learning
  lens. \emph{Nature Reviews Chemistry} \textbf{2021}, \emph{5}, 388--405\relax
\mciteBstWouldAddEndPuncttrue
\mciteSetBstMidEndSepPunct{\mcitedefaultmidpunct}
{\mcitedefaultendpunct}{\mcitedefaultseppunct}\relax
\EndOfBibitem
\bibitem[Molinero and Moore(2009)Molinero, and Moore]{mw_2009}
Molinero,~V.; Moore,~E.~B. Water Modeled As an Intermediate Element between
  Carbon and Silicon. \emph{The Journal of Physical Chemistry B} \textbf{2009},
  \emph{113}, 4008--4016\relax
\mciteBstWouldAddEndPuncttrue
\mciteSetBstMidEndSepPunct{\mcitedefaultmidpunct}
{\mcitedefaultendpunct}{\mcitedefaultseppunct}\relax
\EndOfBibitem
\bibitem[Moore and Molinero(2011)Moore, and Molinero]{nature_ice}
Moore,~E.~B.; Molinero,~V. Structural transformation in supercooled water
  controls the crystallization rate of ice. \emph{Nature} \textbf{2011},
  \emph{479}, 506–508\relax
\mciteBstWouldAddEndPuncttrue
\mciteSetBstMidEndSepPunct{\mcitedefaultmidpunct}
{\mcitedefaultendpunct}{\mcitedefaultseppunct}\relax
\EndOfBibitem
\bibitem[Holten \latin{et~al.}(2013)Holten, Limmer, Molinero, and
  Anisimov]{mw_LL}
Holten,~V.; Limmer,~D.~T.; Molinero,~V.; Anisimov,~M.~A. {Nature of the
  anomalies in the supercooled liquid state of the mW model of water}.
  \emph{The Journal of Chemical Physics} \textbf{2013}, \emph{138},
  174501\relax
\mciteBstWouldAddEndPuncttrue
\mciteSetBstMidEndSepPunct{\mcitedefaultmidpunct}
{\mcitedefaultendpunct}{\mcitedefaultseppunct}\relax
\EndOfBibitem
\bibitem[Xu and Molinero(2011)Xu, and Molinero]{mw_LL_conf}
Xu,~L.; Molinero,~V. Is There a Liquid–Liquid Transition in Confined Water?
  \emph{The Journal of Physical Chemistry B} \textbf{2011}, \emph{115},
  14210--14216\relax
\mciteBstWouldAddEndPuncttrue
\mciteSetBstMidEndSepPunct{\mcitedefaultmidpunct}
{\mcitedefaultendpunct}{\mcitedefaultseppunct}\relax
\EndOfBibitem
\bibitem[Moore \latin{et~al.}(2010)Moore, de~la Llave, Welke, Scherlis, and
  Molinero]{mw_nanoice}
Moore,~E.~B.; de~la Llave,~E.; Welke,~K.; Scherlis,~D.~A.; Molinero,~V.
  Freezing{,} melting and structure of ice in a hydrophilic nanopore.
  \emph{Phys. Chem. Chem. Phys.} \textbf{2010}, \emph{12}, 4124--4134\relax
\mciteBstWouldAddEndPuncttrue
\mciteSetBstMidEndSepPunct{\mcitedefaultmidpunct}
{\mcitedefaultendpunct}{\mcitedefaultseppunct}\relax
\EndOfBibitem
\bibitem[Factorovich \latin{et~al.}(2015)Factorovich, Molinero, and
  Scherlis]{mw_hydrosurface}
Factorovich,~M.~H.; Molinero,~V.; Scherlis,~D.~A. Hydrogen-Bond Heterogeneity
  Boosts Hydrophobicity of Solid Interfaces. \emph{Journal of the American
  Chemical Society} \textbf{2015}, \emph{137}, 10618--10623\relax
\mciteBstWouldAddEndPuncttrue
\mciteSetBstMidEndSepPunct{\mcitedefaultmidpunct}
{\mcitedefaultendpunct}{\mcitedefaultseppunct}\relax
\EndOfBibitem
\bibitem[Factorovich \latin{et~al.}(2014)Factorovich, Molinero, and
  Scherlis]{mw_vp}
Factorovich,~M.~H.; Molinero,~V.; Scherlis,~D.~A. {A simple grand canonical
  approach to compute the vapor pressure of bulk and finite size systems}.
  \emph{The Journal of Chemical Physics} \textbf{2014}, \emph{140},
  064111\relax
\mciteBstWouldAddEndPuncttrue
\mciteSetBstMidEndSepPunct{\mcitedefaultmidpunct}
{\mcitedefaultendpunct}{\mcitedefaultseppunct}\relax
\EndOfBibitem
\bibitem[Gadea \latin{et~al.}(2020)Gadea, Perez~Sirkin, Molinero, and
  Scherlis]{mw_bubbles}
Gadea,~E.~D.; Perez~Sirkin,~Y.~A.; Molinero,~V.; Scherlis,~D.~A.
  Electrochemically Generated Nanobubbles: Invariance of the Current with
  Respect to Electrode Size and Potential. \emph{The Journal of Physical
  Chemistry Letters} \textbf{2020}, \emph{11}, 6573--6579\relax
\mciteBstWouldAddEndPuncttrue
\mciteSetBstMidEndSepPunct{\mcitedefaultmidpunct}
{\mcitedefaultendpunct}{\mcitedefaultseppunct}\relax
\EndOfBibitem
\bibitem[Jacobson \latin{et~al.}(2010)Jacobson, Hujo, and
  Molinero]{mw_clathrates}
Jacobson,~L.~C.; Hujo,~W.; Molinero,~V. Amorphous Precursors in the Nucleation
  of Clathrate Hydrates. \emph{Journal of the American Chemical Society}
  \textbf{2010}, \emph{132}, 11806--11811\relax
\mciteBstWouldAddEndPuncttrue
\mciteSetBstMidEndSepPunct{\mcitedefaultmidpunct}
{\mcitedefaultendpunct}{\mcitedefaultseppunct}\relax
\EndOfBibitem
\bibitem[DeMille and Molinero(2009)DeMille, and Molinero]{mw_ions}
DeMille,~R.~C.; Molinero,~V. Coarse-grained ions without charges: Reproducing
  the solvation structure of NaCl in water using short-ranged potentials.
  \emph{The Journal of Chemical Physics} \textbf{2009}, \emph{131},
  034107\relax
\mciteBstWouldAddEndPuncttrue
\mciteSetBstMidEndSepPunct{\mcitedefaultmidpunct}
{\mcitedefaultendpunct}{\mcitedefaultseppunct}\relax
\EndOfBibitem
\bibitem[Perez~Sirkin \latin{et~al.}(2016)Perez~Sirkin, Factorovich, Molinero,
  and Scherlis]{mw_vp_electrolyte}
Perez~Sirkin,~Y.~A.; Factorovich,~M.~H.; Molinero,~V.; Scherlis,~D.~A. Vapor
  Pressure of Aqueous Solutions of Electrolytes Reproduced with Coarse-Grained
  Models without Electrostatics. \emph{Journal of Chemical Theory and
  Computation} \textbf{2016}, \emph{12}, 2942--2949\relax
\mciteBstWouldAddEndPuncttrue
\mciteSetBstMidEndSepPunct{\mcitedefaultmidpunct}
{\mcitedefaultendpunct}{\mcitedefaultseppunct}\relax
\EndOfBibitem
\bibitem[Lu \latin{et~al.}(2019)Lu, Barnett, and Molinero]{mw_membrane}
Lu,~J.; Barnett,~A.; Molinero,~V. Effect of Polymer Architecture on the
  Nanophase Segregation, Ionic Conductivity, and Electro-Osmotic Drag of Anion
  Exchange Membranes. \emph{The Journal of Physical Chemistry C} \textbf{2019},
  \emph{123}, 8717--8726\relax
\mciteBstWouldAddEndPuncttrue
\mciteSetBstMidEndSepPunct{\mcitedefaultmidpunct}
{\mcitedefaultendpunct}{\mcitedefaultseppunct}\relax
\EndOfBibitem
\bibitem[Mochizuki and Molinero(2018)Mochizuki, and Molinero]{mw_antifreeze}
Mochizuki,~K.; Molinero,~V. Antifreeze Glycoproteins Bind Reversibly to Ice via
  Hydrophobic Groups. \emph{Journal of the American Chemical Society}
  \textbf{2018}, \emph{140}, 4803--4811\relax
\mciteBstWouldAddEndPuncttrue
\mciteSetBstMidEndSepPunct{\mcitedefaultmidpunct}
{\mcitedefaultendpunct}{\mcitedefaultseppunct}\relax
\EndOfBibitem
\bibitem[Chan \latin{et~al.}(2019)Chan, Cherukara, Narayanan, Loeffler,
  Benmore, Gray, and Sankaranarayanan]{cg_ML}
Chan,~H.; Cherukara,~M.~J.; Narayanan,~B.; Loeffler,~T.~D.; Benmore,~C.;
  Gray,~S.~K.; Sankaranarayanan,~S. K. R.~S. Machine learning coarse grained
  models for water. \emph{Nat. Commun.} \textbf{2019}, \emph{10}\relax
\mciteBstWouldAddEndPuncttrue
\mciteSetBstMidEndSepPunct{\mcitedefaultmidpunct}
{\mcitedefaultendpunct}{\mcitedefaultseppunct}\relax
\EndOfBibitem
\end{mcitethebibliography}

\end{document}